\documentclass[aps,pra,twocolumn,amsfonts,amssymb,amsmath]{revtex4}
\usepackage{graphicx} 
\usepackage{array}
\usepackage{epic}
\usepackage{eepic}
\usepackage{SIunits}
\newcommand{\alphaR}[0]{\alpha_{\mathrm{R}}}
\newcommand{\alphaI}[0]{\alpha_{\mathrm{I}}}
\newcommand{\dhole}[0]{d_{\mathrm{hole}}}
\newcommand{\dholec}[0]{d_{\mathrm{hole,c}}}
\newcommand{\dholes}[0]{d_{\mathrm{hole,s}}}
\newcommand{\dholei}[0]{d_{\mathrm{hole},i}}
\newcommand{\Gammah}[0]{\Gamma_{\mathrm{h}}}
\newcommand{\Gammahole}[0]{\Gamma_{\mathrm{hole}}}
\newcommand{\Gammaholec}[0]{\Gamma_{\mathrm{hole,c}}}
\newcommand{\Gammaholes}[0]{\Gamma_{\mathrm{hole,s}}}
\newcommand{\Gammaholei}[0]{\Gamma_{\mathrm{hole},i}}
\newcommand{\Gammainh}[0]{\Gamma_{\mathrm{inh}}}
\newcommand{\omegam}[0]{\omega_{\mathrm{m}}}
\renewcommand{\Re}[0]{\mathrm{Re}}
\renewcommand{\Im}[0]{\mathrm{Im}}
\newcommand{\ket}[1]{\left|#1\right>}
\newcommand{\bra}[1]{\left<#1\right|}
\newcommand{\ketbra}[2]{\left|#1\right>\negthickspace\left<#2\right|}
\newcommand{\kete}[0]{\left|\mathrm{e}\right>}
\newcommand{\ketg}[0]{\left|\mathrm{g}\right>}
\newcommand{\ketr}[0]{\left|\mathrm{r}\right>}
\newcommand{\ketrone}[0]{\left|\mathrm{r1}\right>}
\newcommand{\ketrtwo}[0]{\left|\mathrm{r2}\right>}
\newcommand{\ketc}[0]{\left|\mathrm{c}\right>}
\newcommand{\ketf}[0]{\left|\mathrm{f}\right>}
\newcommand{\rhoe}[0]{\rho_{\mathrm{e}}}
\newcommand{\rhog}[0]{\rho_{\mathrm{g}}}
\newcommand{\rhor}[0]{\rho_{\mathrm{r}}}
\newcommand{\beg}[0]{b_{\mathrm{eg}}}
\newcommand{\ber}[0]{b_{\mathrm{er}}}
\newcommand{\berone}[0]{b_{\mathrm{er1}}}
\newcommand{\bertwo}[0]{b_{\mathrm{er2}}}
\newcommand{\bec}[0]{b_{\mathrm{ec}}}
\newcommand{\bef}[0]{b_{\mathrm{ef}}}
\newcommand{\Trg}[0]{T_{\mathrm{rg}}}
\newcommand{\Trone}[0]{T_{\mathrm{r1}}}
\newcommand{\Trtwo}[0]{T_{\mathrm{r2}}}
\newcommand{\Tgr}[0]{T_{\mathrm{gr}}}
\newcommand{\Tcg}[0]{T_{\mathrm{cg}}}
\newcommand{\Tfc}[0]{T_{\mathrm{fc}}}
\newcommand{\Tcf}[0]{T_{\mathrm{cf}}}
\renewcommand{\vec}[1]{\mathbf{#1}}
\newcommand{\etac}[0]{\eta_{\mathrm{c}}}
\newcommand{\etas}[0]{\eta_{\mathrm{s}}}
\newcommand{\etapl}[0]{\eta_+}
\newcommand{\etami}[0]{\eta_-}
\newcommand{\phic}[0]{\phi_{\mathrm{c}}}
\newcommand{\phis}[0]{\phi_{\mathrm{s}}}
\newcommand{\phipl}[0]{\phi_+}
\newcommand{\phimi}[0]{\phi_-}
\newcommand{\PrYSO}[0]{Pr$^{3+}$:Y$_2$SiO$_5\,$}
\newcommand{\ErYSO}[0]{Er$^{3+}$:Y$_2$SiO$_5\,$}
\newcommand{\EuYSO}[0]{Eu$^{3+}$:Y$_2$SiO$_5\,$}
\newcommand{\TmYAG}[0]{Tm$^{3+}$:Y$_3$Al$_5$O$_{12}\,$}
\newcommand{\figfont}[0]{\fontsize{6}{6pt}\selectfont}
\newcommand{\std}[0]{\mathrm{std}}
\newcommand{\var}[0]{\mathrm{Var}}
\newcommand{\abs}[0]{\mathrm{abs}}
\newcommand{\psif}[0]{\psi_{\mathrm{f}}}
\newcommand{\densf}[0]{\hat{\rho}_{\mathrm{f}}}

\begin{document}

\title{Laser stabilization using spectral hole burning}

\author{L.~Rippe, B.~Julsgaard, A.~Walther, S.~Kr\"{o}ll}
\affiliation{Department of Physics, Lund Institute of Technology,
  P.O.~Box 118, SE-22100 Lund, Sweden}

\date{\today }

\begin{abstract}
  We have frequency stabilized a Coherent CR699-21 dye laser to a
  transient spectral hole on the 606 nm transition in \PrYSO. A
  frequency stability of 1 kHz has been obtained on the 10 {\micro}s
  timescale together with a long-term frequency drift below 1 kHz/s. 
  RF magnetic fields are used to repopulate the hyperfine levels
  allowing us to control the dynamics of the spectral hole. A detailed
  theory of the atomic response to laser frequency errors has been
  developed which allows us to design and optimize the laser
  stabilization feedback loop, and specifically we give a stability
  criterion that must be fulfilled in order to obtain very low drift
  rates. The laser stability is sufficient for performing quantum
  gate experiments in \PrYSO. 
\end{abstract}

\pacs{42.60.-v, 32.80.-t, 42.50.Md, 06.30.Ft}

\maketitle

\hyphenation{theo-rem}


\section{Introduction}
\label{sec:introduction}
Quantum information processing and the ability to experimentally
manipulate quantum systems has evolved significantly over the past
decade, and the search for new and potentially scalable qubit systems
is ongoing. One of these schemes employs rare-earth-ion-doped crystals
\cite{Ohlsson2002, Roos2004, Longdell2004, Longdell2004a,
  Wesenberg2006} where, among others, the materials \PrYSO and \EuYSO
are good candidates with very good coherence properties
\cite{Equall1994, Equall1995, Fraval2004, Fraval2005}.  Qubits are
represented as hyperfine levels $\ket{0}$ and $\ket{1}$, and they are
manipulated using an optical transition to an excited level $\kete$.
This optical manipulation requires very stable lasers in order to
obtain high-fidelity gate operations. In order not to be limited by
laser performance, the laser must remain phase coherent on a timescale
comparable to the optical coherence time, $T_2$, of the atomic
material.  Our experimental focus is on \PrYSO which has an optical
transition at 605.977 nm with coherence time as long as $T_2 = 152$
{\micro}s \cite{Equall1995}, which is accessible with dye lasers only.
The narrow optical transitions are hidden inside a much broader
inhomogeneous profile, but they can be isolated by optical pumping
methods \cite{Nilsson2004, Rippe2005}, taking advantage of the very
long hyperfine level lifetime (being $\approx$100 seconds in \PrYSO
and much longer in a moderate magnetic field). These initialization
techniques employ several cycles of population transfer with
subsequent relaxation and may last for times of the order of several
hundreds of milliseconds.  This also places some long-term stability
requirements on the laser performance.

Frequency stabilization of lasers is a science in itself. Previously,
dye lasers have been stabilized using optical cavities
\cite{Drever1983, Hough1984, Helmcke1987, Houssin1988, Kallenbach1989,
  Steiner1989, Zhu1992, Zhu1993, Sellars1995, Young1999}, and recently
the method using spectral hole burning has been demonstrated with
semi-conductor lasers \cite{Sellin1999, Strickland2000, Pryde2001,
  Pryde2002, Sellin2001, Bottger2001, Bottger2003} and with a
Ti:sapphire laser \cite{Merkel2000}. We demonstrate for the first time
a dye laser stabilized to a spectral hole. The method employing
spectral holes has a number of advantages over optical cavities.
Firstly, the sensitivity to vibrations requires that the atomic medium
moves much less than an optical wavelength in an optical coherence
time.  However, for a high-finesse cavity this sensitivity is
essentially multiplied by the average number of round trips made by a
photon in the cavity, which can be several orders of magnitude. In
addition, using the same hole-burning crystal for laser stabilization
and for further experiments may reduce the sensitivity to vibrations
significantly.  Secondly, to make a phase-coherent laser on the
timescale of $T_2$ requires a physical reference system with a phase
memory of comparable duration. Hence, when performing experiments on
e.g.~\PrYSO, with $T_2 = 152$ {\micro}s, the material itself will
automatically be sufficient for laser stabilization. A similar round
trip time in a cavity is possible, but technically very challenging. A
cavity also requires mode matching, and optical isolators are required
to prevent the directly back-reflected field from entering the laser
cavity.  The main drawbacks of spectral holes include the requirement
of cryogenic cooling, and for transient hole systems the long-term
stability presents a challenge since the spectral hole position may
change over time.

In the following sections we present a detailed theory of the dynamics
of spectral holes and on the response of the atomic medium to errors
in the laser frequency. This helped us to design the feedback hardware
and in realizing the laser stabilization experimentally. In doing so,
we also introduced an ``RF-eraser'', which consists of RF magnetic
fields controlling the hyperfine level lifetimes. In this manner, we
can adjust and optimize the spectral hole dynamics for best laser
performance. We also demonstrate experimental support for the
theoretical calculations with emphasis on the problem of laser drift. 
The RF magnetic fields are very useful in this context.  Apart from
the theoretical understanding, it is also very important to build
fast, low-noise, and low-offset feedback electronics.  We have
designed analog electronic circuits using the best components
available on the market for the fastest parts of the regulation
system. The most important design considerations are given in this
paper, while the technical details are available elsewhere
\cite{Rippe2006, GroupHomepage}.


\section{Theoretical description of ion locking}
\label{sec:theory}
The entire theoretical understanding of laser stabilization to
spectral holes in inhomogeneous broadened transitions is one of the
key results of this paper and is introduced in this section. Section
\ref{sec:basic_idea} presents simple arguments to explain the basic
idea and to compare the procedure of locking to spectral holes with
the method of locking to optical cavities. We then proceed to
Secs.~\ref{sec:2-level-Maxwell-Bloch}
and~\ref{sec:2-level-with-reservior} where we introduce the
Maxwell-Bloch equations, written in a form convenient for
inhomogeneously broadened transitions and then calculate the
propagation effects of laser light through the atomic material in the
presence of spectral holes. With these results we are able to
calculate the error signals suitable for the laser stabilization
scheme using the Pound-Drever-Hall method \cite{Drever1983} in
Sec.~\ref{sec:error-signals}. This section provides a general
understanding of the response of atoms to errors in the laser
frequency, which will put us in a position to design the electronics
hardware for the stabilization feedback loop.  In
Sec.~\ref{sec:laser-drift} we analyze the possibility of linear drift
errors and their prevention, and finally in
Sec.~\ref{sec:general_remarks_error_signal} we summarize our
theoretical calculations.

The intention of this paper is to enable the reader to understand the
details of laser stabilization to spectral holes with special emphasis
on the feedback system design. Experimentally, we focus on a
particular atomic system consisting of ions in a \PrYSO crystal.
Instead of developing a general theory we make a few approximations
suitable for this particular system. However, we present the
calculations in such detail that it should be clear how to extend the
theory to other systems.

In order to maintain the physical understanding we will restrict
ourselves to analytical derivations and make approximations, rather
than numerical simulations, when the calculations become difficult.
Our theory is quantitatively accurate for many practical systems.

For readers not interested in all the technical details we will now
summarize the most important theoretical results. After the simple
discussion of laser stabilization schemes in Sec.~\ref{sec:basic_idea}
we proceed to Sec.~\ref{sec:2-level-Maxwell-Bloch} where we state in
Eqs.~(\ref{eq:MB_linear_Fourier}-\ref{eq:MB_linear_phase}) the
important propagation equations for different frequency components of
the laser light through matter. In
Sec.~\ref{sec:2-level-with-reservior} these are applied to the special
case of a spectral hole in a broad inhomogeneous profile in
Eqs.~\eqref{eq:alphaR_hole} and~\eqref{eq:alphaI_hole} which, together
with the hole shape parameters of
Eqs.~(\ref{eq:def_dhole_gammahole}-\ref{eq:def_R_G_three-level}), form
the workhorse for all calculations of error signals. In
Sec.~\ref{sec:error-signals} the error signals are calculated; the
main result being the transfer function~\eqref{eq:Transfer_function}
connecting errors in laser frequency to modulations in the detected
laser power. These calculations also indicate the optimum parameter
choices for best performance. To mention one example, the modulation
index $m = 0.56$ in general optimizes the magnitude of the error
signal which is different from the optimum result $m=1.08$ known from
cavity locking \cite{Black2001}. However, additional constraints on
the modulation index may be necessary when we take the possibility of
laser drift into account in the rather technical
section~\ref{sec:laser-drift}.  Here we calculate the
correction~\eqref{eq:error_signal_drift} to the error signal if the
laser is drifting linearly with time. This specifically leads to a
criterion~\eqref{eq:drift_criterion} which must be fulfilled in order
to prevent laser drift. A practical version of this criterion is
illustrated in Fig.~\ref{fig:thresholds}.

\subsection{Stabilization - an overview}
\label{sec:basic_idea}
Before entering into the detailed theoretical description of laser
locking to spectral holes, we will give a general picture of the
stabilization mechanisms. We discuss the concept of frequency locking
and phase locking, and to this end we start by briefly discussing how
stabilization to a cavity works. An extensive and pedagogical
description can be found in \cite{Black2001}.

In stabilization setups with optical cavities the laser light is sent
into a high-finesse cavity and the back-reflected light is detected.
The aim is to maintain the laser frequency at the cavity resonance at
all times, in which case the reflected field is ideally zero (all
light is transmitted through the cavity). Let us assume that the
cavity mirrors reflect 99.99\% of the light intensity. The fact that
no light is reflected in the ideal case is due to the fact that the
immediately reflected laser field at the incoming mirror interferes
destructively with the 0.01\% leakage of the $10^4$ times more intense
intra-cavity field.

Let us now assume that the incident light suddenly changes phase on a
timescale much shorter than the cavity build-up time, $\tau =
\frac{L}{Tc}$ (here $T = 0.01\%$ is the mirror transmission, $c$ is
the speed of light, and $L$ the cavity length. For $L = 10$ cm we have
$\tau \approx 3$ {\micro}s). The immediately reflected beam then
changes its phase and does not interfere destructively with the cavity
leakage field, which retains its original phase for a time of the
order of $\tau$. The intensity is then non-zero and an error can be
detected. The immediate phase difference (including the sign) of the
two fields can be measured using the Pound-Drever-Hall method
\cite{Drever1983}. If the phase change persists for a time longer than
$\tau$ the reflected field returns to zero.

We also consider the case where the frequency is changed from the
cavity resonance. On a timescale longer than $\tau$ this field will
build up inside the cavity, but since the resonance condition is not
met, the cavity length is not equal to an exact integer multiple of
$\lambda/2$, where $\lambda$ is the wave length of the light. Hence,
each round trip in the cavity is slightly advanced or delayed in
phase, and the total contribution will be phase shifted, with a sign
depending on the sign of the frequency error. When this field leaks
out of the cavity and interferes with the immediately reflected field,
we can again, in a similar fashion to above, measure the frequency
error (including the sign). Note, that the non-zero reflection
persists over time and is not limited to a duration $\tau$.

The above two examples illustrate the fact that on long timescales the
detection and correction of frequency errors is possible, while phase
errors can only be detected for a limited time given by the
\emph{phase memory time} $\tau$. We shall now argue that this picture
also holds for locking to spectral holes. In this scenario the laser
light is incident on an inhomogeneously broadened transition and the
laser will burn a spectral hole in the absorbing background. To some
extent, this spectral hole can be thought of as the inverse of the
pedagogically simpler case with no background absorption and the
presence of atoms at a single frequency only. We adopt this picture
below.

It is well known that when scanning a laser across an atomic resonance
the absorption will change (proportionally to
$\frac{\Gammah^2}{4}/(\frac{\Gammah^2}{4}+\Delta^2)$ where $\Gammah$
is the full width half maximum (FWHM) line-width and $\Delta$ the
detuning). At the same time, the index of refraction will change
proportionally to
$\Delta\frac{\Gammah}{2}/(\frac{\Gammah^2}{4}+\Delta^2)$ which, for
small detunings $\Delta$, causes a phase shift of the light
proportional to the detuning. This effect can be detected by methods
similar to the Pound-Drever-Hall method \cite{Drever1983} and is used
for measuring frequency errors.

The above picture describes a steady state which requires
modifications on short time scales. If the phase of the incoming light
changes abruptly, it takes of the order of the optical coherence time
$T_2 = \frac{2}{\Gammah}$ before the atoms have reached a new steady
state ($\Gammah$ is here measured in rad/sec). Meanwhile, the atoms
keep radiating at the original phase, which allows the phase error to
be measured but only for a limited time, $T_2$.

We see that the methods of optical cavities and spectral holes are
very similar. In a common picture we can think of both as frequency
filters where the phase memory is set by the inverse line-width and
the filtering gives rise to slowly varying fields. The possibility of
detecting phase errors relies on the interference between an
immediately changing field and the slowly varying field. For the
optical cavity the immediately changing field is the directly
reflected beam, while for spectral holes it is the remaining part of
the incoming light transmitted through the atoms. The slowly varying
field is the intra-cavity field in the case of optical cavities, while
it is the light emitted coherently by atoms in the case of spectral
holes. The Pound-Drever-Hall method \cite{Drever1983} allows for the
detection of the above mentioned interference in both cases, but there
is one important difference. The modulation sidebands required for
this scheme are not stored in the optical cavity (they are not
resonant) and hence do not experience the frequency filtering.  For
the method of spectral holes these sidebands create side holes which
must be taken into account to understand the complete picture.

\subsection{Two-level atoms and Maxwell-Bloch equations}
\label{sec:2-level-Maxwell-Bloch}
For the detailed theoretical calculations we start with an ensemble of
inhomogeneously broadened two-level atoms. We allow laser light to
propagate through these along the $z$-direction. With a large beam
cross section a one-dimensional theory is sufficient, and the
Maxwell-Bloch equations can be written (see e.g.~\cite{Allen1975,
  Milonni1988, Mandel1995}):
\begin{gather}
  \label{bloch_uv}
    \frac{\partial}{\partial t}(u-iv) = -(\frac{\Gammah}{2}+i\Delta)
      (u-iv) - i\Omega w, \\
  \label{bloch_w}
    \frac{\partial}{\partial t}w = \frac{i}{2}
      \left[\Omega(u+iv)-\Omega^*(u-iv)\right]
      -\frac{1}{T_1}(1+w), \\
\label{eq:MB}
    \left(\frac{\partial}{\partial z} + \frac{n}{c}\frac{\partial}{\partial t}
     \right) \Omega = \frac{i\alpha_0}{2\pi}\int_{-\infty}^{\infty}
     g(\Delta)(u-iv)d\Delta. 
\end{gather}
Here $(u,v,w)$ is the usual Bloch-vector which depends on time $t$,
position $z$, and detuning $\Delta$ (from a chosen reference point).
The electric field is described in terms of the complex Rabi frequency
$\Omega(z,t) = \mu\mathcal{E}(z,t)/\hbar$, where $\mathcal{E}$ is the
complex electric field and $\mu$ is the dipole moment along the
direction of the field (we consider only a single linear polarization
mode). $\Gammah = 2/T_2$ is the FWHM homogeneous line-width of the
atoms in rad/sec, $T_1$ and $T_2$ are the life and coherence times of
the optical transition, respectively. In Eq.~(\ref{eq:MB}) $n$ is the
refractive index of non-absorbing background atoms, and $g(\Delta)$ is
a dimensionless function describing the inhomogeneous distribution of
atoms such that $g(\Delta)$ is proportional to the number of atoms
with transition frequency $\Delta$. We use an unconventional but
experimentally convenient normalization such that if the absorption
length measured with a weak laser field at frequency $\Delta_0$
amounts to $\alpha_0$, then $g(\Delta_0)$ must be unity. The integral
over $\Delta$ in Eq.~\eqref{eq:MB} effectively adds the contribution
of the polarization from all the atoms to the electric field $\Omega$
at position $z$ and time $t$.

Eqs.~(\ref{bloch_uv}-\ref{eq:MB}) are in general difficult to solve
analytically. However, for our specific needs regarding laser
stabilization we will make a number of approximations in the
following.  We start by noting that in Eq.~(\ref{eq:MB}) the term
$\frac{n}{c}\frac{\partial\Omega}{\partial t}$ is only relevant when
describing very fast changes, on the time scale $L/c$ where $L$ is the
length of the sample. We therefore neglect this term completely.
\subsubsection{Linear regime of Maxwell-Bloch equations}
\label{sec:linear_Maxwell_Bloch}
The next approximation is to consider
Eqs.~(\ref{bloch_uv}-\ref{eq:MB}) in the linear regime where, $w
\approx -1$ for all atoms, i.e.~the probability of being in the
excited state is small. In Sec.~\ref{sec:2-level-with-reservior} we
discuss the validity of this approximation. Inserting (with $w = -1$)
the integral form $u(z,t)-iv(z,t) = i\int_{-\infty}^t
e^{-(\frac{\Gammah}{2}+i\Delta)(t-t')}\Omega(z,t')dt'$ of
Eq.~(\ref{bloch_uv}) into Eq.~\eqref{eq:MB} and expressing the
electric field $\Omega$ in terms of its Fourier components,
$\Omega(z,t) = \int_{-\infty}^{\infty}\Omega(z,\omega)e^{-i\omega
  t}d\omega$, it follows that Eq.~(\ref{eq:MB}) can be written in
Fourier space as:
\begin{equation}
\label{eq:MB_linear_Fourier}
  \begin{split}
  \frac{\partial}{\partial z}\Omega(z,\omega)
   &= -\frac{\alpha_0}{2\pi}\int_{-\infty}^{\infty}
   \frac{g(\Delta)d\Delta}
        {\frac{\Gammah}{2}+i(\Delta-\omega)}\Omega(z,\omega)\\
   &\equiv -\frac{\alphaR(\omega)+ i\alphaI(\omega)}{2}\Omega(z,\omega). 
  \end{split}
\end{equation}
where we define $\alphaR(\omega)$ and $\alphaI(\omega)$ as the real
and imaginary absorption lengths, respectively. For a single frequency
component of the field $\Omega(z,\omega) =
A(z,\omega)e^{-i\phi(z,\omega)}$ with real amplitude $A$ and phase
$\phi$ we have the relation:
\begin{align}
\label{eq:MB_linear_amplitude}
  \frac{\partial A(z,\omega)}{\partial z} &=
     -\frac{\alphaR(\omega)}{2} A(z,\omega), \\
\label{eq:MB_linear_phase}
  \frac{\partial\phi(z,\omega)}{\partial z} &=
     +\frac{\alphaI(\omega)}{2}. 
\end{align}
Eqs.~(\ref{eq:MB_linear_Fourier}-\ref{eq:MB_linear_phase}) will be the
workhorse for many calculations in the following sections. Our goal is
to model the frequency variations of the incoming laser field,
propagate this field through the atomic medium via
Eqs.~\eqref{eq:MB_linear_amplitude} and~\eqref{eq:MB_linear_phase},
and finally derive an error signal useful for frequency stabilization
based on the outgoing field. We will reach this goal in
Sec.~\ref{sec:error-signals}, but before that we introduce a model
which describes the effect of hole burning in terms of the shape
function, $g(\Delta)$.
\subsubsection{Example: Inhomogeneous profiles}
\label{sec:example_inh_profile}
Let us give a simple and useful example of the above equations. 
Consider a Lorentzian-shaped inhomogeneous profile, $g(\Delta) =
\frac{\Gammainh^2}{4}/\left(\frac{\Gammainh^2}{4} + \Delta^2\right)$. 
Using the residue theorem we find (for $\Gammainh \gg \Gammah$):
\begin{equation}
\label{eq:alphaRI_lorentz_inh}
  \alphaR(\omega) = \alpha_0 \frac{\frac{\Gammainh^2}{4}}
                    {\frac{\Gammainh^2}{4} + \omega^2}, \quad
  \alphaI(\omega) = \alpha_0 \frac{\omega\frac{\Gammainh}{2}}
                    {\frac{\Gammainh^2}{4} + \omega^2}. 
\end{equation}
Here we have chosen the frequency zero point to be at the center of
the profile. We see that the absorption length at this frequency is
$\alpha_0$, consistent with $g(\Delta) = 1$ for $\Delta = 0$. 

We could also consider a Gaussian-shaped profile, $g(\Delta) =
\exp\left(-4\ln(2)\Delta^2/\Gammainh^2\right)$, with FWHM $\Gammainh$.
Inserting this into Eq.~(\ref{eq:MB_linear_Fourier}) we find with help
from \cite{Abramowitz1972}, $\frac{\partial\Omega(z,\omega)}{\partial
  z} = -\frac{\alpha_0 w(Z)}{2}\Omega(z,\omega)$, where $Z =
\frac{2\sqrt{\ln 2}}{\Gammainh}(\omega+i\frac{\Gammah}{2})$ and $w(Z)$
is the error function for complex arguments. Thus $\alphaR =
\alpha_0\Re\{w(Z)\}$ and $\alphaI = \alpha_0\Im\{w(Z)\}$, known as the
Voigt profile \cite{Milonni1988}.  In general, the shape of the
inhomogeneous profile varies depending on the broadening mechanism
\cite{Stoneham1969, Wesenberg2004d}.

\subsection{Two-level atoms with a reservoir state}
\label{sec:2-level-with-reservior}
The calculations in the previous sections need to be refined in order
to describe the effect of spectral hole burning. So, in addition to
the ground $\ketg$ and excited $\kete$ states we add a third reservoir
state $\ketr$ (see Fig.~\ref{fig:different_atomic_levels}(b)) and
write the Bloch equations for these:
\begin{align}
  \label{bloch3_uv}
  \frac{\partial}{\partial t}(u-iv) &= -(\frac{\Gammah}{2}+i\Delta)
    (u-iv) - i\Omega(\rho_e - \rho_g), \\
  \label{bloch3_rhoe}
  \frac{\partial\rho_e}{\partial t} &= \frac{i}{4}
    \left[\Omega(u+iv)-\Omega^*(u-iv)\right]
    -\frac{1}{T_1}\rho_e,\\
  \label{bloch3_rhog}
  \notag
  \frac{\partial\rho_g}{\partial t} &= -\frac{i}{4}
    \left[\Omega(u+iv)-\Omega^*(u-iv)\right] \\
      &\quad +\frac{\beg}{T_1}\rho_e -\frac{1}{\Tgr}\rho_g 
             + \frac{1}{\Trg}\rho_r,\\
  \label{bloch3_rhor}
    \frac{\partial\rho_r}{\partial t} &= \frac{\ber}{T_1}\rho_e 
          +\frac{1}{\Tgr}\rho_g - \frac{1}{\Trg}\rho_r. 
\end{align}
We assume the optical field, $\Omega$, only couples to the transition
$\ketg \rightarrow \kete$ and hence $u$ and $v$ still refer to this
transition, and only the factor $w = \rhoe-\rhog$ appears in the
driving term in Eq.~(\ref{bloch3_uv}), as was the case in
Eq.~(\ref{bloch_uv}). We add the possibility of decays from the
excited state to the reservoir state. The branching ratios from
$\kete$ to $\ketg$ and $\kete$ to $\ketr$ are denoted $\beg$ and
$\ber$, respectively. We also model relaxation between the $\ketg$ and
$\ketr$ levels. The timescale for decays from $\ketg$ to $\ketr$ is
$\Tgr$, which in general need not be the same as the timescale $\Trg$
in the opposite direction. For the homogeneous line-width, $\Gammah$,
we now have:
\begin{equation}
  \frac{\Gammah}{2} = \frac{1}{T_2} = \frac{1}{T_2^{(0)}} + \frac{1}{2\Tgr},
\end{equation}
where $T_2^{(0)}$ is the coherence time of the optical transition
$\ketg \rightarrow \kete$ in the absence of ground state relaxation,
and the term $\frac{1}{2\Tgr}$ takes the finite lifetime of the state
$\ketg$ into account.

\subsubsection{Separation of timescales}
\label{sec:separate_time_scales}
Our next step from Eqs.~(\ref{bloch3_uv}-\ref{bloch3_rhor}) is to use
steady-state solutions for $\rhoe$, $\rhog$, and $\rhor$ while $u$ and
$v$ still are allowed to vary in time according to
Eq.~(\ref{bloch3_uv}).  This is a good approximation since in our
specific case we have naturally different timescales for the ground
state populations and the optical coherence, $\Trg,\Tgr \gg T_2$.
Furthermore, when the laser is actively stabilized to a line-width
narrower than $\Gammah$, it is a good approximation to assume a zeroth
order starting point, $\Omega = \Omega_0 e^{-i\Delta_0 t}$, where the
laser is running perfectly at a monochromatic frequency, $\Delta_0$.
If the variations from this starting point are small, the populations
will always be close to their steady-state values.

So, with $\Omega = \Omega_0 e^{-i\Delta_0 t}$ we calculate the
steady-state solutions of Eqs.~(\ref{bloch3_uv}-\ref{bloch3_rhor}).
With a little work we obtain and expression for the population
difference $\rhog - \rhoe$ ($\rhoe$, $\rhog$, and $\rhor$ are given in
detail in Eq.~(\ref{eq:rho_egr_SS})):
\begin{equation}
\label{eq:rhog-rhoe_SS}
  \rhog - \rhoe = G\left(1-\dhole\frac{\frac{\Gammahole^2}{4}}
         {\frac{\Gammahole^2}{4} + (\Delta - \Delta_0)^2}\right),
\end{equation}
where $\dhole$ is the relative hole depth and $\Gammahole$ is the FWHM
of the hole. These parameters can be written:
\begin{equation}
\label{eq:def_dhole_gammahole}
  \dhole = \frac{(1+R)\frac{s_0}{2}}{1 + (1+R)\frac{s_0}{2}}, \quad
  \Gammahole = \Gammah\sqrt{1 + (1+R)\frac{s_0}{2}},
\end{equation}
where $s_0$ is the resonant saturation parameter:
\begin{equation}
  \label{eq:def_s0}
  s_0 = |\Omega_0|^2 T_1 T_2,
\end{equation}
and for our particular case of
Eqs.~(\ref{bloch3_uv}-\ref{bloch3_rhor}) we have:
\begin{equation}
\label{eq:def_R_G_three-level}
  R = \frac{1+ \frac{\ber\Trg}{T_1}}{1 + \frac{\Trg}{\Tgr}}, \quad
  G = \frac{1}{1 + \frac{\Trg}{\Tgr}}. 
\end{equation}
The saturation parameter $s_0$ is a measure of the probability of an
atom being in the excited state $\kete$ at resonance. In steady state
at $\Delta = \Delta_0$ we have $\frac{\rhoe}{\rhog} =
\frac{s_0}{2}/\left(1 + \frac{s_0}{2}\right)$. The parameter $G$ is a
measure of the fraction of atoms in the ground state $\ketg$ in
equilibrium in the absence of the laser light or when the detuning,
$\Delta - \Delta_0$, is large. The parameter $R$ is a measure of how
likely it is for an atom to be trapped in the reservoir state. The
essence of $R$ is in the term $\ber\Trg/T_1$, which is the ratio of
the rate $\ber/T_1$ from $\kete$ into the reservoir state $\ketr$ and
the rate $1/\Trg$ out of the reservoir state. From
Eq.~(\ref{eq:def_dhole_gammahole}) it is clear that a spectral hole
can be deep and broad for different reasons: Firstly, if the laser
field is strong with a high saturation parameter $s_0$, although $R$
is small, and secondly, if the trapping parameter $R$ is large even a
weak field with $s_0 \ll 1$ is capable of digging a deep, wide hole.

Now, we wish to employ Eq.~(\ref{eq:MB})
or~(\ref{eq:MB_linear_Fourier}) together with
Eqs.~(\ref{bloch3_uv}-\ref{bloch3_rhor}). We insert the steady-state
value of Eq.~(\ref{eq:rhog-rhoe_SS}) into Eq.~(\ref{bloch3_uv})
written in integral form. Since the steady-state value is
time-independent we may perform the same steps as those leading to
Eq.~(\ref{eq:MB_linear_Fourier}).  We will incorporate the value of
$\rhog-\rhoe$ into the $g(\Delta)$ shape function and just pretend
that we never left the linear approximation,
Eq.~(\ref{eq:MB_linear_Fourier}), of a two-level system. This is done
correctly when:
\begin{equation}
\label{eq:g_hole_only}
  g(\Delta) = \frac{\rhog-\rhoe}{G} =
    1-\frac{\dhole\frac{\Gammahole^2}{4}}
    {\frac{\Gammahole^2}{4} + (\Delta - \Delta_0)^2}. 
\end{equation}
The division by $G$ (the fraction of atoms in $\ketg$ far off
resonance) assures that $g(\Delta)$ is correctly normalized to unity
away from the spectral hole, i.e.~$\alpha_0$ is the absorption length
for a weak laser field in the absence of the spectral hole.

Let us retrace our steps so far and underline the approximations made.
We have reached the two important
equations~(\ref{eq:MB_linear_Fourier}) and~(\ref{eq:g_hole_only}). The
$g(\Delta)$ function for a spectral hole describes how many atoms
actually participate in the active two-level transition
$\ketg\rightarrow\kete$. We included the effect of saturation where
atoms can also populate the excited state $\kete$ (which
mathematically also creates a hole in $\rhog-\rhoe$). However, since
we assumed the populations $\rhoe$, $\rhog$, and $\rhor$ to be
essentially constant in time, we have restricted ourselves to
solutions where the laser field does not deviate much from a perfect
field, $\Omega = \Omega_0 e^{-i\Delta_0 t}$ (we have linearized the
theory around this zeroth order solution). Note, that the field
$\Omega$ can still have fast variations in e.g.~its phase, as long as
the phase excursions are not too large. Since both population trapping
in the reservoir state and the effect of saturation (leading to
population trapping in the excited state) are incorporated into the
single parameter $g(\Delta)$, we effectively model the three-level
equations~(\ref{bloch3_uv}-\ref{bloch3_rhor}) with our initial linear
two-level system with low saturation, as described by
Eq.~\eqref{eq:MB_linear_Fourier}.

Using Eq.~(\ref{eq:g_hole_only}) also requires another approximation.
We note that $\dhole$ and $\Gammahole$ depend on the resonant
saturation parameter, $s_0$. If the optical depth, $\alpha_0 L$, of
the atomic sample is large, the saturation parameter will depend on
$z$, and the use of a $z$-independent $g(\Delta)$ will be incorrect.
However, if the laser field burns holes, the attenuation will be less
than $\alpha_0 L$. Practically, the equations will be applicable for
$\alpha_0 L$ not too much greater than unity.

Finally, we point out that $\dhole$ and $\Gammahole$, as defined in
Eq.~(\ref{eq:def_dhole_gammahole}), refer to the structure in the
population, not to the depth and width which would be measured in an
absorption experiment. From~\eqref{eq:def_dhole_gammahole} we always
have the relation:
\begin{equation}
\label{eq:relate_dhole_Gammahole}
  1 - \dhole = \left(\frac{\Gammah}{\Gammahole}\right)^2. 
\end{equation}

\subsubsection{Absorption and phase shift from a spectral hole}
\label{sec:absorption_phase_spec_hole}
Let us now employ Eqs.~(\ref{eq:MB_linear_Fourier})
and~(\ref{eq:g_hole_only}) to calculate the attenuation and phase
shift of a laser field in the presence of a spectral hole. We take for
the $g(\Delta)$ function:
\begin{equation}
  g(\Delta) = \frac{\frac{\Gammainh^2}{4}}{\frac{\Gammainh^2}{4} + \Delta^2}
   \left(1-\frac{\dhole\frac{\Gammahole^2}{4}}
    {\frac{\Gammahole^2}{4} + (\Delta - \Delta_0)^2}\right),
\end{equation}
i.e.~we have a spectral hole burned at frequency $\Delta_0$ into an
inhomogeneously broadened Lorentzian profile with width $\Gammainh$
centered at $\Delta = 0$. Inserting this into
Eq.~(\ref{eq:MB_linear_Fourier}) we find, with the help from the
residue theorem,
\begin{widetext}
\begin{align}
\label{eq:alphaR_hole}
  \alphaR(\omega) &= \alpha_0\left[\frac{\frac{\Gammainh^2}{4}}
    {\frac{\Gammainh^2}{4}+\omega^2}-\frac{\frac{\Gammainh^2}{4}}
     {\frac{\Gammainh^2}{4}
   +\Delta_0^2}\cdot\frac{\frac{\Gammahole(\Gammahole+\Gammah)}{4}
    \dhole}{\frac{(\Gammahole+\Gammah)^2}{4}
    +(\Delta_0-\omega)^2}\right] 
   \rightarrow 
    \alpha_0\left[1-\frac{\frac{\Gammahole(\Gammahole+\Gammah)}{4}
    \dhole}{\frac{(\Gammahole+\Gammah)^2}{4}
    +(\Delta_0-\omega)^2}\right], \\
\label{eq:alphaI_hole}
  \alphaI(\omega) &= \alpha_0\left[\frac{\omega\frac{\Gammainh}{2}}
    {\frac{\Gammainh^2}{4}
    +\omega^2} + \frac{\frac{\Gammainh^2}{4}}
    {\frac{\Gammainh^2}{4} +\Delta_0^2}\cdot
    \frac{\frac{\Gammahole}{2}(\Delta_0-\omega)\dhole}
      {\frac{(\Gammahole+\Gammah)^2}{4}+(\Delta_0-\omega)^2}\right]
    \rightarrow
    \alpha_0\frac{\frac{\Gammahole}{2}(\Delta_0-\omega)\dhole}
      {\frac{(\Gammahole+\Gammah)^2}{4}+(\Delta_0-\omega)^2}, 
\end{align}
\end{widetext}
with $\Gammahole$ and $\dhole$ defined in
Eq.~\eqref{eq:def_dhole_gammahole}. In the first terms we assumed that
$\Gammainh \gg \Gammah$. The arrows indicate the limit when $\Gammainh
\rightarrow \infty$, i.e.~when we neglect the effect of the possibly
very wide inhomogeneous background.  Note, we had to include the
inhomogeneous profile in this calculation since otherwise $g(\Delta)
\rightarrow 1$ for $\Delta \rightarrow \infty$, making the integral in
Eq.~(\ref{eq:MB_linear_Fourier}) divergent.  Note, also that the
factors containing $\Gammainh$ are given by
Eq.~\eqref{eq:alphaRI_lorentz_inh} and they may be replaced by the
corresponding results for a Gaussian profile.

If we compare Eqs.~(\ref{eq:g_hole_only}) and~(\ref{eq:alphaR_hole})
we see that in an absorption measurement with a weak field (not
changing the populations further) the measured width and depth of the
hole are related to $\Gammahole$ and $\dhole$ by:
\begin{equation}
\label{eq:relate_gamma_d_meas}
  \Gammahole^{\mathrm{(meas)}} = \Gammahole + \Gammah, \quad
  \dhole^{\mathrm{(meas)}} = \frac{\Gammahole\dhole}{\Gammahole + \Gammah}. 
\end{equation}

\subsection{Calculation of error signals}
\label{sec:error-signals}
Now, let us turn to the calculation of real error signals used in the
locking procedure. We will start simply by connecting the Rabi
frequency, $\Omega$, used in the theoretical calculations, with real
experimental parameters, and we will also state the well known results
from Pound-Drever-Hall frequency modulation \cite{Drever1983,
  Bjorklund1983}.

To this end, we assume the laser field propagating into the locking
system is perfectly monochromatic, with $E^{(\mathrm{in})}(t) =
\Re\{\mathcal{E}_0 e^{-i\Delta_0 t}\}$ in the rotating frame. The
electric field is connected to the incoming intensity through the
relation $I^{(\mathrm{in})} = |\mathcal{E}_0|^2/2\mu_0 c$, which again
is connected to the incoming power, $P$, and cross-sectional area $A$ by
$I^{(\mathrm{in})} = P^{(\mathrm{in})}/A$. Since we have defined the
Rabi frequency as $\Omega = \mu\mathcal{E}/\hbar$, where $\mathcal{E}$
is the complex electric field inside the atomic sample, we can
equivalently describe our monochromatic incoming laser field in terms
of the complex Rabi frequency:
\begin{equation}
\label{eq:Omega0_and_power}
  \Omega(0,t) = \Omega_0 e^{-i\Delta_0 t}
  \quad\text{with }
  \Omega_0 = \frac{\mu}{\hbar}\sqrt{\frac{2\mu_0 c P}{n A}}. 
\end{equation}
Here the refractive index $n$ takes into account the fact that the
electric field is different inside and outside the atomic sample.

Next, we phase modulate this incoming laser field at frequency
$\omegam$ such that:
\begin{equation}
\label{eq:intro_modulation}
  \begin{split}
  \Omega(0,t) &= \Omega_0 e^{-i[\Delta_0 t + m\sin(\omegam t)]} \\
    &\approx \Omega_0(J_0 + J_1 e^{-i\omegam t}
      - J_1 e^{i\omegam t})e^{-i\Delta_0 t},
  \end{split}
\end{equation}
where $m$ is the modulation index, which is assumed to be small in the
second line, such that only the first sideband is significant. $J_0$
and $J_1$ are the Bessel functions of the first kind with $m$ as the
argument. 

It is now time to pass this modulated field through the atomic sample.
We shall simply assume that for the carrier, upper sideband, and lower
sideband, respectively, there are constant phase shifts of $\phic$,
$\phipl$, and $\phimi$ and constant transmission coefficients $\etac$,
$\etapl$, and $\etami$ of the amplitudes, such that the outgoing Rabi
frequency can be written:
\begin{equation}
\label{eq:propagate_static_amp_phase}
  \begin{split}
  \Omega(L,t) = \Omega_0\{&\etac J_0 e^{-i\phic} 
    + \etapl J_1  e^{-i(\omegam t + \phipl)} \\
      - &\etami J_1 e^{i(\omegam t - \phimi)}\}e^{-i\Delta_0 t}. 
  \end{split}
\end{equation}
These $\eta$ and $\phi$ parameters may be calculated from
Eqs.~\eqref{eq:alphaR_hole} and~\eqref{eq:alphaI_hole}. When the above
field hits a photo-detector, they will generate a photo-current
proportional to the power $P^{(\mathrm{out})}(t)$ incident on the
detector. This power is time-dependent and can be calculated from the
above equation together with the absolute square of
Eq.~\eqref{eq:Omega0_and_power} to give:
\begin{widetext}
\begin{equation}
  \begin{split}
  \label{eq:Pdet_static_model}
  P^{(\mathrm{out})}(t) = P^{(\mathrm{in})}&\left\{
    \left[J_0^2\etac^2 + J_1^2(\etapl^2 + \etami^2)\right] + J_0 J_1 \etac
    \left[(\etapl + \etami)\sin\left(\phic-\frac{\phipl + \phimi}{2}\right)
      \sin\left(\omegam t + \frac{\phipl - \phimi}{2}\right)\right.\right. \\
    & \qquad\qquad\qquad\qquad\qquad\qquad\qquad +\!\left.\left. 
   (\etapl - \etami)\cos\left(\phic-\frac{\phipl+\phimi}{2}\right)
    \cos\left(\omegam t + \frac{\phipl - \phimi}{2}\right)\right]\right\}. 
  \end{split}
\end{equation}
\end{widetext}
It takes some trigonometric relations to reach this, and we have
neglected the higher order contribution oscillating at $2\omegam$. 

The first term in square brackets is a DC term describing the average
power reaching the detector.  The second term in square brackets is
that useful and important for laser stabilization and closely
resembles the well known result from \cite{Bjorklund1983}. Here, this
term is not expressed in the most useful form for calculations, but it
clarifies some important qualitative properties of the error signal at
different demodulation phases for our particular setup. 

First of all, the term $\frac{\phipl-\phimi}{2}$ is always rather
small. For a broad inhomogeneous profile ($\Gammainh \gg \omegam$) in
the absence of hole burning the phase changes little for a frequency
variation of $\omegam$, and the phase difference $\phipl-\phimi$ is
small. In the presence of a spectral hole the phase may change
dramatically with small frequency changes, but the two side holes at
frequencies $\Delta_0 \pm \omegam$ are almost identical and the phase
difference remains small.  In any case, we always have $\frac{\phipl -
  \phimi}{2} \lesssim \alpha_0 L \omegam / \Gammainh \ll 1$ (unless
$\alpha_0 L$ is ridiculously large). As a result, we need only
consider the two terms in the rightmost square brackets above to be
pure sine and cosine terms, respectively. Note, some work has been
performed on narrow inhomogeneous profiles in the case when $\Gammainh
\gg \omegam$ does not hold \cite{Bottger2007}.

Secondly, the term $\phic-\frac{\phipl + \phimi}{2}$ depends strongly
on the situation. If the atomic contribution to the phase varies
slowly versus frequency on a scale broader than $\omegam$, this terms
is small. For instance, for an inhomogeneously broadened profile with
width $\Gammainh$ and no hole-burning effects, we must have
$\phic-\frac{\phipl + \phimi}{2} = O([\omegam/\Gammainh]^2)$, since
$\phic = \frac{\phipl + \phimi}{2}$ to first order. In this case the
$\sin(\omegam t)$-term in Eq.~(\ref{eq:Pdet_static_model}) is second
order in $\omegam/\Gammainh$. At the same time, we have in general
that $\etapl-\etami$ varies to first order in $\omegam/\Gammainh$,
while $\cos(\phic-\frac{\phipl + \phimi}{2}) \approx 1$. Hence the
$\cos(\omegam t)$ term is much larger than the $\sin(\omegam t)$ term,
which can be neglected.

The situation changes when hole burning is present. Then, in general,
the difference between $\phic$ and $\phi_{\pm}$ can be quite large,
and both $\cos(\phic-\frac{\phipl + \phimi}{2})$ and
$\sin(\phic-\frac{\phipl + \phimi}{2})$ can be on the order of unity.
At the same time, $\etapl + \etami$ can be on the order of unity while
$\etapl - \etami$ is of the order of $\omegam/\Gammainh$.  Hence the
$\sin(\omegam t)$ term is on the order of unity and dominates the
weaker $\cos(\omegam t)$ term.

Experimentally, we may demodulate the signal given by
Eq.~\eqref{eq:Pdet_static_model} by mixing it with a local oscillator
in order to obtain a useful error signal for the laser stabilization
system. When choosing the phase in this procedure there may be unknown
phase shifts in amplifiers, etc., but the ``cosine'' or ``sine''
quadrature can be found by maximizing or minimizing the error signal
when hole burning is not present (for rare-earth-ion-doped crystals
this can be done at an elevated temperature).

\subsubsection{Harmonic analysis}
\label{sec:harm_analysis}
Above we have calculated the detected power in a case when the input
light is perfectly monochromatic. In reality the input light will vary
in amplitude and frequency over time. The amplitude variations are
relatively easy to measure and correct for; we briefly discuss how we
do this experimentally in Sec.~\ref{sec:Exp_setup_feedback_sys}. Hence,
in the following we concentrate on frequency errors. We may imagine
that the transmission coefficients $\etac$, $\etapl$, and $\etami$ and
phase shifts $\phic$, $\phipl$, and $\phimi$ in
Eq.~\eqref{eq:Pdet_static_model} vary in time, which calls for further
calculations. In general, a convenient method is to assume the
incoming laser field to be of the form:
\begin{equation}
\label{eq:intro_harmonic_error}
  \Omega(0,t) = \Omega_0 e^{-i(\Delta_0 t + \epsilon\sin(\omega t))},
\end{equation}
i.e.~we have an almost single-frequency laser at $\Delta_0$, but with
an additional small harmonic disturbance of the phase with frequency
$\omega$ and magnitude $\epsilon$, which we assume to be much less
than unity. We note that the instantaneous laser frequency,
$\omega^{\mathrm{inst}}_0$, is then given by:
\begin{equation}
\label{eq:harm_error_inst_freq}
  \omega^{\mathrm{inst}}_0 = \frac{\partial \phi(t)}{\partial t} = \Delta_0+
   \epsilon\omega\cos(\omega t). 
\end{equation}
The above model is valid when the laser is running in the frequency
stabilized mode with a narrow line-width. After phase modulating the
beam at frequency $\omegam$ the light field becomes:
\begin{equation}
\label{eq:general_harm_modulation}
  \begin{split}
  \Omega(0,t) &\approx \Omega_0 
     (1+\frac{\epsilon}{2}e^{-i\omega t} - \frac{\epsilon}{2}e^{i\omega t})\\
     &\quad\times(J_0 + J_1 e^{-i\omegam t}
      - J_1 e^{i\omegam t})e^{-i\Delta_0 t},
  \end{split}
\end{equation}
which is just Eq.~\eqref{eq:intro_modulation} together with the
harmonic error of Eq.~\eqref{eq:intro_harmonic_error} for $\epsilon
\ll 1$.

We now have nine frequency components which will propagate through our
atomic medium, and we will treat these in a similar manner as in
Eq.~\eqref{eq:propagate_static_amp_phase}. To this end we make the
following assumptions. Since $\epsilon$ is small there is so little
energy in the $\omega$-sidebands that these have no hole-burning
effects. Consequently, we will have three spectral holes at
frequencies $\Delta_0$ and $\Delta_0 \pm\omegam$. For these holes we
have $\Gammahole \ll \omegam$ such that the holes can be treated
independently when utilizing Eqs.~\eqref{eq:alphaR_hole}
and~\eqref{eq:alphaI_hole}. We also neglect the effect of the
inhomogeneous profile ($\Gammainh \gg \omegam$). As a consequence, the
two spectral holes at frequencies $\Delta_0 \pm\omegam$ become
identical and we introduce a common absorption coefficient, $\etas =
\etapl = \etami$ and phase shift, $\phis = \phipl = \phimi$ where ``s''
refers to the sidebands at frequency $\omegam$.

Introducing the hole widths $\Gammaholec$, $\Gammaholes$ and hole
depths $\dholec$, $\dholes$ for the carrier and sidebands,
respectively, we may define:
\begin{align}
\label{eq:abs_omega_sideband}
  \eta_i(\omega) &= \exp\left(-\frac{\alpha_0 L}{2}
      \left[1-\frac{\frac{\Gammaholei(\Gammaholei+\Gammah)}{4}\dholei}
      {\frac{(\Gammaholei+\Gammah)^2}{4}+\omega^2}\right]\right), \\
\label{eq:phi_omega_sideband}
  \phi_i(\omega) &= -\frac{\alpha_0 L}{2}
     \frac{\frac{\Gammaholei}{2}\dholei\omega}
     {\frac{(\Gammaholei+\Gammah)^2}{4}+\omega^2},
\end{align}
where $i = \mathrm{c,s}$, and $\omega$ now refers to the positive
$\omega$-sideband of the harmonic error, i.e.~$\etac(\omega)$ is the
absorption of the frequency component $\Delta_0 + \omega$. We note
that $\eta_i(-\omega) = \eta_i(\omega)$ and $\phi_i(-\omega) =
-\phi_i(\omega)$. Eqs.~\eqref{eq:abs_omega_sideband}
and~\eqref{eq:phi_omega_sideband} have been derived directly from
Eqs.~\eqref{eq:alphaR_hole} and~\eqref{eq:alphaI_hole} with
appropriate frequency substitutions. For instance, for the hole at
$\Delta_0 + \omegam$ we must substitute $\Delta_0 \rightarrow \Delta_0
+ \omegam$ in Eqs.~\eqref{eq:alphaR_hole} and~\eqref{eq:alphaI_hole}
but, on the other hand, for the $\omega$-sideband we must use
$\Delta_0 + \omegam + \omega$ instead of $\omega$ as argument in
Eqs.~\eqref{eq:alphaR_hole} and~\eqref{eq:alphaI_hole}. Consequently,
Eqs.~\eqref{eq:abs_omega_sideband} and~\eqref{eq:phi_omega_sideband}
depend only on $\omega$. After a little algebra, the outgoing field
can be written:
\begin{equation}
  \begin{split}
    \Omega(L,&t) = J_0\Omega_0[\etac(0)-i\epsilon\etac(\omega)
     \sin(\omega t + \phic(\omega))]e^{-i\Delta_0 t} \\
     +&J_1\Omega_0[\etas(0)-i\epsilon\etas(\omega)
     \sin(\omega t + \phis(\omega))]e^{-i(\Delta_0+\omegam) t} \\
     -&J_1\Omega_0[\etas(0)-i\epsilon\etas(\omega)
     \sin(\omega t + \phis(\omega))]e^{-i(\Delta_0-\omegam) t}. 
  \end{split}
\end{equation}
Assume that this field impinges on a photo-detector and let us
calculate the power versus time, as we did in
Eq.~\eqref{eq:Pdet_static_model}. The term oscillating at frequency
$\omegam$ can be written:
\begin{equation}
\label{eq:Pdet_harmonic_model}
  \begin{split}
  P^{(\mathrm{out})}_{\omegam}&(t) = 4P^{(\mathrm{in})}J_0 J_1 \sin(\omegam t) \times \\
  \Re&\negthickspace\left\{ 
    \frac{\etac(\omega)\etas(0)e^{i\phic(\omega)}-\etac(0)\etas(\omega)
      e^{i\phis(\omega)}}{i\omega}\cdot \epsilon \omega e^{i\omega t}\right\},
  \end{split}
\end{equation}
where we have neglected terms of order $O(\epsilon^2)$. Firstly, we
note that this is proportional to $J_0 J_1$ and oscillates as
$\sin(\omegam t)$. In the example discussed around
Eq.~\eqref{eq:Pdet_static_model} we would have found similar behavior
if we had assumed the side holes to be identical (with $\phipl =
\phimi$ and $\etapl = \etami$). Secondly, in the curly brackets the
real part of the factor $\epsilon\omega e^{i\omega t}$ is just the
instantaneous frequency of the incoming laser (relative to
$\Delta_0$), see Eq.~\eqref{eq:harm_error_inst_freq}. Hence, the
fraction in the curly brackets acts as a transfer function mapping
this harmonic frequency excursion onto the measured power (and
eventually the output voltage) on the detector. This is similar to the
way in which a complex impedance $Z(\omega)$ maps a complex current
$I(\omega)$ onto a complex voltage $V(\omega) = Z(\omega)I(\omega)$
for individual Fourier components in electrical engineering. Hence,
the transfer function $T(\omega)$ is directly applicable for purposes
of feedback loop design for the laser stabilization system.

\subsubsection{Low-frequency errors}
\label{sec:low_freq_errors}
Before we proceed to utilize Eq.~\eqref{eq:Pdet_harmonic_model}, we
will make a small adjustment. We have assumed that the spectral holes
burned at $\Delta_0$ and $\Delta_0 \pm \omegam$ are static. This is a
poor approximation if we consider harmonic errors which vary slower
than the hole lifetime, $\Trg$, so instead we will make a more general
assumption that the actual hole center, $\tilde{\Delta}_0(t)$, is
time-dependent, and is given by the history of the instantaneous
frequency of Eq.~\eqref{eq:harm_error_inst_freq} averaged back in time
with a characteristic memory time, $\Trg$:
\begin{equation}
\label{eq:harm_low_freq_averaging}
  \begin{split}
  \tilde{\Delta}_0(t) &= \frac{1}{\Trg}\int_{-\infty}^t \omega^{\mathrm{inst}}_0(t')
    e^{-(t-t')/\Trg} dt' \\
   &= \Delta_0+\Re\left\{\frac{1}{1+i\omega\Trg}\cdot
      \epsilon\omega e^{i\omega t}\right\}. 
  \end{split}
\end{equation}
We also assume that the hole shape is otherwise unchanged. This is
only a good model for small frequency excursions. Now, instead of
calculating the frequency excursion from the hole center as
$\omega_0^{\mathrm{inst}} - \Delta_0$ we will use:
\begin{equation}
\label{eq:eff_freq_excursion}
  \omega_0^{\mathrm{inst}} - \tilde{\Delta}_0(t) = 
    \Re\left\{\frac{i\omega\Trg}{1+i\omega\Trg}\cdot
    \epsilon\omega e^{i\omega t}\right\}. 
\end{equation}
This is the effective instantaneous frequency excursion from the hole
center, which becomes small compared to
Eq.~\eqref{eq:harm_error_inst_freq} for low frequencies. To obtain the
transmitted power, $P^{(\mathrm{out})}_{\omegam}(t)$, the complex
frequency excursion in the curly brackets of
Eq.~\eqref{eq:eff_freq_excursion} can therefore replace the factor
$\epsilon\omega e^{i\omega t}$ in Eq.~\eqref{eq:Pdet_harmonic_model}
yielding:
\begin{equation}
\label{eq:Transfer_function}
\begin{split}
  T(\omega) &= \frac{\etac(\omega)\etas(0)e^{i\phic(\omega)}-\etac(0)\etas(\omega)
      e^{i\phis(\omega)}}{i\omega + \frac{1}{\Trg}}, \\
  P^{(\mathrm{out})}_{\omegam}(t) &= 4P^{(\mathrm{in})}J_0 J_1
  \Re\negthickspace\left\{T(\omega)\cdot \epsilon \omega e^{i\omega t}\right\}
   \cdot\sin(\omegam t). 
\end{split}
\end{equation}
For high frequencies this change plays no role, and hence
Eq.~\eqref{eq:Transfer_function} is a very useful model for the atomic
response to harmonic errors in laser frequency on all timescales.
However, the extra ad hoc term $\frac{1}{\Trg}$ in the denominator is
far from giving the full picture of laser stability at low
frequencies. This is discussed further in Sec.~\ref{sec:laser-drift}.
\subsubsection{Evaluating the transfer function}
\label{sec:evaluate_trans_func}
The transfer function in Eq.~\eqref{eq:Transfer_function} can be
evaluated by using the expressions in
Eqs.~\eqref{eq:abs_omega_sideband} and~\eqref{eq:phi_omega_sideband}.
Below we will consider different regimes of this function.

Firstly, we note that by inserting
Eq.~\eqref{eq:relate_dhole_Gammahole}
into~\eqref{eq:abs_omega_sideband} the transmission coefficient at the
hole centers can be written $\eta_i(0) = \exp(-\frac{\alpha_0
  L}{2}\frac{\Gammah}{\Gammaholei})$. Secondly, it is clear from the
denominator of $T(\omega)$ in Eq.~\eqref{eq:Transfer_function} that
the inverse hole lifetime, $\Trg^{-1}$, is one characteristic
frequency, and from Eqs.~\eqref{eq:abs_omega_sideband}
and~\eqref{eq:phi_omega_sideband} that the hole width, $\Gammaholei$,
is another characteristic frequency.

Now, let us evaluate the transfer function in the regime $\omega \ll
\Gammaholei$. Here we may use the approximation $\eta_i(\omega)
\approx \eta_i(0)$ and:
\begin{equation}
  \phi_i(\omega) \approx -\alpha_0 L \frac{\omega}{\Gammah}\cdot
  f\left(\frac{\Gammaholei}{\Gammah}\right),
\end{equation}
where the function $f$ is defined by:
\begin{equation}
\label{eq:f_function}
  f(x) = \frac{x-1}{x(x+1)}, 
  \quad x_{\mathrm{c}} = \frac{\Gammaholec}{\Gammah}, 
  \quad x_{\mathrm{s}} = \frac{\Gammaholes}{\Gammah}. 
\end{equation}
\begin{figure}[t]
  \centering
  \includegraphics{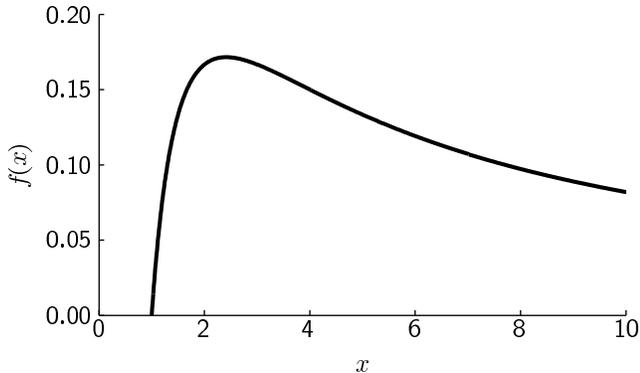}
  \caption{The function $f(x)$ defined in Eq.~\eqref{eq:f_function}. 
    At $x = 2.41$ it attains its maximum value of 0.172. Physically,
    $f(x)$ is proportional to the slope of $\phi_i(\omega)$ at $\omega
    = 0$ in Eq.~\eqref{eq:phi_omega_sideband}.} 
  \label{fig:f_function}
\end{figure}
This function $f(x)$ is shown in Fig.~\ref{fig:f_function}. If $\omega
\ll \Gammaholei$ we will always have $\phi_i(\omega) \ll \alpha_0 L$,
which means practically that the exponentials in
Eq.~\eqref{eq:Transfer_function} can be approximated as
$e^{i\phi_i(\omega)} \approx 1+i\phi_i(\omega)$. We then find the
transfer function:
\begin{equation}
\label{eq:Trans_func_MF}
  T(\omega) \approx -\frac{\alpha_0 L}{\Gammah}\frac{i\omega\Trg}{1+i\omega\Trg}
  e^{-\frac{\alpha_0 L}{2} \left(\frac{1}{x_{\mathrm{c}}}+\frac{1}{x_{\mathrm{s}}}\right)}
   [f(x_{\mathrm{c}})-f(x_{\mathrm{s}})]. 
\end{equation}
In the central regime, $\Trg^{-1} \ll \omega \ll \Gammaholei$, the
term $\frac{i\omega\Trg}{1+i\omega\Trg}$ is unity and the transfer
function is real. In the low-frequency limit, $\omega\ll \Trg^{-1}$,
the transfer function becomes imaginary and proportional to $\omega$.

In the high-frequency limit, $\omega \gg \Gammahole$, we have
$\eta_i(\omega) = \exp(-\frac{\alpha_0 L}{2})$, $\phi_i(\omega) =
-\frac{\alpha_0 L \Gammaholei\dholei}{4\omega} \ll 1$, and $\omega \gg
\frac{1}{\Trg}$. Then we obtain:
\begin{equation}
\label{eq:Trans_func_HF}
  T(\omega) \approx -\frac{1}{i\omega}e^{-\frac{\alpha_0 L}{2}}
    [e^{-\frac{\alpha_0 L}{2x_{\mathrm{c}}}}-e^{-\frac{\alpha_0 L}{2x_{\mathrm{s}}}}]. 
\end{equation}
\begin{figure}[t]
  \centering
  \includegraphics{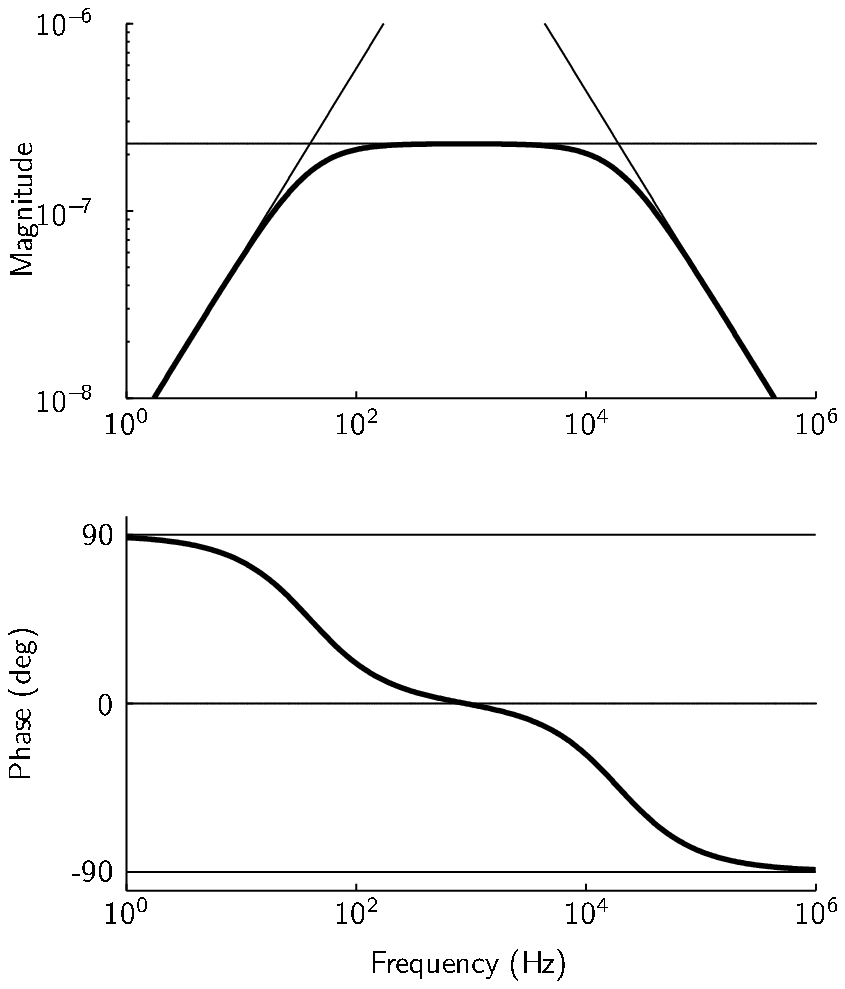}
  \caption{The magnitude and phase of the transfer function
    $T(\omega)$ calculated from Eq.~\eqref{eq:Transfer_function}
    (heavy lines). The light lines show the asymptotic cases discussed
    around Eqs.~\eqref{eq:Trans_func_MF} and~\eqref{eq:Trans_func_HF}.
    Parameters used are $T_1 = 150$ {\micro}s, $T_2 = 18$ {\micro}s,
    $\Trg = \Tgr = 4$ ms, $\ber = 0.5$, $\Omega_0 = 2\pi\cdot 1$ kHz,
    and $m = 0.40$, giving $\Gammah = 2\pi\cdot 17.5$ kHz,
    $\Gammaholec\approx 2\pi\cdot 21$ kHz and $\Gammaholes\approx
    2\pi\cdot 18$ kHz. These parameters are close to useful working
    values, as we shall see in Sec.~\ref{sec:laser-characterization}.}
  \label{fig:Transfer_function_example}
\end{figure}
The transfer function $T(\omega)$ has been plotted in
Fig.~\ref{fig:Transfer_function_example} for a choice of reasonable
experimental parameters (see the figure caption). It is clear that
there are three distinct regimes, as calculated above.  Assuming that
the terms in the square brackets in Eqs.~\eqref{eq:Trans_func_MF}
and~\eqref{eq:Trans_func_HF} are positive, the transfer function is a
negative real number times $i\omega$, 1, and $\frac{1}{i\omega}$ for
the low-, medium-, and high-frequency regimes, respectively. This
behavior is clearly seen in the magnitude of $T(\omega)$ shown on the
upper plot in Fig.~\ref{fig:Transfer_function_example}. The fact that
the transfer function is real at medium frequencies means that the
error signal $\propto\Re\{T(\omega)e^{i\omega t}\}$ will oscillate in
phase with the actual frequency error $\propto\Re\{e^{i\omega t}\}$. 
For high frequencies, the extra $\frac{1}{i}$ factor makes the error
signal oscillate as $\Re\{e^{i[\omega t-\pi/2]}\}$, i.e.~the response
is $90^{\circ}$ delayed. This is shown as the phase reaching
$-90^{\circ}$ in the lower plot in
Fig.~\ref{fig:Transfer_function_example}. For low frequencies the
situation is the opposite; the phase is advanced by $90^{\circ}$. This
behavior of the gain and phase has been previously reported in
experiments and numerical simulations \cite{Pryde2001,Pryde2002}. 

In our calculations we always assume that the power in the carrier
beam is higher than in either of the sidebands, leading to
$\Gammaholec > \Gammaholes$. Then, according to the definition in
Eq.~\eqref{eq:f_function}, $x_{\mathrm{c}}$ will be larger than
$x_{\mathrm{s}}$, and the term in the square brackets in
Eq.~\eqref{eq:Trans_func_HF} will be positive, as we assumed above. 
For Eq.~\eqref{eq:Trans_func_MF}, however, we can have a situation
where $f(x_{\mathrm{c}}) < f(x_{\mathrm{s}})$ if, e.g.~$2.41 <
x_{\mathrm{s}} < x_{\mathrm{c}}$, according to
Fig.~\ref{fig:f_function}. In this case there is a $270^{\circ}$ phase
shift between the medium- and high-frequency regimes which in practice
means that the sign of the error signal cannot be chosen correctly for
all frequency components in a closed feedback loop. Physically, the
sign change occurs when the slope of $\phic(\omega)$ around $\omega =
0$ equals the slope of $\phis(\omega)$ in
Eq.~\eqref{eq:phi_omega_sideband} and we must assure that this is
never the case. Note also that prior to and in the initialization of
the laser locking feedback loop the spectral holes are broad and
shallow (since the laser is broadband). The carrier hole will be
deeper than the sideband holes while the widths are roughly the same,
limited by the broad laser line-width. This in turn assures that the
slope of $\phi(\omega)$ for the carrier is larger than for the
sidebands.  We must choose the right parameters such that the sign
will remain correct when the feedback loop is closed and the laser
line width narrows. 

We should also comment on our theoretical approximation that $\epsilon
\ll 1$ in Eq.~\eqref{eq:general_harm_modulation}. According to
Eq.~\eqref{eq:harm_error_inst_freq} this means that the instantaneous
frequency excursion at frequency $\omega$ must be much smaller than
$\epsilon\omega$. At first sight, this seems to limit the usefulness
of the calculations at low frequencies. However, for $\omega <
\Gammahole$ the linearity of~\eqref{eq:phi_omega_sideband} versus
$\omega$ enables us to relax this condition on $\epsilon$.  In
practice, the theory is valid when $\epsilon\omega < \Gammahole$.

\subsubsection{Parameter choice in general}
\label{sec:parameter_choice_theory}
The theoretical observations above enable us to discuss the optimum
parameters in general. Further estimations associated with our
particular setup will be given in Sec.~\ref{sec:system-design}.

We start by focusing on how to obtain a large error signal. The main
reason for this is to obtain a high signal-to-noise ratio in the error
signal. Our first observation is the fact that the detected power
given in Eq.~\eqref{eq:Pdet_harmonic_model} is proportional to
$P^{(\mathrm{in})}$. It is no surprise that more light gives a higher
signal at the detector, but it is wrong to just naively increase the
incoming light power $P^{(\mathrm{in})}$ and expect a better
performance. Doing so will increase $s_0$ in Eq.~\eqref{eq:def_s0} and
in turn the hole widths $\Gammaholei$ in
Eq.~\eqref{eq:def_dhole_gammahole}.  However, increasing the incoming
power and at the same time increasing the beam area, $A$, leading to
an unchanged intensity, will always help.  Hence, it is a good idea to
use an atomic sample with a large area orthogonal to the direction of
beam propagation.

Next we observe in Eq.~\eqref{eq:Pdet_harmonic_model} that the front
factor $J_0 J_1$ attains its maximum value of 0.339 when the
modulation index is $m = 1.08$. This value is often used in laser
stabilization setups utilizing optical cavities \cite{Black2001}.
However, as opposed to the resonance lines in a cavity, the shape of
spectral holes depends on the optical power. If the carrier and
sideband powers were equal the holes would be identical, i.e.~$\etac =
\etas$ and $\phic = \phis$, leading to a zero error signal according
to Eq.~\eqref{eq:Pdet_harmonic_model}. For this reason the optimum
modulation index is lower than 1.08, leading to a more asymmetric
power distribution between the carrier and the sidebands. We have
searched our parameter space with numerical methods while optimizing
the signal in Eq.~\eqref{eq:Pdet_harmonic_model}. The result is that
$m = 0.56$ is a good choice. However, there are further complications
regarding the laser stability which suggest that $m$ should be even
lower. This will be discussed in Sec.~\ref{sec:laser-drift}.

Regarding the hole widths $\Gammaholec$ and $\Gammaholes$ it is clear
from Eq.~\eqref{eq:Trans_func_MF} and Fig.~\ref{fig:f_function} that
$x_{\mathrm{c}} = \Gammaholec/\Gammah$ should not be much greater than
2.41, since a higher value simply makes the $f$-function decrease
again.  Also, we would like to make $x_{\mathrm{s}} =
\Gammaholes/\Gammah$ small in order to decrease $f(x_{\mathrm{s}})$.
We can do this by lowering the modulation index $m$. Making $m$ too
small will also lower the factor $J_0 J_1$ and this is why we found
$m=0.56$ to be the optimum choice seen solely from the point of view
of optimizing the error signal. However, the magnitude of the error
signal is not everything.  The narrower the hole widths $\Gammaholei$,
the longer the duration of the atomic phase memory and hence
potentially better phase stability of the laser can be obtained. We
should also note that a given width, $\Gammaholei$, can be obtained in
different ways according to Eq.~\eqref{eq:def_dhole_gammahole}. One
could choose a high intensity (high $s_0$) and a short hole lifetime
$\Trg$ (low $R$ according to~\eqref{eq:def_R_G_three-level}) if
adjustable. On the other hand, a low intensity and a long hole
lifetime could give the same result. In general, the latter will give
the better long-term stability of the spectral hole.

To estimate the optimum optical density $\alpha_0 L$ let us assume
that $x_{\mathrm{c}} \approx 2$ and $x_{\mathrm{s}} \approx 1$. This
is not far from optimum given the discussion above. Inserting this
into either Eq.~\eqref{eq:Trans_func_MF} or~\eqref{eq:Trans_func_HF}
leads to the ballpark estimate $\alpha_0 L \approx 1.15$,
corresponding to a background intensity transmission of $e^{-\alpha_0
  L} \approx 32\%$.  Note, this is on the edge of our approximation
that $\alpha_0 L$ should not be too large for quantitatively correct
results.

\subsection{Laser drift}
\label{sec:laser-drift}
In the previous sections we have calculated the error signals for
laser locking based on the linearized model with the time-independent
distribution function $g(\Delta)$. For slowly varying errors on
timescales slower than the hole lifetime, $\Trg$, we presented in
Sec.~\ref{sec:low_freq_errors} an ad hoc model to account for the loss
of gain.  However, this does not really illustrate the real challenges
in long-term stability of the laser frequency. In the present section
we will show that under certain conditions there is a solution to the
equations where the laser is locked, but the frequency is drifting
linearly with time. Below in Sec.~\ref{sec:drift_model} we will
re-calculate the shape function $g(\Delta)$ and then the absorption
lengths $\alphaR$ and $\alphaI$ in the presence of laser drift. This
requires that we reconsider Eqs.~(\ref{bloch3_uv}-\ref{bloch3_rhor})
in more detail.  Based on these calculations, we will in
Sec.~\ref{sec:error_signal_drift} derive corrections to the error
signal from laser drift.

\subsubsection{The drift model}
\label{sec:drift_model}
We consider a situation where the incoming laser field is given by
\begin{equation}
\label{eq:Omega_in_drift}
  \Omega(0,t) = \Omega_0 e^{-i(\Delta_0 + \frac{\beta t}{2})t}
  \quad\Rightarrow\quad
  \omega_{\mathrm{inst}} = \Delta_0 + \beta t,
\end{equation}
i.e.~the instantaneous frequency $\omega_{\mathrm{inst}}$ varies
linearly with time, where $\beta$ is the drift rate in
$\mathrm{rad/s}^2$. We assume that the drift is much less than
$\Gammah$ during an optical coherence time $T_2$ (i.e.~$\beta \ll
\Gammah^2$). Then the populations $\rhoe$ and $\rhog$ can be safely
taken as constants on the timescale of $T_2$, and the coherences $u$
and $v$ will follow the incoming field adiabatically. We find:
\begin{equation}
  u-iv = \frac{-i\Omega_0 e^{-i(\Delta_0 + \frac{\beta t}{2})t}(\rhoe-\rhog)}
    {\frac{\Gammah}{2} + i[\Delta-\Delta_0 -\beta t]}
    + O(\frac{\Omega_0\beta}{\Gammah^3}). 
\end{equation}
Inserting this into Eqs.~(\ref{bloch3_rhoe}-\ref{bloch3_rhor}) gives
the population equations:
\begin{align}
  \label{bloch_rhoe_drift}
  \frac{\partial\rho_e}{\partial t} &= -\frac{1}{T_1}
    \left(\frac{s(t)}{2}[\rhoe-\rhog] + \rhoe\right) ,\\
  \label{bloch_rhog_drift}
  \frac{\partial\rho_g}{\partial t} &= \frac{1}{T_1}
    \left(\frac{s(t)}{2}[\rhoe-\rhog] + \beg\rhoe\right)
     -\frac{1}{\Tgr}\rho_g + \frac{1}{\Trg}\rho_r,\\
  \label{bloch_rhor_drift}
    \frac{\partial\rho_r}{\partial t} &= \frac{\ber}{T_1}\rho_e 
          +\frac{1}{\Tgr}\rho_g - \frac{1}{\Trg}\rho_r,
\end{align}
where $s(t)$ is the time-dependent saturation parameter
\begin{equation}
\label{eq:s_timedep_drift}
  s(t) = \frac{s_0 \frac{\Gammah^2}{4}}{\frac{\Gammah^2}{4}
         + (\Delta - \Delta_0 - \beta t)^2}
\end{equation}
and $s_0$ is still given by Eq.~\eqref{eq:def_s0}. Of course, the
above equations can be solved by numerical methods. However, our aim
is to derive an intuitive condition for the presence of a linear laser
drift. This is best done analytically.  First, we note that the
population equations~(\ref{bloch_rhoe_drift}-\ref{bloch_rhor_drift})
depend only on $t$ and $\Delta$ in the combination of $\Delta - \beta
t$.  We will change variables to $\Delta$. We will also make a series
expansion in the dimensionless parameter $\xi = \frac{\beta
  \Trg}{\Gammahole}$.  This parameter is a measure of how far the
laser drifts during a hole lifetime, $\Trg$, compared to the width of
the hole, $\Gammahole$. We multiply the population equations by $\xi$
and find:
\begin{align}
  \label{Eq_rhoe_drift}
  \xi\frac{\partial\rho_e}{\partial\Delta} &= \frac{\Trg}{T_1\Gammahole}
    \left(\frac{s(\Delta)}{2}[\rhoe-\rhog] + \rhoe\right) ,\\
  \notag
  \xi\frac{\partial\rho_g}{\partial\Delta} &= -\frac{\Trg}{T_1\Gammahole}
    \left(\frac{s(\Delta)}{2}[\rhoe-\rhog] + \beg\rhoe\right) \\
  \label{Eq_rhog_drift}
    &\quad +\frac{\Trg}{\Tgr}\frac{\rho_g}{\Gammahole} 
           -\frac{\rho_r}{\Gammahole},\\
  \label{Eq_rhor_drift}
    \xi\frac{\partial\rho_r}{\partial\Delta} &= 
    -\frac{\Trg\ber}{T_1\Gammahole}\rho_e 
          - \frac{\Trg}{\Tgr}\frac{\rho_g}{\Gammahole} 
          + \frac{\rho_r}{\Gammahole}. 
\end{align}
The change in sign arises from the change of variables, and the
saturation parameter now depends on $\Delta$ as:
\begin{equation}
\label{def:s_vs_Delta}
  s(\Delta) = \frac{s_0\frac{\Gammah^2}{4}}
                   {\frac{\Gammah^2}{4} + (\Delta-\Delta_0)^2}. 
\end{equation}
The above equations describe the populations when the laser has
instantaneous frequency $\Delta_0$ while the drift rate is $\beta$.
The reason for multiplying by $\xi$ becomes apparent when we now
define the series expansion $\rhoe = \rhoe^{(0)} + \xi\rhoe^{(1)} +
\xi^2\rhoe^{(2)} + \ldots$, and similarly for $\rhog$ and $\rhoe$.
Inserting these into Eqs.~(\ref{Eq_rhoe_drift}-\ref{Eq_rhor_drift})
leads to iterative equations for the different expansion coefficients:
\begin{equation}
\label{eq:drift_rho_egr_iteration}
  \begin{bmatrix}
    \partial\rhoe^{(j)}/\partial\Delta \\
    \partial\rhor^{(j)}/\partial\Delta \\
    0
  \end{bmatrix}
  = \frac{1}{\Gammahole}\vec{A}(\Delta)
  \begin{bmatrix}
    \rhoe^{(j+1)} \\
    \rhog^{(j+1)} \\
    \rhor^{(j+1)}
  \end{bmatrix}
\end{equation}
where the dimensionless matrix $\vec{A}$ is given by:
\begin{equation}
\label{def:vecA}
  \vec{A} = 
  \begin{bmatrix}
    \frac{Trg}{T_1}(\frac{s}{2}+1) & -\frac{\Trg}{T_1}\frac{s}{2} & 0 \\
    -\frac{Trg}{T_1}\ber & -\frac{\Trg}{\Tgr} & 1 \\
    1 & 1 & 1
  \end{bmatrix}. 
\end{equation}
We have omitted $\Delta$ in $s$ and $\vec{A}$ for brevity. Note, that
the first and second rows in this matrix reflect
Eqs.~\eqref{Eq_rhoe_drift} and~\eqref{Eq_rhor_drift}, respectively,
for the individual terms in the series expansion. A solution
fulfilling these two equations also satisfies
Eq.~\eqref{Eq_rhog_drift} automatically. The third row in $\vec{A}$
ensures that $\rhoe^{(j)} + \rhog^{(j)} + \rhor^{(j)} = 0$ for $j \ge
1$. Together with the condition $\rhoe^{(0)} + \rhog^{(0)} +
\rhor^{(0)} = 1$ the total population is always conserved independent
of $\xi$.

In order to find $\rhoe^{(1)}$, $\rhog^{(1)}$, and $\rhor^{(1)}$, we
need to calculate the inverse of $\vec{A}$ and we must know the zeroth
order steady-state solutions $\rhoe^{(0)}$, $\rhog^{(0)}$, and
$\rhor^{(0)}$. With $G$ and $R$ defined as in
Eq.~\eqref{eq:def_R_G_three-level} we derive
\begin{widetext}
\begin{gather}
\label{eq:vecA_inverse}
  \vec{A}^{-1} = \frac{G}{\frac{s}{2}(1+R)+1}\frac{T_1}{\Trg}
  \begin{bmatrix}
    1+\frac{\Trg}{\Tgr} & -\frac{\Trg}{T_1}\frac{s}{2} & 
    \frac{\Trg}{T_1}\frac{s}{2} \\
    -1-\frac{\Trg\ber}{\Tgr} & -\frac{\Trg}{T_1}(\frac{s}{2}+1) & 
    \frac{\Trg}{T_1}(\frac{s}{2}+1) \\
    \frac{\Trg\ber}{T_1}-\frac{\Trg}{\Tgr} & \frac{\Trg}{T_1}(s+1) &
    \frac{\Trg}{T_1}[\frac{\Trg}{\Tgr}(\frac{s}{2}+1)
                    +\frac{\Trg\ber}{T_1}\frac{s}{2}]
  \end{bmatrix}, \\
\label{eq:rho_egr_SS}
  \rhoe^{(0)} = \frac{G\cdot\frac{s}{2}}{\frac{s}{2}(1+R)+1}, \quad
  \rhog^{(0)} = \frac{G\cdot[\frac{s}{2}+1]}{\frac{s}{2}(1+R)+1}, \quad
  \rhor^{(0)} = \frac{G\cdot[\frac{\Trg}{\Tgr}(\frac{s}{2}+1)
               +\frac{\Trg\ber}{T_1}\frac{s}{2}]}{\frac{s}{2}(1+R)+1}. 
\end{gather}
\end{widetext}
Here Eq.~\eqref{eq:rho_egr_SS} is simply the steady-state solution of
Eqs.~(\ref{bloch3_rhoe}-\ref{bloch3_rhor}) or solutions of
Eqs.~(\ref{Eq_rhoe_drift}-\ref{Eq_rhor_drift}) when $\xi = 0$.  We
also note that the zeroth order populations can be found simply by
taking
\begin{equation}
  \begin{bmatrix}
    \rhoe^{(0)} \\ \rhog^{(0)} \\ \rhor^{(0)}
  \end{bmatrix}
  = \vec{A}^{-1}
  \begin{bmatrix}
    0 \\ 0 \\ 1
  \end{bmatrix}. 
\end{equation}
Taking the derivative of $\rhoe^{(0)}$ and $\rhor^{(0)}$ with respect
to $\Delta$ in Eq.~\eqref{eq:rho_egr_SS} and multiplying
Eq.~\eqref{eq:drift_rho_egr_iteration} from the left by
$\Gammahole\vec{A}^{-1}$ we obtain expressions for $\rhoe^{(1)}$,
$\rhog^{(1)}$, and $\rhor^{(1)}$. We remember that the distribution
function is given by $g(\Delta) = \frac{1}{G}(\rhog - \rhoe) =
\frac{1}{G}(\rhog^{(0)} - \rhoe^{(0)}) + \frac{\xi}{G}(\rhog^{(1)} -
\rhoe^{(1)}) + O(\xi^2)$. The term linear in $\xi$ is our correction,
$g_{\mathrm{drift}}(\Delta)$, to the shape function $g(\Delta)$ due to
the drift. Carrying out the calculation we find:
\begin{widetext}
  \begin{gather}
\label{eq:g_drift}
    g_{\mathrm{drift}}(\Delta) 
    = \frac{\beta T_1}{2(1+\frac{\Trg}{\Tgr})^2}
    \frac{s_0\Gammah^2(\Delta-\Delta_0)[\frac{\Gammah^2}{4}+(\Delta-\Delta_0)^2]}
       {\left[\frac{\Gammahole^2}{4}+(\Delta-\Delta_0)^2\right]^3}\left(
    \left[1+\frac{3}{2}\frac{\Trg}{\Tgr}+\frac{1}{2}\frac{\Trg^2}{\Tgr^2}\right]
    + \frac{\Trg}{2T_1}\left[\ber(1+\frac{\Trg}{T_1})
           -\beg\frac{\Trg}{\Tgr}\right]\right) \\
\label{eq:alphaI_drift}
   \Rightarrow\quad\alphaI(\Delta_0) = -\alpha_0
   \left(\frac{\beta\Trg}{\Gammahole}\right)
   \dhole\times\frac{\frac{T_1}{\Trg}\left[1+\frac{3}{2}\frac{\Trg}{\Tgr}
      +\frac{1}{2}\frac{\Trg^2}{\Tgr^2}\right] + 
      \frac{1}{2}\left[\ber(1+\frac{\Trg}{T_1})-\beg\frac{\Trg}{\Tgr}\right]}
    {\left(1+\frac{\Trg}{\Tgr}\right)
     \left(2+\frac{\Trg}{\Tgr}+\frac{\ber\Trg}{T_1}\right)}, 
  \end{gather}
\end{widetext}
where $\Gammahole$ and $\dhole$ are defined in
Eq.~\eqref{eq:def_dhole_gammahole}. The second line is derived with
help from the residue theorem by inserting
$g_{\mathrm{drift}}(\Delta)$ into Eq.~\eqref{eq:MB_linear_Fourier} and
setting $\omega = \Delta_0$, i.e.~we have calculated the imaginary
absorption length $\alphaI$ experienced by the laser beam at frequency
$\Delta_0$ drifting at rate $\beta$. The first-order drift
contribution to $\alphaR(\Delta_0)$ is zero, which is easily seen by
symmetry.

\subsubsection{Error signal from drift}
\label{sec:error_signal_drift}
Expression~\eqref{eq:alphaI_drift} may look a little complicated, but
all we really need to understand is the factor to the left of the
$\times$-sign. This factor says, that $\alphaI$ should be calculated
by taking the background $\alpha_0$ times ``how far we climbed up the
hole'' (this is $\xi = \frac{\beta\Trg}{\Gammahole}$) times the
relative depth, $\dhole$, of the hole. The rightmost fraction in
Eq.~\eqref{eq:alphaI_drift} is merely a constant independent of laser
power. Hence, this constant is the same for the center and side holes.
In the, not so uncommon, case when $\ber\Trg \gg T_1$ (leading to $R
\gg 1$) the factor is approximately equal to
$[2(1+\frac{\Trg}{\Tgr})]^{-1}$.  For brevity we make this
approximation below remembering we can always return to the more
accurate expression if necessary.

In order to calculate the error signal from a drifting laser, we now
add a phase modulation at frequency $\omegam$ with modulation index
$m$ to the incoming field of Eq.~\eqref{eq:Omega_in_drift} in a
completely analogous way to Eq.~\eqref{eq:intro_modulation} for a
static incoming field. This again leads to a carrier and two sideband
fields and we may employ Eq.~\eqref{eq:Pdet_static_model} to calculate
the error signal. When the drift is slow ($\xi \ll 1$) the absorption
coefficients $\etac$, $\etapl$ and $\etami$ will be given by $\eta_i =
\exp(-\frac{\alpha_0 L}{2}\frac{\Gammah}{\Gammaholei})$, and the phase
shift will be given by $\frac{L}{2}$ times
Eq.~\eqref{eq:alphaI_drift}. We then obtain:
\begin{equation}
\label{eq:error_signal_drift}
  \begin{split}
    P^{(\mathrm{out})}_{\omegam}(t) &= -P^{(\mathrm{in})}J_0 J_1 \frac{\alpha_0 L}{2}
    e^{-\frac{\alpha_0 L}{2}\!\left(\frac{\Gammah}{\Gammaholec}
      +\frac{\Gammah}{\Gammaholes}\right)} \\
    &\times\frac{\beta\Trg}{1+\frac{\Trg}{\Tgr}}
    \left[\frac{\dholec}{\Gammaholec}-\frac{\dholes}{\Gammaholes}\right]
    \sin(\omegam t),
  \end{split}
\end{equation}
where the approximation $\sin(\phic-\phis) \approx \phic-\phis$ is
employed. Comparing this expression to the low-frequency version in
Eq.~\eqref{eq:Trans_func_MF} we find most importantly that the
difference in $f$-functions has been replaced by the difference in the
ratios $\dholei/\Gammaholei$. As discussed previously, there is a risk
of obtaining the wrong sign for the error signal. The difference in
square brackets in Eq.~\eqref{eq:error_signal_drift} must be positive
for zero drift with $\beta = 0$ to be a stable solution. If this is
not the case, there will be a bi-stable solution with positive or
negative non-zero values of $\beta$. To calculate the magnitude
$|\beta|$ of this drift rate requires a complete calculation of the
non-linear problem of
Eqs.~(\ref{bloch_rhoe_drift}-\ref{eq:s_timedep_drift}) which is
outside the scope of this work.

We wish to operate the laser stabilization system without drift, and
to this end we derive a stability criterion based on convenient
experimental parameters. The drift problems occur when the hole depths
are too large, as we shall see below. Given the modulation index, $m$,
we will determine a threshold value that the \emph{measured} center
hole depth, $\dholec^{(\mathrm{meas})}$, should not exceed. The hole
width and depth can be measured by scanning a laser across the hole
and monitoring the absorption, as we will show in
Sec.~\ref{sec:laser-characterization}. Also, for practical values of
the modulation index, the carrier has the major fraction of the
optical power, and the total light transmission of the locking beam
itself can be utilized to obtain an approximate estimate of
$\dholec^{(\mathrm{meas})}$.

\begin{figure}[t]
  \centering
  \includegraphics{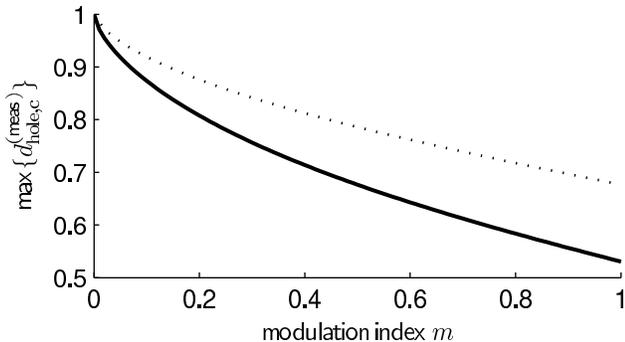}
  \caption{Thresholds for stable laser operation. If
    $\dholec^{(\mathrm{meas})}$ is below the solid line the condition
    in Eq.~\eqref{eq:drift_criterion} is met and the zero-drift
    solution is stable. In order for the low-frequency part of the
    laser locking to have the correct sign ($f(x_{\mathrm{c}}) -
    f(x_{\mathrm{s}}) > 0$ in Eq.~\eqref{eq:Trans_func_MF}) we require
    the less stringent condition that $\dholec^{(\mathrm{meas})}$ is
    below the dotted line. Between the two lines the laser can be
    locked in a linearly drifting mode.}
  \label{fig:thresholds}
\end{figure}
We first observe that $\Gammaholes^2-\Gammah^2 =
y(\Gammaholec^2-\Gammah^2)$, where we define $y = J_1^2/J_0^2$ as the
power ratio between one of the first sidebands and the carrier beam.
To see this, use Eq.~(\ref{eq:def_dhole_gammahole}) remembering that
$s_0$ has to be multiplied by $J_0^2$ and $J_1^2$ to calculate the
correct saturation parameter for the carrier and sidebands,
respectively. In addition, with Eq.~(\ref{eq:relate_dhole_Gammahole})
and the threshold condition $\dholec/\Gammaholec >
\dholes/\Gammaholes$, we derive $(\Gammaholes/\Gammaholec)^2 >
y^{2/3}$. Next, we replace $\Gammaholes^2$ by
$y(\Gammaholec^2-\Gammah^2)+\Gammah^2$, and we are left with a
condition depending on only the two parameters $y$ and
$\Gammaholec/\Gammah$. A combination of
Eqs.~(\ref{eq:relate_dhole_Gammahole})
and~(\ref{eq:relate_gamma_d_meas}) will show that the measured hole
depth can be written $\dholec^{(\mathrm{meas})} = 1 -
\Gammah/\Gammaholec$, and we finally reach our criterion:
\begin{equation}
\label{eq:drift_criterion}
  \dholec^{(\mathrm{meas})} < 1-\sqrt{\frac{y^{2/3}-y}{1-y}},
  \qquad y = \frac{J_1^2}{J_0^2}. 
\end{equation}
Given the modulation index, $m$, we can calculate $y$ numerically and
plot the threshold value as a function of $m$. This has been done in
Fig.~\ref{fig:thresholds} (solid line).

We may also derive a criterion for ensuring the correct positive sign
of the factor $[f(x_{\mathrm{c}}) - f(x_{\mathrm{s}})]$ in
Eq.~\eqref{eq:Trans_func_MF} assuring that the low frequency part of
the error signal is correct. There is no simple analytical expression
for this, but a numerical result is shown by the dotted line in
Fig.~\ref{fig:thresholds}. Note, that it is easier to fulfill the
criterion for correct low-frequency behavior than the criterion for no
linear drift. This is an important observation which shows that all
the calculations regarding the drift model are worthwhile and
necessary to obtain a complete understanding of laser stability. It is
indeed possible that $\dholec^{(\mathrm{meas})}$ is in between the
dotted and solid lines in Fig.~\ref{fig:thresholds}, in which case the
laser stabilization system is apparently locked but still the laser is
drifting linearly.

We conclude this section by pointing out that the drift calculations
can be performed in a similar manner for a simple two-level system in
absence of a reservoir state $\ketr$, but the results can readily be
guessed by setting $\Trg = 0$, $\Tgr = \infty$, $\ber = 1$, and $\beg =
0$. Then the three-level case will reduce to the two-level case and
Eq.~\eqref{eq:alphaI_drift} will reduce to $\alphaI(\Delta_0) =
-\frac{\alpha_0}{2}\frac{\beta T_1}{\Gammahole}\dhole$.
\subsection{General remarks on the calculations}
\label{sec:general_remarks_error_signal}
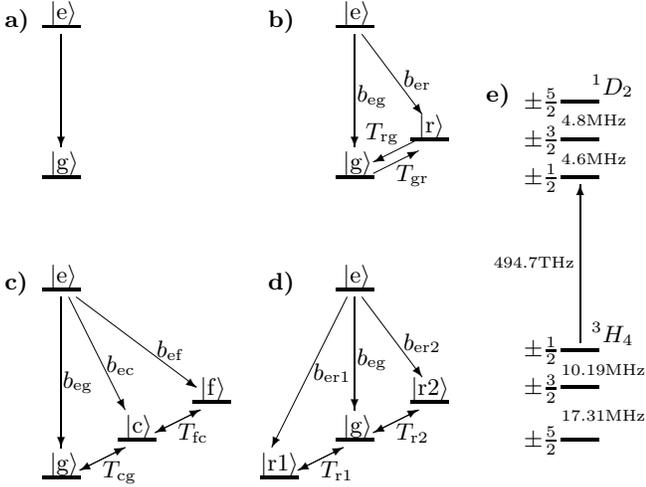
\begin{figure}[t]
  \centering
  \setlength{\unitlength}{1mm}
  \begin{picture}(85,65)
    \put(0 ,60){\bf a)}
    \put(35,60){\bf b)}
    \put(0 ,25){\bf c)}
    \put(35,25){\bf d)}
    \put(64,50){\bf e)}
    \linethickness{1pt}
    \put(5,60){\line(1,0){5}}  
    \put(6,61){$\kete$}        
    \put(5,25){\line(1,0){5}}  
    \put(6,26){$\kete$}        
    \put(44,60){\line(1,0){5}}  
    \put(45,61){$\kete$}        
    \put(44,25){\line(1,0){5}}  
    \put(45,26){$\kete$}        
    \linethickness{1pt}
    \put(5,40){\line(1,0){5}}  
    \put(6,41){$\ketg$}        
    \put(5,0){\line(1,0){5}}   
    \put(6,1){$\ketg$}         
    \put(44,40){\line(1,0){5}}  
    \put(45,41){$\ketg$}        
    \put(44,5){\line(1,0){5}}   
    \put(45,6){$\ketg$}         
    \linethickness{1pt}
    \put(54,45){\line(1,0){5}}  
    \put(55,46){$\ketr$}        
    \put(54,10){\line(1,0){5}}  
    \put(54,11){$\ketrtwo$}     
    \put(34,0){\line(1,0){5}}   
    \put(34,1){$\ketrone$}      
    \put(15,5){\line(1,0){5}}   
    \put(16,6){$\ketc$}         
    \put(25,10){\line(1,0){5}}  
    \put(26,11){$\ketf$}        
    \linethickness{0.3pt}
    \put(7.5,59){\vector(0,-1){15}} 
    \put(46.5,59){\vector(0,-1){15}} 
    \put(47.5,59){\vector(3,-4){8}}  
    \put(7.5,24){\vector(0,-1){20}} 
    \put(8.5,24){\vector(1,-2){7.5}}
    \put(9.5,24){\vector(4,-3){16}} 
    \put(46.5,24){\vector(0,-1){15}} 
    \put(47.5,24){\vector(3,-4){8}}  
    \put(45.5,24){\vector(-1,-2){10}}
    \linethickness{0.5pt}
    \put(10,1){\vector(2,1){6}}    
    \put(16,4){\vector(-2,-1){6}}  
    \put(20,6){\vector(2,1){6}}    
    \put(26,9){\vector(-2,-1){6}}  
    \put(39,1){\vector(2,1){6}}    
    \put(45,4){\vector(-2,-1){6}}  
    \put(49,6){\vector(2,1){6}}    
    \put(55,9){\vector(-2,-1){6}}  
    \put(49,40.5){\vector(2,1){6}} 
    \put(55,45){\vector(-2,-1){6}} 
    \put(13,0){$\Tcg$}     
    \put(23,5){$\Tfc$}     
    \put(42,0){$\Trone$}   
    \put(52,5){$\Trtwo$}   
    \put(52,39.5){$\Tgr$}  
    \put(48,45){$\Trg$}    
    \put(7.7,12){$\beg$}  
    \put(13.5,14){$\bec$}  
    \put(20.2,16){$\bef$}  
    \put(41,13){$\berone$} 
    \put(46.8,15){$\beg$}  
    \put(53,17){$\bertwo$} 
    \put(46.8,50){$\beg$}  
    \put(53,52){$\ber$}    
    \linethickness{1pt}
    \put(74,17){\line(1,0){5}}  
    \put(74,12){\line(1,0){5}}  
    \put(74,5){\line(1,0){5}}   
    \put(69,16.1){$\pm\frac{1}{2}$}
    \put(69,11.1){$\pm\frac{3}{2}$}
    \put(69,4.1){$\pm\frac{5}{2}$}
    \put(74,7.5){\figfont 17.31MHz}
    \put(74,13.7){\figfont 10.19MHz}
    \put(78,18){$^3H_4$}
    \put(74,40){\line(1,0){5}}  
    \put(74,45){\line(1,0){5}}  
    \put(74,50){\line(1,0){5}}  
    \put(69,39.1){$\pm\frac{1}{2}$}
    \put(69,44.1){$\pm\frac{3}{2}$}
    \put(69,49.1){$\pm\frac{5}{2}$}
    \put(74,41.7){\figfont 4.6MHz}
    \put(74,46.8){\figfont 4.8MHz}
    \put(78,51){$^1D_2$}
    \linethickness{0.5pt}
    \put(76.5,18){\vector(0,1){21}} 
    \put(65,28){\figfont 494.7THz}
  \end{picture}
  \caption{Different level schemes used in this paper. We define
    timescales for relaxation between ground state levels and
    branching ratios from the excited state. The excited state
    lifetime is always denoted $T_1$. {\bf (a)} The most naive
    scenario with two levels, considered in
    Sec.~\ref{sec:2-level-Maxwell-Bloch} and in the first row of
    Tab.~\ref{tab:G_R}. {\bf (b)} Our basic model for all
    calculations, described in Secs.~\ref{sec:2-level-with-reservior}
    and~\ref{sec:laser-drift}.  Rows two to four in Tab.~\ref{tab:G_R}
    refer to this case.  Note, we may have different relaxation
    timescales $\Trg \ne \Tgr$. {\bf (c)} and {\bf (d)} Different
    schemes with three ground states coupled as shown with RF-magnetic
    fields.  Hence the timescale is the same in two opposite
    directions. These cases are reflected by rows five and six in
    Tab.~\ref{tab:G_R}, respectively. {\bf (e)} The real \PrYSO level
    scheme.}
  \label{fig:different_atomic_levels}
\end{figure}


%
%
\begin{table}[t]
\begin{tabular}{|m{20mm}|r@{$\,\,$}l|r@{$\,\,$}l|} \hline
  Case    &\multicolumn{2}{c|}{\emph{$G$}}   
          &\multicolumn{2}{c|}{\emph{$R$}}  \\ \hline
  Two-level                     & $G$&$=1$                           
                                & $R$&$=1$  \\ \hline
  Three-level                   & $G$&$=\frac{1}{1+\frac{\Trg}{\Tgr}}$  
                                & $R$&$=\frac{1+\frac{\ber\Trg}{T_1}}
                                            {1+\frac{\Trg}{\Tgr}}$  \\ \hline
  Three-level, natural decay    & $G$&$=1$ 
                                & $R$&$=1+\frac{\ber\Trg}{T_1}$ \\ \hline
  Three-level, RF eraser        & $G$&$=\frac{1}{2}$  
                                & $R$&$=\frac{1}{2}[1+
                                     \frac{\ber\Trg}{T_1}]$ \\ \hline
  Four-level, RF eraser (1)     & $G$&$=\frac{1}{3}$ 
                                & $R$&$=\frac{1}{3}[1+\frac{2\bec\Tcg+\bef
                                    \left(2\Tcg+\Tfc\right)}{T_1}]$ \\ \hline
  Four-level, RF eraser (2)     & $G$&$=\frac{1}{3}$ 
                                & $R$&$=\frac{1}{3}[1+\frac{\Trone\berone+
                                              \Trtwo\bertwo}{T_1}]$ \\ \hline
\end{tabular}

\caption{The value of $G$ and $R$ for the different setups shown in 
  Fig.~\ref{fig:different_atomic_levels}.  The first row gives the relations
  for the two-level atom (Fig.~\ref{fig:different_atomic_levels}(a)) and the 
  second row describes
  the two-level plus reservoir state system considered in 
  Sec.~\ref{sec:2-level-with-reservior} 
  (Fig.~\ref{fig:different_atomic_levels}(b)). The third and fourth rows are
  special cases of the second row. In the third row we assume $\Tgr = \infty$, 
  which describes a one-way natural decay from states $\ketr$ to $\ketg$. 
  In the fourth row we assume $\Trg = \Tgr$ which describes the case when an 
  RF magnetic field couples the otherwise uncoupled states $\ketr$ and 
  $\ketg$. The fifth and sixth rows correspond to the cases shown in 
  Fig.~\ref{fig:different_atomic_levels}(c) and
  Fig.~\ref{fig:different_atomic_levels}(d), respectively, where there are two
  reservoir states. These four-level cases are presented since they
  resemble our experimental case using \PrYSO as the atomic medium.} 
\label{tab:G_R}
\end{table}
Up until now we have considered a two-level system with a single
reservoir state to model the trapping of atoms in the hole-burning
process. This is a simple system which allows for not too complicated
analytical solutions, thereby maintaining the physical understanding.
This simple system is actually found in \TmYAG \cite{Strickland2000,
  Merkel2000}, and the even simpler pure two-level system is found in
\ErYSO \cite{Sellin2001, Bottger2001} and Er:KTP \cite{Pryde2002}.

However, more complicated cases exist. For our experiments with \PrYSO
there are three ground state levels and three excited state levels,
see Fig.~\ref{fig:different_atomic_levels}(e), and all nine possible
transitions exist, due to the inhomogeneous broadening, and in
principle play a role. If one wishes to extend the theory to cover
this we must in Eq.~\eqref{eq:MB_linear_Fourier} replace $\alpha_0
g(\Delta)$ by $\sum_j\!\alpha_{0,j}g_j(\Delta)$, where $j$ identifies
the individual transition, $g_j(\Delta)$ is the shape function
calculated with the appropriate saturation parameter given the
transition strength, and $\alpha_{0,j}$ is the absorption length for
the individual atomic species. The total absorption length must
fulfill $\alpha_0 = \sum_j \alpha_{0,j}$.

To calculate $g_j(\Delta)$ for the case with three distinct ground
states we consider the physical systems shown in
Fig.~\ref{fig:different_atomic_levels}(c,d).  The active optical
transition is still $\ketg \rightarrow \kete$, but two reservoir
states are present. In Fig.~\ref{fig:different_atomic_levels}(c) an
asymmetric case is shown where the reservoir states are labeled
$\ketc$ and $\ketf$ for ``close'' and ``far'', respectively,
describing their position in the RF pumping scheme relative to the
state $\ket{g}$. In Fig.~\ref{fig:different_atomic_levels}(d) the
symmetric case is shown where the reservoir states are labeled
$\ketrone$ and $\ketrtwo$. It is a simple matter to generalize the
methods of Sec.~\ref{sec:2-level-with-reservior} to three ground state
levels.  Doing so, we calculate steady-state populations in order to
derive the distribution function $g_j(\Delta)$. This function is found
to have exactly the same form as Eq.~\eqref{eq:def_dhole_gammahole},
apart from new values of $R$ and $G$, which are given in
Tab.~\ref{tab:G_R} (fifth and sixth rows). As shown in
Fig.~\ref{fig:different_atomic_levels}(c,d) we assume the same
timescale in both directions between two ground states,
e.g.~$\Tcf=\Tfc$, etc.  This is valid in our experimental case with
RF-magnetic fields coupling the adjacent levels.

Hence, for a single atomic species, adding more ground reservoir
states only changes the spectral holes quantitatively, but
qualitatively we still have a Lorentzian-shaped hole fulfilling
Eq.~\eqref{eq:def_dhole_gammahole}, as for the simple case of
two-levels plus a single reservoir state. However, for multiple atomic
species the contribution of different transitions with different
strengths may lead us to sum up quite different $g_j(\Delta)$
functions with a non-Lorentzian shape as the result.  Instead of
performing a complete quantitative examination of this, we tried in
the experiments to keep this difference small, in order to mimic the
three-level system and demonstrate the qualitative features of the
theoretical calculations.  The results in Tab.~\ref{tab:G_R} will help
us do this.

We also wish to remind the reader that our theory generally assumes
perfect lasers or perfect lasers with harmonic errors. In real life
this is not the case, but our approximations are still quite good if
the stabilization system maintains a narrow line-width in the laser.
If the laser line width, for example, is 1 kHz and the hole width is
20 kHz, there will be some kind of folding effect of the order of 5\%.
Also, we have assumed that $\alpha_0 L$ is not too large. If we, for
example, set $\alpha_0 L \approx 1$ and assume a measured hole depth
of around 50\%, the transmission of the carrier beam is $e^{-1/2}
\approx 60\%$.  This means that the saturation parameter varies by
40\% over the sample, and we can approximately take this into account
by lowering the saturation parameter to 80\% of the value calculated
from Eqs.~\eqref{eq:def_s0} and~\eqref{eq:Omega0_and_power}. In this
manner (for the two levels plus a single reservoir state) we should be
able to keep the theory quantitatively correct within around 10\%,
while all the qualitative features should hold true.


\section{System design}
\label{sec:system-design}
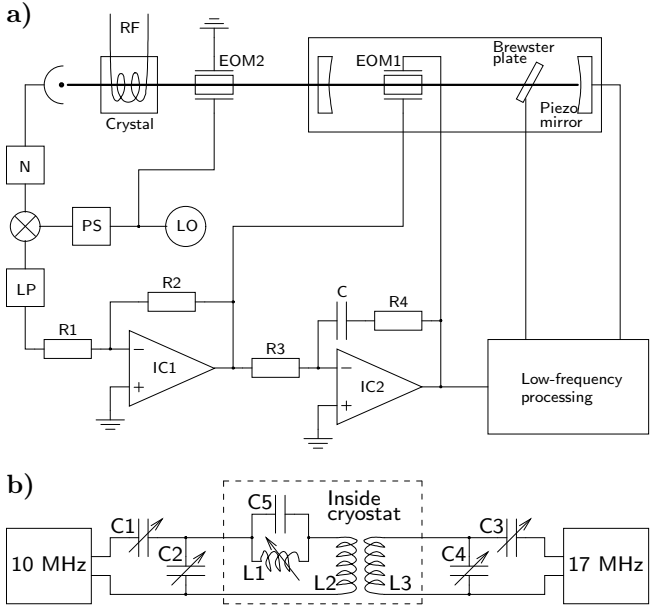
\begin{figure}[t]
  \centering
\setlength{\unitlength}{0.00054681in}
\begingroup\makeatletter\ifx\SetFigFont\undefined%
\gdef\SetFigFont#1#2#3#4#5{%
  \reset@font\fontsize{#1}{#2pt}%
  \fontfamily{#3}\fontseries{#4}\fontshape{#5}%
  \selectfont}%
\fi\endgroup%
{\renewcommand{\dashlinestretch}{30}
\begin{picture}(6144,5685)(0,-10)
\path(3162,2442)(3162,1722)(3972,2082)(3162,2442)
\path(3207,2262)(3297,2262)
\path(3207,1902)(3297,1902)
\path(3252,1947)(3252,1857)
\path(2847,1587)(3117,1587)
\path(2892,1542)(3072,1542)
\path(2937,1497)(3027,1497)
\path(2982,1587)(2982,1677)
\path(2982,2712)(3162,2712)
\path(3162,2892)(3162,2532)
\path(3252,2892)(3252,2532)
\path(3252,2712)(3432,2712)
\path(3522,2802)(3972,2802)(3972,2622)
	(3522,2622)(3522,2802)
\path(3432,2712)(3522,2712)
\path(3972,2712)(4062,2712)
\path(2352,2352)(2802,2352)(2802,2172)
	(2352,2172)(2352,2352)
\path(2262,2262)(2352,2262)
\path(2802,2262)(2892,2262)
\path(1812,5052)(2172,5052)(2172,4872)
	(1812,4872)(1812,5052)
\path(2172,5097)(1812,5097)
\path(1812,4827)(2172,4827)
\path(1992,4827)(1992,4692)
\path(1992,5097)(1992,5232)
\path(3612,5052)(3972,5052)(3972,4872)
	(3612,4872)(3612,5052)
\path(3972,5097)(3612,5097)
\path(3612,4827)(3972,4827)
\path(3792,4827)(3792,4692)
\path(3792,5097)(3792,5232)
\path(5457,5232)(5592,5232)(5592,4692)(5457,4692)
\path(5457,5232)(5457,5231)(5458,5228)
	(5459,5220)(5461,5206)(5464,5187)
	(5467,5165)(5471,5140)(5474,5114)
	(5477,5090)(5480,5067)(5482,5046)
	(5484,5027)(5485,5010)(5486,4993)
	(5487,4978)(5487,4962)(5487,4946)
	(5486,4931)(5485,4914)(5484,4897)
	(5482,4878)(5480,4857)(5477,4834)
	(5474,4810)(5471,4784)(5467,4759)
	(5464,4737)(5461,4718)(5459,4704)
	(5458,4696)(5457,4693)(5457,4692)
\path(1857,5502)(2127,5502)
\path(1902,5547)(2082,5547)
\path(1947,5592)(2037,5592)
\path(1992,5502)(1992,5412)
\path(3117,5232)(2982,5232)(2982,4692)(3117,4692)
\path(3117,5232)(3117,5231)(3116,5228)
	(3115,5220)(3113,5206)(3110,5187)
	(3107,5165)(3103,5140)(3100,5114)
	(3097,5090)(3094,5067)(3092,5046)
	(3090,5027)(3089,5010)(3088,4993)
	(3087,4978)(3087,4962)(3087,4946)
	(3088,4931)(3089,4914)(3090,4897)
	(3092,4878)(3094,4857)(3097,4834)
	(3100,4810)(3103,4784)(3107,4759)
	(3110,4737)(3113,4718)(3115,4704)
	(3116,4696)(3117,4693)(3117,4692)
\path(1362,2982)(1812,2982)(1812,2802)
	(1362,2802)(1362,2982)
\path(1272,2892)(1362,2892)
\path(1812,2892)(1902,2892)
\path(372,2532)(822,2532)(822,2352)
	(372,2352)(372,2532)
\path(282,2442)(372,2442)
\path(822,2442)(912,2442)
\path(867,1767)(1137,1767)
\path(912,1722)(1092,1722)
\path(957,1677)(1047,1677)
\path(1002,1767)(1002,1857)
\path(1184,2618)(1184,1898)(1994,2258)(1184,2618)
\path(1229,2438)(1319,2438)
\path(1229,2078)(1319,2078)
\path(1274,2123)(1274,2033)
\path(1272,822)(1272,462)
\path(1362,822)(1362,462)
\path(1272,642)(1182,642)
\path(1362,642)(1452,642)
\path(1182,507)(1497,822)
\path(1433.360,715.934)(1497.000,822.000)(1390.934,758.360)
\path(1902,282)(1542,282)
\path(1902,372)(1542,372)
\path(1722,282)(1722,192)
\path(1722,372)(1722,462)
\path(1587,192)(1902,507)
\path(1838.360,400.934)(1902.000,507.000)(1795.934,443.360)
\path(4782,822)(4782,462)
\path(4872,822)(4872,462)
\path(4782,642)(4692,642)
\path(4872,642)(4962,642)
\path(4692,507)(5007,822)
\path(4943.360,715.934)(5007.000,822.000)(4900.934,758.360)
\path(4602,282)(4242,282)
\path(4602,372)(4242,372)
\path(4422,282)(4422,192)
\path(4422,372)(4422,462)
\path(4287,192)(4602,507)
\path(4538.360,400.934)(4602.000,507.000)(4495.934,443.360)
\path(2442,417)(2352,417)
\path(2802,417)(2892,417)
\path(2802,237)(2487,642)
\path(2584.353,565.696)(2487.000,642.000)(2536.992,528.860)
\path(2442,417)(2441,419)(2440,423)
	(2438,428)(2436,435)(2436,442)
	(2438,451)(2442,462)(2446,468)
	(2449,475)(2452,481)(2455,487)
	(2457,492)(2459,498)(2461,503)
	(2462,508)(2464,513)(2467,516)
	(2471,518)(2475,518)(2481,514)
	(2487,507)(2494,493)(2501,477)
	(2506,460)(2510,444)(2514,428)
	(2517,413)(2520,398)(2523,383)
	(2526,368)(2529,352)(2531,338)
	(2532,327)(2531,323)(2528,324)
	(2523,329)(2518,335)(2512,342)
	(2506,350)(2500,358)(2494,367)
	(2490,377)(2486,389)(2485,402)
	(2487,417)(2492,432)(2499,445)
	(2506,457)(2513,467)(2519,476)
	(2525,485)(2531,492)(2537,499)
	(2545,505)(2555,510)(2565,511)
	(2577,507)(2587,496)(2596,482)
	(2602,465)(2606,449)(2609,433)
	(2611,417)(2613,401)(2614,385)
	(2617,369)(2619,352)(2621,338)
	(2622,327)(2621,323)(2618,324)
	(2613,329)(2608,335)(2602,342)
	(2596,350)(2590,358)(2584,367)
	(2580,377)(2576,389)(2575,402)
	(2577,417)(2582,432)(2589,445)
	(2596,457)(2603,467)(2609,476)
	(2615,485)(2621,492)(2627,499)
	(2635,505)(2645,510)(2655,511)
	(2667,507)(2677,496)(2686,482)
	(2692,465)(2696,449)(2699,433)
	(2701,417)(2703,401)(2704,385)
	(2707,369)(2709,352)(2711,338)
	(2712,327)(2711,323)(2708,324)
	(2703,329)(2698,335)(2692,342)
	(2686,350)(2680,358)(2674,367)
	(2670,377)(2666,389)(2665,402)
	(2667,417)(2672,432)(2679,445)
	(2686,455)(2693,464)(2699,471)
	(2705,477)(2711,483)(2717,488)
	(2725,494)(2735,500)(2745,505)
	(2757,507)(2767,505)(2776,499)
	(2783,490)(2788,480)(2792,469)
	(2795,456)(2798,444)(2800,433)
	(2801,425)(2802,420)(2802,417)
\path(2577,687)(2577,1047)
\path(2667,1047)(2667,687)
\path(2667,867)(2757,867)
\path(2577,867)(2487,867)
\path(2487,867)(2352,867)(2352,417)
\path(2892,417)(2892,867)(2757,867)
\thicklines
\put(2352,642){\ellipse{8}{8}}
\put(2892,642){\ellipse{8}{8}}
\thinlines
\path(2892,642)(3162,642)
\path(3612,642)(3611,642)(3607,643)
	(3599,643)(3586,645)(3570,646)
	(3552,647)(3534,648)(3515,647)
	(3496,646)(3477,642)(3468,640)
	(3461,637)(3454,634)(3449,632)
	(3444,630)(3440,628)(3437,627)
	(3434,626)(3431,624)(3428,623)
	(3426,622)(3424,621)(3422,619)
	(3421,617)(3420,615)(3420,613)
	(3421,609)(3423,606)(3426,601)
	(3432,597)(3446,590)(3462,583)
	(3479,578)(3495,574)(3511,570)
	(3526,567)(3541,564)(3556,561)
	(3571,558)(3587,555)(3601,553)
	(3612,552)(3616,553)(3615,556)
	(3610,561)(3604,566)(3597,572)
	(3589,578)(3581,584)(3572,590)
	(3562,594)(3550,598)(3537,599)
	(3522,597)(3507,592)(3494,585)
	(3482,578)(3472,571)(3463,565)
	(3454,559)(3447,553)(3440,547)
	(3434,539)(3429,529)(3428,519)
	(3432,507)(3443,497)(3457,488)
	(3474,482)(3490,478)(3506,475)
	(3522,473)(3538,471)(3554,470)
	(3570,467)(3587,465)(3601,463)
	(3612,462)(3616,463)(3615,466)
	(3610,471)(3604,476)(3597,482)
	(3589,488)(3581,494)(3572,500)
	(3562,504)(3550,508)(3537,509)
	(3522,507)(3507,502)(3494,495)
	(3482,488)(3472,481)(3463,475)
	(3454,469)(3447,463)(3440,457)
	(3434,449)(3429,439)(3428,429)
	(3432,417)(3443,407)(3457,398)
	(3474,392)(3490,388)(3506,385)
	(3522,383)(3538,381)(3554,380)
	(3570,377)(3587,375)(3601,373)
	(3612,372)(3616,373)(3615,376)
	(3610,381)(3604,386)(3597,392)
	(3589,398)(3581,404)(3572,410)
	(3562,414)(3550,418)(3537,419)
	(3522,417)(3507,412)(3494,405)
	(3482,398)(3472,391)(3463,385)
	(3454,379)(3447,373)(3440,367)
	(3434,359)(3429,349)(3428,339)
	(3432,327)(3443,317)(3457,308)
	(3474,302)(3490,298)(3506,295)
	(3522,293)(3538,291)(3554,290)
	(3570,287)(3587,285)(3601,283)
	(3612,282)(3616,283)(3615,286)
	(3610,291)(3604,296)(3597,302)
	(3589,308)(3581,314)(3572,320)
	(3562,324)(3550,328)(3537,329)
	(3522,327)(3507,322)(3494,315)
	(3482,308)(3472,301)(3463,295)
	(3454,289)(3447,283)(3440,277)
	(3434,269)(3429,259)(3428,249)
	(3432,237)(3443,227)(3457,218)
	(3474,212)(3490,208)(3506,205)
	(3522,203)(3538,201)(3554,200)
	(3570,197)(3587,195)(3601,193)
	(3612,192)(3616,193)(3615,196)
	(3610,201)(3604,206)(3597,212)
	(3589,218)(3581,224)(3572,230)
	(3562,234)(3550,238)(3537,239)
	(3522,237)(3507,232)(3494,225)
	(3482,218)(3472,211)(3463,205)
	(3454,199)(3447,193)(3440,187)
	(3434,179)(3429,169)(3428,159)
	(3432,147)(3443,137)(3458,128)
	(3477,121)(3497,116)(3519,112)
	(3542,109)(3564,106)(3583,104)
	(3598,103)(3607,102)(3611,102)(3612,102)
\path(3162,642)(3163,642)(3167,643)
	(3175,643)(3188,645)(3204,646)
	(3222,647)(3240,648)(3259,647)
	(3278,646)(3297,642)(3306,640)
	(3313,637)(3320,634)(3325,632)
	(3330,630)(3334,628)(3337,627)
	(3340,626)(3343,624)(3346,623)
	(3348,622)(3350,621)(3352,619)
	(3353,617)(3354,615)(3354,613)
	(3353,609)(3351,606)(3348,601)
	(3342,597)(3328,590)(3312,583)
	(3295,578)(3279,574)(3263,570)
	(3248,567)(3233,564)(3218,561)
	(3203,558)(3187,555)(3173,553)
	(3162,552)(3158,553)(3159,556)
	(3164,561)(3170,566)(3177,572)
	(3185,578)(3193,584)(3202,590)
	(3212,594)(3224,598)(3237,599)
	(3252,597)(3265,593)(3277,587)
	(3287,581)(3297,575)(3305,570)
	(3312,564)(3320,559)(3326,554)
	(3332,549)(3338,542)(3343,535)
	(3346,526)(3346,517)(3342,507)
	(3331,497)(3317,488)(3300,482)
	(3284,478)(3268,475)(3252,473)
	(3236,471)(3220,470)(3204,467)
	(3187,465)(3173,463)(3162,462)
	(3158,463)(3159,466)(3164,471)
	(3170,476)(3177,482)(3185,488)
	(3193,494)(3202,500)(3212,504)
	(3224,508)(3237,509)(3252,507)
	(3265,503)(3277,497)(3287,491)
	(3297,485)(3305,480)(3312,474)
	(3320,469)(3326,464)(3332,459)
	(3338,452)(3343,445)(3346,436)
	(3346,427)(3342,417)(3331,407)
	(3317,398)(3300,392)(3284,388)
	(3268,385)(3252,383)(3236,381)
	(3220,380)(3204,377)(3187,375)
	(3173,373)(3162,372)(3158,373)
	(3159,376)(3164,381)(3170,386)
	(3177,392)(3185,398)(3193,404)
	(3202,410)(3212,414)(3224,418)
	(3237,419)(3252,417)(3265,413)
	(3277,407)(3287,401)(3297,395)
	(3305,390)(3312,384)(3320,379)
	(3326,374)(3332,369)(3338,362)
	(3343,355)(3346,346)(3346,337)
	(3342,327)(3331,317)(3317,308)
	(3300,302)(3284,298)(3268,295)
	(3252,293)(3236,291)(3220,290)
	(3204,287)(3187,285)(3173,283)
	(3162,282)(3158,283)(3159,286)
	(3164,291)(3170,296)(3177,302)
	(3185,308)(3193,314)(3202,320)
	(3212,324)(3224,328)(3237,329)
	(3252,327)(3265,323)(3277,317)
	(3287,311)(3297,305)(3305,300)
	(3312,294)(3320,289)(3326,284)
	(3332,279)(3338,272)(3343,265)
	(3346,256)(3346,247)(3342,237)
	(3331,227)(3317,218)(3300,212)
	(3284,208)(3268,205)(3252,203)
	(3236,201)(3220,200)(3204,197)
	(3187,195)(3173,193)(3162,192)
	(3158,193)(3159,196)(3164,201)
	(3170,206)(3177,212)(3185,218)
	(3193,224)(3202,230)(3212,234)
	(3224,238)(3237,239)(3252,237)
	(3265,233)(3277,227)(3287,221)
	(3297,215)(3305,210)(3312,204)
	(3320,199)(3326,194)(3332,189)
	(3338,182)(3343,175)(3346,166)
	(3346,157)(3342,147)(3331,137)
	(3316,128)(3297,121)(3277,116)
	(3255,112)(3232,109)(3210,106)
	(3191,104)(3176,103)(3167,102)
	(3163,102)(3162,102)
\put(2307,912){\makebox(0,0)[lb]{{\SetFigFont{8}{9.6}{\sfdefault}{\mddefault}{\updefault}C5}}}
\put(2262,237){\makebox(0,0)[lb]{{\SetFigFont{8}{9.6}{\sfdefault}{\mddefault}{\updefault}L1}}}
\put(2937,147){\makebox(0,0)[lb]{{\SetFigFont{8}{9.6}{\sfdefault}{\mddefault}{\updefault}L2}}}
\put(3657,147){\makebox(0,0)[lb]{{\SetFigFont{8}{9.6}{\sfdefault}{\mddefault}{\updefault}L3}}}
\put(552.000,4962.000){\arc{360.000}{1.5708}{4.7124}}
\put(1722,3612){\ellipse{382}{382}}
\put(192,3612){\ellipse{284}{284}}
\thicklines
\put(1002,2442){\ellipse{8}{8}}
\put(2982,2262){\ellipse{8}{8}}
\put(4152,2712){\ellipse{8}{8}}
\put(2172,2892){\ellipse{8}{8}}
\put(1272,3612){\ellipse{8}{8}}
\put(2172,2262){\ellipse{8}{8}}
\put(4152,2082){\ellipse{8}{8}}
\put(537,4961){\ellipse{28}{28}}
\put(1722,642){\ellipse{8}{8}}
\put(1722,102){\ellipse{8}{8}}
\put(4422,642){\ellipse{8}{8}}
\put(4422,102){\ellipse{8}{8}}
\thinlines
\path(2982,2712)(2982,2262)
\path(3162,1902)(2982,1902)(2982,1677)
\path(3162,2262)(2892,2262)
\path(2172,2262)(2262,2262)
\path(4062,2712)(4152,2712)(4152,2082)(3972,2082)
\path(642,3792)(1002,3792)(1002,3432)
	(642,3432)(642,3792)
\path(4602,1632)(6132,1632)(6132,2532)
	(4602,2532)(4602,1632)
\path(2892,5412)(5682,5412)(5682,4512)
	(2892,4512)(2892,5412)
\path(5064,5178)(5127,5146)(4936,4795)
	(4872,4827)(5064,5178)
\path(327,3612)(642,3612)
\path(1002,3612)(1542,3612)
\path(1272,3612)(1272,4152)(1992,4152)(1992,4692)
\path(912,5232)(1452,5232)(1452,4692)
	(912,4692)(912,5232)
\path(192,4962)(372,4962)
\path(1992,5412)(1992,5232)
\path(3792,5232)(4152,5232)(4152,2712)
\path(4152,2082)(4602,2082)
\path(4961,4831)(4962,2532)
\path(5592,4962)(5862,4962)(5862,2532)
\path(98,3517)(291,3712)
\path(96,3706)(290,3517)
\thicklines
\path(5457,4962)(597,4962)
\thinlines
\path(1002,2442)(1002,2892)(1272,2892)
\path(912,2442)(1182,2442)
\path(1182,2082)(1002,2082)(1002,1857)
\path(3792,4692)(3792,3612)(2172,3612)
	(2172,2262)(1992,2262)
\path(1902,2892)(2172,2892)
\path(822,462)(1002,462)(1002,642)(1182,642)
\path(822,282)(1002,282)(1002,102)
	(1722,102)(1722,192)
\path(5322,282)(5142,282)(5142,102)
	(4422,102)(4422,192)
\path(5322,462)(5142,462)(5142,642)(4962,642)
\path(1722,462)(1722,642)
\path(4422,462)(4422,642)
\path(5322,732)(6132,732)(6132,12)
	(5322,12)(5322,732)
\path(12,732)(822,732)(822,12)
	(12,12)(12,732)
\path(12,3207)(372,3207)(372,2847)
	(12,2847)(12,3207)
\path(372,4017)(12,4017)(12,4377)
	(372,4377)(372,4017)
\path(192,4962)(192,4377)
\path(192,4017)(192,3747)
\path(192,3477)(192,3207)
\path(192,2847)(192,2442)(282,2442)
\path(1452,642)(2352,642)
\path(1722,102)(3162,102)
\path(3612,102)(4422,102)
\path(4692,642)(3612,642)
\dashline{60.000}(2082,1182)(3972,1182)(3972,12)
	(2082,12)(2082,1182)
\path(1002,5637)(1002,5635)(1002,5632)
	(1002,5626)(1002,5616)(1002,5602)
	(1002,5586)(1002,5565)(1001,5542)
	(1001,5516)(1001,5488)(1001,5459)
	(1001,5428)(1001,5397)(1002,5365)
	(1002,5333)(1002,5300)(1003,5266)
	(1003,5232)(1004,5197)(1005,5162)
	(1006,5126)(1007,5090)(1009,5055)
	(1012,5001)(1015,4959)(1017,4928)
	(1018,4907)(1019,4895)(1019,4888)
	(1019,4885)(1019,4883)(1020,4879)
	(1021,4873)(1024,4864)(1028,4852)
	(1034,4836)(1042,4822)(1048,4815)
	(1054,4810)(1060,4808)(1065,4807)
	(1070,4807)(1074,4809)(1077,4811)
	(1081,4813)(1084,4816)(1086,4819)
	(1089,4823)(1092,4826)(1095,4831)
	(1099,4835)(1102,4841)(1107,4847)
	(1111,4854)(1116,4862)(1121,4871)
	(1126,4882)(1132,4901)(1136,4920)
	(1138,4935)(1139,4948)(1140,4959)
	(1139,4968)(1139,4977)(1138,4988)
	(1137,5000)(1135,5014)(1133,5029)
	(1130,5044)(1128,5050)(1126,5054)
	(1124,5058)(1121,5061)(1119,5064)
	(1117,5066)(1115,5067)(1113,5068)
	(1111,5069)(1109,5069)(1107,5070)
	(1106,5070)(1104,5070)(1102,5070)
	(1101,5069)(1099,5069)(1097,5068)
	(1095,5067)(1093,5066)(1091,5065)
	(1089,5062)(1087,5060)(1085,5057)
	(1082,5053)(1080,5048)(1078,5042)
	(1075,5027)(1072,5012)(1071,4998)
	(1069,4987)(1068,4977)(1067,4969)
	(1067,4960)(1067,4950)(1068,4938)
	(1070,4922)(1074,4903)(1081,4882)
	(1090,4863)(1098,4848)(1106,4836)
	(1111,4828)(1116,4823)(1119,4819)
	(1123,4815)(1128,4811)(1135,4807)
	(1145,4802)(1158,4798)(1173,4796)
	(1188,4798)(1201,4802)(1211,4806)
	(1219,4810)(1224,4813)(1228,4817)
	(1232,4820)(1237,4826)(1243,4834)
	(1251,4845)(1259,4860)(1268,4880)
	(1274,4902)(1278,4922)(1280,4939)
	(1281,4952)(1280,4963)(1279,4973)
	(1278,4982)(1276,4992)(1275,5005)
	(1272,5019)(1270,5035)(1266,5051)
	(1264,5057)(1262,5061)(1260,5065)
	(1258,5068)(1256,5071)(1254,5073)
	(1252,5074)(1250,5076)(1248,5076)
	(1247,5077)(1245,5077)(1244,5078)
	(1242,5078)(1241,5078)(1239,5078)
	(1238,5077)(1237,5077)(1235,5076)
	(1234,5075)(1232,5074)(1230,5073)
	(1228,5071)(1226,5069)(1224,5066)
	(1222,5063)(1220,5059)(1218,5054)
	(1216,5048)(1212,5032)(1210,5016)
	(1208,5001)(1206,4989)(1205,4978)
	(1204,4969)(1203,4959)(1202,4948)
	(1203,4935)(1205,4919)(1209,4900)
	(1216,4880)(1222,4868)(1228,4858)
	(1235,4849)(1240,4841)(1246,4835)
	(1250,4829)(1254,4824)(1258,4820)
	(1262,4816)(1266,4813)(1269,4810)
	(1274,4807)(1279,4805)(1285,4804)
	(1291,4805)(1298,4807)(1306,4812)
	(1314,4820)(1323,4835)(1330,4850)
	(1335,4863)(1338,4872)(1340,4878)
	(1340,4881)(1340,4884)(1340,4888)
	(1340,4894)(1341,4907)(1343,4927)
	(1345,4958)(1348,5001)(1352,5055)
	(1354,5090)(1356,5126)(1357,5162)
	(1358,5198)(1359,5233)(1360,5267)
	(1361,5300)(1361,5333)(1361,5365)
	(1362,5397)(1362,5429)(1362,5459)
	(1362,5488)(1362,5516)(1362,5542)
	(1362,5565)(1362,5586)(1362,5603)
	(1362,5616)(1362,5626)(1362,5632)
	(1362,5635)(1362,5637)
\put(1632,3567){\makebox(0,0)[lb]{{\SetFigFont{6}{7.2}{\sfdefault}{\mddefault}{\updefault}LO}}}
\put(732,3567){\makebox(0,0)[lb]{{\SetFigFont{6}{7.2}{\sfdefault}{\mddefault}{\updefault}PS}}}
\put(2487,2397){\makebox(0,0)[lb]{{\SetFigFont{6}{7.2}{\sfdefault}{\mddefault}{\updefault}R3}}}
\put(3162,2937){\makebox(0,0)[lb]{{\SetFigFont{6}{7.2}{\sfdefault}{\mddefault}{\updefault}C}}}
\put(3657,2847){\makebox(0,0)[lb]{{\SetFigFont{6}{7.2}{\sfdefault}{\mddefault}{\updefault}R4}}}
\put(4647,5277){\makebox(0,0)[lb]{{\SetFigFont{6}{7.2}{\sfdefault}{\mddefault}{\updefault}Brewster}}}
\put(5097,4557){\makebox(0,0)[lb]{{\SetFigFont{6}{7.2}{\sfdefault}{\mddefault}{\updefault}mirror}}}
\put(5097,4692){\makebox(0,0)[lb]{{\SetFigFont{6}{7.2}{\sfdefault}{\mddefault}{\updefault}Piezo}}}
\put(1092,5457){\makebox(0,0)[lb]{{\SetFigFont{6}{7.2}{\sfdefault}{\mddefault}{\updefault}RF}}}
\put(4917,1902){\makebox(0,0)[lb]{{\SetFigFont{6}{7.2}{\sfdefault}{\mddefault}{\updefault}processing}}}
\put(957,4512){\makebox(0,0)[lb]{{\SetFigFont{6}{7.2}{\sfdefault}{\mddefault}{\updefault}Crystal}}}
\put(3387,2037){\makebox(0,0)[lb]{{\SetFigFont{6}{7.2}{\sfdefault}{\mddefault}{\updefault}IC2}}}
\put(4647,5142){\makebox(0,0)[lb]{{\SetFigFont{6}{7.2}{\sfdefault}{\mddefault}{\updefault}plate}}}
\put(1497,3027){\makebox(0,0)[lb]{{\SetFigFont{6}{7.2}{\sfdefault}{\mddefault}{\updefault}R2}}}
\put(507,2577){\makebox(0,0)[lb]{{\SetFigFont{6}{7.2}{\sfdefault}{\mddefault}{\updefault}R1}}}
\put(1407,2217){\makebox(0,0)[lb]{{\SetFigFont{6}{7.2}{\sfdefault}{\mddefault}{\updefault}IC1}}}
\put(2037,5142){\makebox(0,0)[lb]{{\SetFigFont{6}{7.2}{\sfdefault}{\mddefault}{\updefault}EOM2}}}
\put(3342,5142){\makebox(0,0)[lb]{{\SetFigFont{6}{7.2}{\sfdefault}{\mddefault}{\updefault}EOM1}}}
\put(12,1002){\makebox(0,0)[lb]{{\SetFigFont{10}{12.0}{\rmdefault}{\bfdefault}{\updefault}b)}}}
\put(12,5502){\makebox(0,0)[lb]{{\SetFigFont{10}{12.0}{\rmdefault}{\bfdefault}{\updefault}a)}}}
\put(1452,417){\makebox(0,0)[lb]{{\SetFigFont{8}{9.6}{\sfdefault}{\mddefault}{\updefault}C2}}}
\put(1002,687){\makebox(0,0)[lb]{{\SetFigFont{8}{9.6}{\sfdefault}{\mddefault}{\updefault}C1}}}
\put(4152,417){\makebox(0,0)[lb]{{\SetFigFont{8}{9.6}{\sfdefault}{\mddefault}{\updefault}C4}}}
\put(4512,687){\makebox(0,0)[lb]{{\SetFigFont{8}{9.6}{\sfdefault}{\mddefault}{\updefault}C3}}}
\put(3072,777){\makebox(0,0)[lb]{{\SetFigFont{8}{9.6}{\sfdefault}{\mddefault}{\updefault}cryostat}}}
\put(3072,957){\makebox(0,0)[lb]{{\SetFigFont{8}{9.6}{\sfdefault}{\mddefault}{\updefault}Inside}}}
\put(134,4133){\makebox(0,0)[lb]{{\SetFigFont{6}{7.2}{\sfdefault}{\mddefault}{\updefault}N}}}
\put(109,2957){\makebox(0,0)[lb]{{\SetFigFont{6}{7.2}{\sfdefault}{\mddefault}{\updefault}LP}}}
\put(57,327){\makebox(0,0)[lb]{{\SetFigFont{8}{9.6}{\sfdefault}{\mddefault}{\updefault}10 MHz}}}
\put(5367,327){\makebox(0,0)[lb]{{\SetFigFont{8}{9.6}{\sfdefault}{\mddefault}{\updefault}17 MHz}}}
\put(4917,2082){\makebox(0,0)[lb]{{\SetFigFont{6}{7.2}{\sfdefault}{\mddefault}{\updefault}Low-frequency}}}
\end{picture}
}
\caption{{\bf (a)} Optical and electronic design schematics.
  Abbreviations: N, notch filter; LP, low-pass diplexer; PS, phase
  shifter; LO, local oscillator; EOM, electro-optical modulator. {\bf
    (b)} Electronic schematics of the RF-eraser circuit. See the text
  for component values and details.}
  \label{fig:design-schematics}
\end{figure}


%
Based on our theoretical calculations we are now able to design the
electronic feedback system giving the best performance. Below in
Sec.~\ref{sec:Exp_setup_feedback_sys} we describe our setup for the
feedback system, and in Sec.~\ref{sec:limiting_factors_design} we
describe how to estimate the limitations of the performance.

\subsection{Experimental feedback system setup}
\label{sec:Exp_setup_feedback_sys}
The main building blocks of the feedback system are shown in
Fig.~\ref{fig:design-schematics}(a). Starting from the upper right
corner we show the laser cavity of our Coherent CR699-21 dye laser.
The most important elements controlling the laser frequency in the
commercial version of the laser are a piezo-mounted mirror and a
Brewster plate. These were retained in our setup. In addition we
placed a Linos PM25 electro-optical modulator (EOM 1) inside the
cavity. We fed the two electrodes by two separate amplifier circuits;
IC1 which is fast and IC2 which can deliver high voltage ($\pm 200$
V).

From the laser output the laser beam was directed through a New Focus
model 4002 modulator (EOM 2) applying $\omegam = 2\pi\cdot 50$ MHz
modulation from a local oscillator (Wenzel Associates, Inc.). The
modulated beam was then expanded to cover the entire area of a \PrYSO
crystal with diameter of 19 mm, thickness of 5 mm, and doping
concentration 0.005\%, purchased from Scientific Materials Corp. This
crystal was kept in a cryostat (Oxford Optistat CF-V) operated at 3.0
K.

The electronic circuit design for the RF-eraser coil system, which
enables us to control the hole lifetime, is shown in detail in
Fig.~\ref{fig:design-schematics}(b).  Surrounding the crystal are two
home built sets of coils denoted $L_2$ and $L_3$, each consisting of
two three-turn coils with a diameter of 23 mm and 7 mm separation.
These coils function practically as a transformer, coupling the 10.19
MHz and the 17.31 MHz channel, since they are placed on top of each
other.  In order to effectively reinforce inductance, making resonant
circuits out of the two RF channels a band-stop filter consisting of
$L_1 = 0.76$ {\micro}H and $C_5 = 100$ pF is inserted and adjusted to
block the 17.31 MHz signal.  By adjusting $C_1$ and $C_2$ in
combination we can now tune the resonance frequency for the 10.19 MHz
channel and assure a 50 $\Omega$ impedance match.  Likewise, $C_3$ and
$C_4$ are adjusted to assure the correct resonance frequency and
impedance of the 17.31 MHz channel (the two channels are not
completely independent and adjustments must be made iteratively).  The
effective values of $C_1$ - $C_4$ depend to a large extent on stray
capacitances in the approx.~50 cm leads from the outside adjustable
capacitors to the coils inside the crystal. We used 35 pF adjustable
capacitors for $C_1$ - $C_4$ in addition to 22 pF, 22 pF, and 47 pF
capacitors coupled in parallel to $C_1$ - $C_3$, respectively.

The RF fields are generated as sawtooth sweeps, the 10.19 MHz signal
is 100 kHz wide, and the 17.31 MHz signal is 200 kHz wide. The sweep
time is 0.82 ms, which is comparable to the hyperfine level coherence
time of 0.50 ms and hence the pumping becomes effectively incoherent.
This procedure assures smooth re-population over time of the hyperfine
levels since atoms with different frequencies on the inhomogeneous
hyperfine transition are affected at different times.

After the light has passed the crystal it is measured by a Thorlabs
detector (PDA10BS-AC, bandwidth 150 MHz) and the output is sent
through a notch filter to block possible higher order modulation at
100 MHz, which is known to cause systematic offsets in cavity locking
setups \cite{Salomon1988}. The notch filter also blocks possible
electronic pick-up at 10.19 MHz and 17.31 MHz. The signal is then
mixed with the phase-shifted local oscillator on a Mini-circuits LPD-2
mixer and after low-pass filtering (20 MHz cut-off), the error signal
is produced at DC frequency. Both the notch and low-pass filters are
designed as diplexers with 50 $\Omega$ impedance matching at all
frequencies to avoid reflections (this is particularly important for
the low-pass filter to avoid back-reflections into the mixer
\cite{Hamilton1991}).

Based on the error signal we actuate the laser frequency using EOM 1
which is driven by the pure analog electronics shown in
Fig.~\ref{fig:design-schematics}(a) around IC1 and IC2. The complex
electronic gain from the error signal to the voltage across the
electrodes of EOM 1 is given by:
\begin{equation}
  \label{eq:elec_gain}
  g(\omega) = \frac{R_2}{R_1}\cdot\frac{i\omega(R_3 + R_4)C + 1}{i\omega R_3 C}. 
\end{equation}
We see that there is a characteristic cutoff frequency,
$f_{\mathrm{c}} = \frac{\omega_{\mathrm{c}}}{2\pi} =
\frac{1}{2\pi(R_3+R_4)C}$, which separates this gain into a
low-frequency part proportional to $\frac{1}{i\omega}$ and a
high-frequency part where the gain is constant and real. The resistors
$R_1$ to $R_4$ are adjustable but designed to work in a range such
that the critical frequency, $\omega_{\mathrm{c}}$, can be set equal
to $\Gammahole$. In this manner the electronic gain of
Eq.~\eqref{eq:elec_gain} together with the medium- and high-frequency
part of the atomic response shown in
Fig.~\ref{fig:Transfer_function_example}(a) add up to a total response
proportional to $\frac{1}{i\omega}$. The actual component values used
were $R_1 = 50$ $\Omega$, $R_2 = 50$ k$\Omega$, $R_3 = R_4 = 1$
k$\Omega$, and $C = 6$ nF. The resistors $R_2$ - $R_4$ are adjustable.
The amplifier circuits are based on IC1, which is Analog Devices
amplifier, model AD8021 and IC2 which is an Apex model PA85 amplifier.
Circuits for maintaining low offset levels, for preventing integrator
windup \cite{Astrom2005}, and for signal monitoring, are not shown in
Fig.~\ref{fig:design-schematics}(a).

The output of IC2 is also sent to a low-frequency unit which consists
of a PC processing the input digitally and in turn sending a signal to
the commercial part of the laser. Specifically, the input from IC2 is
sent to a PI regulator (i.e.~an amplifier with gain $g(\omega) \propto
\frac{c_1}{i\omega} + c_2 \propto \frac{1/\tau+i\omega}{i\omega}$). We
chose $\tau$ to be equal to a typical value for the hole lifetime,
$\Trg$. Combining this particular response of the digital processing
with the atomic response, corresponds mathematically to replacing the
denominator $i\omega+\frac{1}{\Trg}$ in the transfer function in
Eq.~\eqref{eq:Transfer_function} by $i\omega$. In other words, the
digital processing together with the low- and medium-frequency
response shown in Fig.~\ref{fig:Transfer_function_example}(a) amounts
to a constant gain response. Together with the low-frequency response,
proportional to $\frac{1}{i\omega}$ of the analog electronic part, the
total response of the entire analog and digital system at all
frequencies becomes proportional to $\frac{1}{i\omega}$. This is known
from feedback theory \cite{Astrom2005} to assure stable operation. The
digitally processed signal is added as a current to the reference
photo-diode behind the stabilization reference cavity. In this manner
the commercial laser control system will detect a frequency error and
actuate the piezo-mounted mirror and the Brewster plate. Finally, to
assure that the working point of the reference cavity remains within
an appropriate range, an integrated version of the digitally processed
signal is sent to the external scan input of the commercial laser
control, which in turn changes the reference cavity length and
maintains the current added to the photo-diode around zero on average.

Apart from a stable frequency it is also important to stabilize the
laser output power. To this end we split off a small fraction of the
laser beam to monitor the laser power. This in turn is used in a
feedback loop to control the amount of pump light sent into the laser
gain medium. Our dye laser is pumped by a Coherent Verdi-V6 laser, and
a Brimrose AOM (model FQM-80-1-.532/WQ) is placed in the pump beam to
adjust the power. Details of the feedback electronics are available in
\cite{Rippe2006, GroupHomepage}.

\subsection{Factors limiting the laser performance}
\label{sec:limiting_factors_design}
Two important design parameters in the setup are the group delay of
signals in the feedback loop and the signal-to-noise ratio (SNR). For
this reason we chose high-speed, low-noise amplifiers, and the group
delay of the filter diplexers was carefully designed so as to be
constant over relevant frequency ranges in order to avoid distortion.
The total feedback loop delay time is below 100 ns. For technical
details see \cite{Rippe2006, GroupHomepage}. Note also, that short
delay times are only possible if the modulation frequency is high.

As discussed in Sec.~\ref{sec:Exp_setup_feedback_sys}, the total
response of the atoms and the electronic feedback system is a gain
function proportional $\frac{1}{i\omega}$. This is purely imaginary,
and corresponds to a constant phase shift of $-90^{\circ}$ for all
frequencies. In reality, there are corrections to this since
amplifiers, mixers, and filters have finite bandwidths. The total
group delay, $\tau$, of the entire system can be modeled as a gain
function, $g_{\mathrm{delay}}(\omega)= e^{-i\omega\tau}$. If the total
feedback gain is too strong we expect oscillations at a frequency
where the feedback phase is $-180^{\circ}$. This will occur around a
frequency for which $e^{-i\omega\tau} = -i$ since we then add
$-90^{\circ}$ to the high-frequency phase shift which is already
$-90^{\circ}$.  For $\tau = 100$ ns we estimate a critical frequency
for oscillations $f_{\mathrm{osc}} =\frac{\omega_{\mathrm{osc}}}{2\pi}
= \frac{1}{4\tau} = 2.5$ MHz which agrees well with our experimental
observations. We discuss the consequences of this delay on laser
stability in Sec.~\ref{sec:phase_stability}. 

Let us describe how to estimate the impact of detector noise.  For our
Thorlabs PDA10BS-AC detector the electronics noise specification is
$-63$ dBm in a 1 MHz bandwidth around the 50 MHz center frequency,
which can be translated into a noise in the output voltage of
$0.16\:\frac{\mu\mathrm{V}}{\sqrt{\mathrm{Hz}}}$. The shot noise
current measured in a bandwidth $B$ (in Hertz) is given by
$i_{\mathrm{SN}} = \sqrt{2eBi_{\mathrm{d}}}$ \cite{Hobbs1997}, where
$i_{\mathrm{d}}$ is the detected current. Knowing the power-to-current
conversion, $C_{I/P} = 0.37$ A/W, and the trans-impedance of the
detector, $C_{U/I} = 1.65\cdot 10^4$ V/A, we may calculate the output
voltage noise given the detected power, $P$.  Adding this to the
electronics noise we obtain $U_{\mathrm{det}}^{\mathrm{noise}} =
0.18\:\frac{\mu\mathrm{V}}{\sqrt{\mathrm{Hz}}} \sqrt{P[\mathrm{mW}] +
  0.77}$. After the photo-detector we mix the output signal with the
50 MHz local oscillator and low-pass filter the result with a
bandwidth $B = 20$ MHz. We can thus calculate the equivalent noise
power given the detected DC power, $P$, as
$P_{\mathrm{det}}^{\mathrm{noise}}=
U_{\mathrm{det}}^{\mathrm{noise}}\sqrt{B}/(C_{I/P}C_{U/I}) = 1.3 \cdot
10^{-4}\mathrm{mW} \sqrt{P[\mathrm{mW}] + 0.77}$. This should be
compared with the amplitude, $P_{\mathrm{det}}^{\mathrm{signal}}= 4P
J_0 J_1\Re\left\{T(\omega)\epsilon\omega e^{i\omega t}\right\}$, of
the 50 MHz modulations in the power from
Eq.~\eqref{eq:Transfer_function}. 

As we shall see, in the experiments described in
Sec.~\ref{sec:laser-characterization} the best laser performance was
observed using a detected power, $P$, of 0.23 mW which corresponds to
the equivalent noise power $P_{\mathrm{det}}^{\mathrm{noise}}=1.3\cdot
10^{-4}\:\mathrm{mW} = 5.7\cdot 10^{-4}P$.  The modulation index, $m$,
is 0.20, giving $J_0 J_1 = 0.1$. In the high frequency regime the
transfer function $T(\omega)$ can be estimated from
Eq.~\eqref{eq:Trans_func_HF}.  Assuming $\alpha_0 L = 0.66$
(corresponding to 52\% transmission) and
$\frac{\Gammah}{\Gammaholec}\approx 0.5$,
$\frac{\Gammah}{\Gammaholes}\approx 1$, we find $T(\omega) \approx
-\frac{0.093}{i\omega}$. The signal can then be estimated as
$P_{\mathrm{det}}^{\mathrm{signal}} \approx 6.4\cdot 10^{-4} P$ for a
harmonic error with phase variations of one degree, i.e.~$\epsilon =
1\cdot 2\pi/360$. This estimate is close to the noise limit, which
means that with a detection bandwidth of 20 MHz we can essentially
detect a 1 degree phase error with a SNR of unity
after an integration time of 50 ns. 

We note that an even better SNR can be obtained with higher powers and
a higher modulation index. However, as we shall see in
Sec.~\ref{sec:Drift_meas}, the presence of laser drift suggests that
we choose the above values. The fact that our \PrYSO crystal has a
diameter of 19 mm allows for a relatively high power without burning
too deep holes causing drift problems.


\section{Laser characterization}
\label{sec:laser-characterization}
In this section we have two main objectives: to characterize the
performance of our stabilized laser and to present some experimental
verification of the theory concerning the drift calculations in
Sec.~\ref{sec:laser-drift}.

The conventional method of characterizing laser performance is to
build two identical lasers, setting up an interference experiment with
these, and calculate the Allen deviation of the observed intensity
beatings on different timescales. Since we only have a single
stabilized laser we cannot apply this simple and useful method.
Instead, we characterized the long-term drift by burning a spectral
hole in an auxiliary crystal and measuring how the shift in laser
frequency relative to this hole. For short timescales (up to the
optical coherence time, $T_2$, of the atoms) we use optical free
induction decay (FID) to measure the phase stability.  We will not
present experimental data on intermediate timescales, but the extremes
above correspond exactly to the important cases for our applications
in quantum information processing.

As already discussed in Sec.~\ref{sec:general_remarks_error_signal},
there are three ground and three excited states in \PrYSO, as shown in
Fig.~\ref{fig:different_atomic_levels}(e). The transition strengths
between these are very different; the strongest and most contributing
ones are $\pm\frac{1}{2}\rightarrow\pm\frac{1}{2}$ (0.55),
$\pm\frac{1}{2}\rightarrow\pm\frac{3}{2}$ (0.38),
$\pm\frac{3}{2}\rightarrow\pm\frac{1}{2}$ (0.40),
$\pm\frac{3}{2}\rightarrow\pm\frac{3}{2}$ (0.60), and
$\pm\frac{5}{2}\rightarrow\pm\frac{5}{2}$ (0.93), while the remaining
four are weak (0.07 or less). The numbers in parentheses are the
relative strengths taken from \cite{Nilsson2004} and we assume that
these are also valid for branching ratios in the decay process. We see
that the ions typically stay within the
$\pm\frac{1}{2},\pm\frac{3}{2}$ space or in the $\pm\frac{5}{2}$
state, and only seldom change between these (the total crossing
probability being 7\%). If this crossing occurs, the RF pumping on the
ground state hyperfine transition
$\pm\frac{5}{2}\rightarrow\pm\frac{3}{2}$ at 17.31 MHz transition will
counteract it. Since the crossing is infrequent, the timescale for
this RF transition $T_{\mathrm{17MHz}}$ can be relatively slow (in
order to maintain a certain hole depth).  On the other hand, an ion
resonant on, e.g.,~the $\pm\frac{1}{2}\rightarrow\pm\frac{1}{2}$
optical transition will decay to the ground $\pm\frac{3}{2}$ state
with a high probability (40\%), and the RF pumping on the hyperfine
transition $\pm\frac{3}{2}\rightarrow\pm\frac{1}{2}$ at 10.19 MHz must
counteract this with a relatively short timescale $T_{\mathrm{10MHz}}$
(to maintain the same hole depth for this atomic species). In all
experiments (apart from some of those illustrated in
Fig.~\ref{fig:CompareCenterSideShape}) we have set $T_{\mathrm{17MHz}}
= 10\cdot T_{\mathrm{10MHz}}$. Hence, by combining a small branching
ratio for the optical decay taking an atom away from the resonant
transition with a long timescale for the RF pumping bringing the atom
back to the resonant transition (or vice versa), we ascertain that the
parameter $R$ is similar for the atomic species resonant on the five
transitions mentioned above (according to the results in
Tab.~\ref{tab:G_R}).  Since, also, the strengths of these five
transitions are not that different, we expect to see experimental
results not too different from the simple three-level model of
Fig.~\ref{fig:different_atomic_levels}(b).
\subsection{Experimental setup}
\label{sec:characterization_exp_setup}
\begin{figure}[t]
  \centering
\setlength{\unitlength}{0.00055556in}
\begingroup\makeatletter\ifx\SetFigFont\undefined%
\gdef\SetFigFont#1#2#3#4#5{%
  \reset@font\fontsize{#1}{#2pt}%
  \fontfamily{#3}\fontseries{#4}\fontshape{#5}%
  \selectfont}%
\fi\endgroup%
{\renewcommand{\dashlinestretch}{30}
\begin{picture}(6019,2829)(0,-10)
\put(386.000,1902.000){\arc{360.000}{1.5708}{4.7124}}
\thicklines
\put(371,1901){\ellipse{28}{28}}
\thinlines
\path(26,1902)(206,1902)
\put(386.000,1100.000){\arc{360.000}{1.5708}{4.7124}}
\thicklines
\put(371,1099){\ellipse{28}{28}}
\thinlines
\path(26,1100)(206,1100)
\put(372.000,290.000){\arc{360.000}{1.5708}{4.7124}}
\thicklines
\put(357,289){\ellipse{28}{28}}
\thinlines
\path(12,290)(192,290)
\put(5988.500,2622.000){\arc{765.000}{2.6516}{3.6315}}
\put(5313.500,2622.000){\arc{765.000}{5.7932}{6.7731}}
\put(3491.000,1092.000){\arc{180.000}{1.5708}{4.7124}}
\put(2546.000,282.000){\arc{180.000}{4.7124}{7.8540}}
\path(1556,2172)(2096,2172)(2096,1632)
	(1556,1632)(1556,2172)
\thicklines
\path(701,2082)(926,2307)
\thinlines
\path(1286,2442)(2636,2442)(2636,2802)
	(1286,2802)(1286,2442)
\path(1286,2622)(26,2622)(26,1902)
\thicklines
\path(2636,2622)(3446,2622)(3446,1902)(431,1902)
\path(3311,1767)(3581,2037)
\path(3581,2487)(3311,2757)
\path(431,1092)(836,1092)(836,2172)(4481,1092)
\path(701,957)(971,1227)
\thinlines
\path(1556,552)(2096,552)(2096,12)
	(1556,12)(1556,552)
\thicklines
\path(2546,282)(431,282)
\path(3446,2622)(4751,2622)
\thinlines
\path(5201,2802)(5381,2802)(5381,2442)
	(5201,2442)(5201,2802)
\thicklines
\path(4751,2622)(5291,2622)(5651,2532)(5966,2532)
\path(4796,2487)(5066,2757)
\path(4796,957)(5066,1227)
\path(4931,2622)(4931,1092)(3491,1092)
\thinlines
\path(4346,1272)(4526,1272)(4526,912)
	(4346,912)(4346,1272)
\thicklines
\path(5966,2667)(5966,2397)
\thinlines
\path(3401,1092)(3400,1092)(3397,1092)
	(3387,1092)(3372,1091)(3350,1091)
	(3324,1090)(3296,1088)(3268,1087)
	(3241,1085)(3216,1083)(3193,1081)
	(3173,1078)(3155,1075)(3138,1071)
	(3123,1067)(3109,1062)(3093,1056)
	(3077,1048)(3062,1040)(3046,1031)
	(3031,1021)(3017,1010)(3003,999)
	(2990,987)(2978,975)(2968,963)
	(2958,952)(2950,941)(2942,930)
	(2936,919)(2931,909)(2926,899)
	(2922,888)(2919,878)(2917,867)
	(2915,855)(2915,843)(2916,832)
	(2918,820)(2922,808)(2926,797)
	(2931,785)(2937,774)(2944,762)
	(2951,750)(2960,737)(2970,724)
	(2981,710)(2993,695)(3006,680)
	(3019,664)(3031,649)(3044,634)
	(3056,619)(3067,605)(3077,592)
	(3086,579)(3094,567)(3100,555)
	(3106,544)(3111,532)(3116,521)
	(3119,509)(3121,498)(3123,487)
	(3123,475)(3123,465)(3121,454)
	(3119,444)(3116,435)(3113,426)
	(3109,417)(3103,407)(3096,396)
	(3088,385)(3079,374)(3068,363)
	(3055,353)(3041,342)(3027,333)
	(3011,324)(2994,317)(2977,310)
	(2959,304)(2944,301)(2928,298)
	(2911,295)(2891,293)(2869,291)
	(2844,289)(2817,287)(2786,286)
	(2755,285)(2723,284)(2694,283)
	(2669,283)(2652,282)(2641,282)
	(2637,282)(2636,282)
\put(1421,2577){\makebox(0,0)[lb]{{\SetFigFont{6}{7.2}{\sfdefault}{\mddefault}{\updefault}Stabilized laser}}}
\put(1421,1452){\makebox(0,0)[lb]{{\SetFigFont{6}{7.2}{\sfdefault}{\mddefault}{\updefault}Locking crystal}}}
\put(3041,732){\makebox(0,0)[lb]{{\SetFigFont{6}{7.2}{\sfdefault}{\mddefault}{\updefault}Fiber}}}
\put(206,507){\makebox(0,0)[lb]{{\SetFigFont{6}{7.2}{\sfdefault}{\mddefault}{\updefault}Det3}}}
\put(206,1317){\makebox(0,0)[lb]{{\SetFigFont{6}{7.2}{\sfdefault}{\mddefault}{\updefault}Det2}}}
\put(206,2127){\makebox(0,0)[lb]{{\SetFigFont{6}{7.2}{\sfdefault}{\mddefault}{\updefault}Det1}}}
\put(1466,642){\makebox(0,0)[lb]{{\SetFigFont{6}{7.2}{\sfdefault}{\mddefault}{\updefault}Extra crystal}}}
\put(5066,2307){\makebox(0,0)[lb]{{\SetFigFont{6}{7.2}{\sfdefault}{\mddefault}{\updefault}AOM1}}}
\put(4256,1317){\makebox(0,0)[lb]{{\SetFigFont{6}{7.2}{\sfdefault}{\mddefault}{\updefault}AOM2}}}
\put(5066,2127){\makebox(0,0)[lb]{{\SetFigFont{6}{7.2}{\sfdefault}{\mddefault}{\updefault}(double pass)}}}
\end{picture}
}
\caption{Experimental setup for characterizing the laser performance.
  Apart from the locking beam needed for the laser stabilization
  system we place an additional probing beam for characterizing the
  locking itself. AOM 1 is in double pass configuration and allows us
  to scan the laser beam frequency without any beam motion. AOM 2
  allows us to shift back to the original stabilized laser frequency
  to characterize the holes when the laser is locked. An extra crystal
  in another cryostat is used for measuring the laser frequency drift
  on long time scales.}
\label{fig:exp_setup_characterization}
\end{figure}
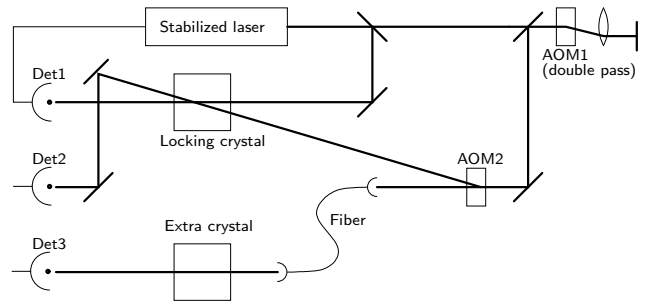

In addition to the laser stabilization system itself, the experimental
setup used to characterize the laser performance is shown in
Fig.~\ref{fig:exp_setup_characterization}. We have the possibility to
send part of the stabilized beam through two acousto-optical
modulators (AOM 1 and 2 in Fig.~\ref{fig:exp_setup_characterization},
both of which are A.~A.~Opto-Electronique modulators).  The first one
(AOM 1) has center a frequency of 200 MHz and a bandwidth of 100 MHz,
and is placed in double-pass configuration. AOM 2 has a center
frequency of 350 MHz and a bandwidth of 200 MHz, and is placed in
single-pass configuration. We may use the zeroth- or first-order
diffracted beams from AOM 2 for experiments on the locking crystal or
on an auxiliary crystal (5 mm thick 0.005\% \PrYSO cooled to 2 K in an
Oxford Spectromag cryostat).  Both AOMs are driven by a Tektronix
AWG520 1 GHz arbitrary waveform generator.

\subsubsection{Drift measurement setup}
\label{sec:setup_drift_meas}
\begin{figure}[t]
  \centering
  \includegraphics{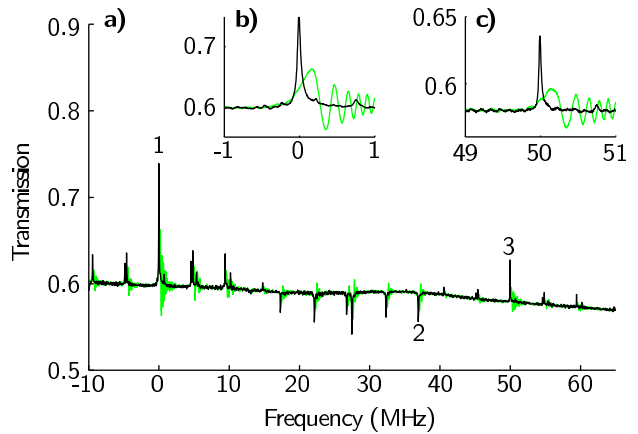}
  \caption{(Color online) {\bf a)} The hole spectrum of the locking
    crystal, shown in black, observed after deconvolving the raw
    transmission, show in green. The peak labeled ``1'' and shown in
    inset b) at zero frequency corresponds to the carrier beam.  This
    beam gives rise to several holes and anti-holes in a frequency
    range of $\pm 36.9$ MHz; the outermost one is labeled ``2''. This
    illustrates that a modulation frequency of $\omegam = 2\pi\cdot
    50$ MHz is desirable in order that the side hole, labeled ``3''
    and shown in inset c), does not interfere with the hole spectrum
    of the carrier beam. The measured line-widths of the peaks shown
    in insets b) and c) are 60 kHz and 36 kHz for the carrier and
    sideband holes, respectively. The raw spectrum spanning $-10$ MHz
    to 70 MHz was collected in 800 {\micro}s.}
  \label{fig:illustrate_deconvolution}
\end{figure}

In the first part of the long term drift experiments we repeatedly
scanned AOM 1 in the frequency range 170 MHz to 210 MHz, which in
double-pass configuration becomes an 80 MHz scan from 340 MHz to 420
MHz. At the same time, AOM 2 is kept fixed at 350 MHz and the
minus-first-order diffracted beam is sent to the locking cryostat.  We
thus scan the probing laser from $-10$ MHz to 70 MHz, and in this way
we are able to measure the absorption around the carrier hole at zero
frequency and one of the side bands at 50 MHz.  The beam is expanded
to almost cover the entire crystal to probe all active ions and to
obtain a weak intensity. In addition, the scan is fast (800
{\micro}s), and accordingly the probing beam does not affect the hole
structure.  The scan is averaged over approximately 100 runs. After
the locking crystal the beam hits a detector (denoted ``Det 2'' in
Fig.~\ref{fig:exp_setup_characterization}) which is a Thorlabs PDB150A
detector set to a bandwidth of 5 MHz. An example of this scanning
signal is shown in Fig.~\ref{fig:illustrate_deconvolution} (green
trace).  Looking at the insets we clearly see ringing effects caused
by the fast scan. It is possible to deconvolve this ringing back to
the original absorption profile (black trace) by the method described
in \cite{Wolf1994, Chang2005}. The scan speed, together with the
detection bandwidth, allows us do resolve structures as narrow as
$\delta f = \text{scan speed/bandwidth} =
\frac{0.1\:\mathrm{MHz}/\micro\mathrm{s}}{5\:\mathrm{MHz}} = 20$ kHz.
This method allows us to compare the center and side holes
simultaneously, as shown in
Fig.~\ref{fig:CompareCenterSideShape}(a,b).  It should be noted, that
in the calculations in \cite{Wolf1994, Chang2005} the absorption is
assumed to be low ($\alpha_0 L \ll 1$), and hence in our case
corrections may be necessary to the reconstructed absorption. Since we
already have a theory that is not exactly correct when approaching
$\alpha_0 L \approx 1$ we shall not pay further attention to this
fact.

In the second part of the long-term drift experiments we used both the
zeroth- and minus-first-order diffracted beams from AOM 2, and the
scan width was narrower. The zeroth-order beam is used to first burn a
spectral hole in the extra crystal (AOM 2 is turned off while doing
this in order not to disturb the locking crystal) and the absorption
from this hole is subsequently measured several times (with AOM 2
turned on).  At the same time, the minus-first-order beam from AOM 2
measures the absorption profile for the center hole in the locking
crystal. This setup enables us to measure the laser frequency drift
and correlate the drift rate with the shape of the center hole.
Examples of holes read out repeatedly for drift measurements are shown
in Fig.~\ref{fig:Drift_Zigzag_Stable}, the detailed analysis of this
will be discussed in Sec.~\ref{sec:Drift_meas}.

\subsubsection{Phase stability measurement setup}
\label{sec:Phase_stab_meas_setup}
\begin{figure}[t]
  \centering
  \includegraphics{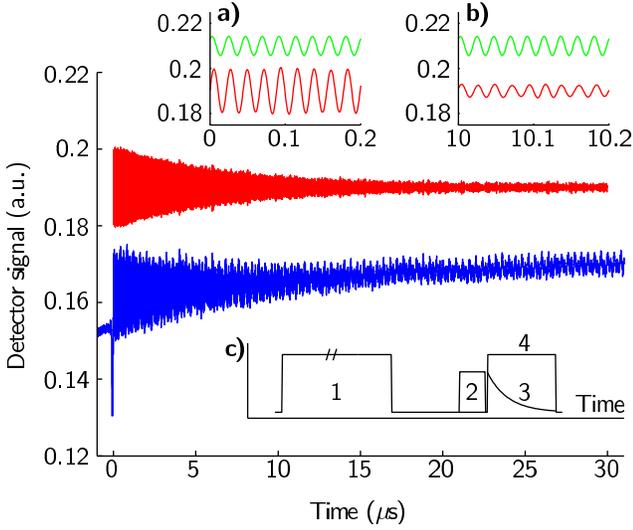}
  \caption{(Color online) Illustration of the phase stability
    measurements. In the main figure the lower (blue) trace shows the
    raw detector signal from the heterodyne FID detection. The upper
    trace (red) shows a filtered version of the lower trace. The lower
    traces (red) in insets {\bf a)} and {\bf b)} show this filtered
    signal in a 200 ns window around time 0 and 10 {\micro}s,
    respectively.  This can be compared to the local oscillator shown
    by the upper traces (green).  We were able to obtain a good
    SNR for the phase difference of these signals.
    Inset {\bf c)} shows the pulse sequence for the experiment: (1) is
    a 10 ms burn pulse scanning between 40 MHz and 50 MHz (relative to
    the AOM double-pass center).  After waiting 100 {\micro}s, pulse
    (2) with a constant frequency of zero and duration 40 {\micro}s
    sets up a coherence in the atoms which leads to the FID at (3).
    Finally, pulse (4) at a frequency of 45 MHz beats with the FID
    signal, leading to the detector signal shown in the main figure
    (lower, blue trace).}
  \label{fig:illustrate_FID}
\end{figure}
In order to measure the laser stability on short timescales we modified
the setup shown in Fig.~\ref{fig:exp_setup_characterization} slightly
such that the zeroth-order diffracted beam from AOM 2 (which is not
used) is sent to the locking crystal with a beam diameter of roughly 2
mm.  AOM 1 is operated around its 200 MHz center frequency and hence,
with the frequency shifted around 400 MHz in double pass configuration,
the probing beam will not interfere with the laser locking system. 

We used optical FID to measure the laser stability and to this end we
programmed the pulse sequence shown in
Fig.~\ref{fig:illustrate_FID}(c) and discussed in the figure caption. 
Referring to this figure, when pulse ``2'' is applied a coherence is
set up in the atomic medium and when pulse ``2'' is turned off the
atoms will keep radiating for a time limited by the optical coherence
time, $T_2$, (which in our case is around 18 {\micro}s) and also by
the inverse bandwidth of the actual coherence. This decaying radiation
``3'' gives a fingerprint of the phase of the laser during pulse
``2''. At the same time, we apply another pulse, ``4'', shifted 45 MHz
in frequency carrying its own phase. The beating of pulses ``3'' and
``4'' hence compares the present phase and the past phase, and this
allows us to calculate the characteristics of the laser, as discussed
in detail in Sec.~\ref{sec:phase_stability}. 

Fig.~\ref{fig:illustrate_FID} illustrates a simple and useful method
\cite{Takeda1982} of obtaining the amplitude and phase of an
oscillating signal, in our case the FID heterodyne signal.  The raw
detector signal is Fourier transformed and filtered in a 40 MHz
bandwidth around the positive 45 MHz component.  The fact that the
negative frequency components are removed leads (after a subsequent
inverse Fourier transform) to a complex representation of the filtered
FID signal in the time domain. The phase and amplitude can be read out
directly from this complex signal.  As an illustration, we show the
filtered version of the FID signal in
Fig.~\ref{fig:illustrate_FID}(a,b) together with a 45 MHz local
oscillator signal derived from the 1 GHz arbitrary waveform generator.
We see that the signals remain in phase over a period of 10 {\micro}s.
This kind of measurement is repeated 100 times, which allows us to
calculate the statistics of the laser stability in a quantitative
manner.

\subsection{Long-term drift}
\label{sec:Drift_meas}

\begin{figure}[t]
  \centering
  \includegraphics{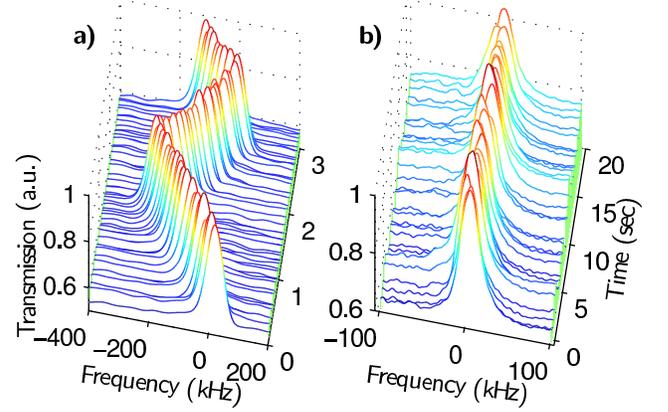}
  \caption{(Color online) Example of drift measurements. {\bf a)}
    Total read-out time is 3 seconds and we see a drift rate of
    roughly 160 kHz/s, while the direction changes occasionally.  {\bf
      b)} Total read out time is 20 s and the drift is 0.3 kHz/s over
    this time.}
  \label{fig:Drift_Zigzag_Stable}
\end{figure}

\begin{figure}[t]
  \centering
  \includegraphics{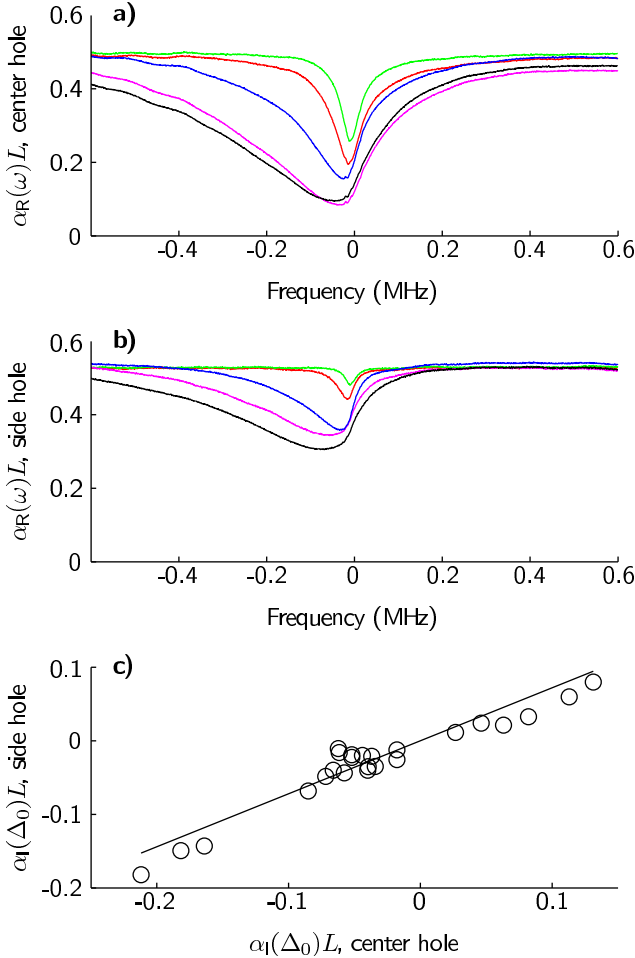}
  \caption{(Color online) In the two upper graphs the measured
    absorption, $\alphaR(\omega)L$, is plotted for comparison under
    different conditions for a number of center holes {\bf (a)} and
    side holes {\bf (b)}. With increasing hole depth, the asymmetry of
    both increases. The imaginary part, $\alphaI(\Delta_0)$, of the
    absorption length at the hole center is a quantitative measure of
    this asymmetry, and can be calculated from $\alphaR(\omega)$ using
    the Kramers-Kr\"onig relations~\eqref{eq:Kramers-Kronig}. In {\bf
      (c)} we see, for several measurements under different
    conditions, a clear linear relationship between this asymmetry for
    the center and side holes.  The straight line is a fit through the
    origin with a slope of 0.72, theoretically we expect a slope of
    unity.}
  \label{fig:CompareCenterSideShape}
\end{figure}

Let us now turn to the experimental results concerning the laser drift
and the hole shapes in the locking crystal. In
Sec.~\ref{sec:laser-drift} we argued that if the spectral holes used
for locking are too deep, the laser may be locked but drifting
linearly in frequency, which in turn will cause the hole shape to be
asymmetric. This is illustrated in
Fig.~\ref{fig:CompareCenterSideShape}(a,b).
From these measured center and side hole shapes, $\alphaR(\omega)L$,
we can calculate the imaginary parts, $\alphaI(\omega)L$, by using the
Kramers-Kr\"onig relations \cite{Jackson1998}.  In our case, these
relations take a slightly simpler form than usual since in
Eq.~\eqref{eq:MB_linear_Fourier} the only $\omega$-dependence is in
the denominator $\frac{\Gammah}{2}+i(\Delta-\omega)$. It can be shown
that:
\begin{equation}
\label{eq:Kramers-Kronig}
  \begin{split}
  \alphaR(\omega_0) &= +\underset{\delta\rightarrow 0}{\lim}
  \frac{1}{\pi}\int_{-\infty}^{\infty}\frac{\alphaI(\omega)(\omega-\omega_0)d\omega}
    {(\omega-\omega_0)^2+\delta^2}, \\
  \alphaI(\omega_0) &= -\underset{\delta\rightarrow 0}{\lim}
  \frac{1}{\pi}\int_{-\infty}^{\infty}\frac{\alphaR(\omega)(\omega-\omega_0)
  d\omega}{(\omega-\omega_0)^2+\delta^2}. 
  \end{split}
\end{equation}
Since $\alphaI(\omega_0)$ is an integral of $\alphaR(\omega)$ times an
odd function in $\omega-\omega_0$, the value of $\alphaI(\Delta_0)$
may be regarded as a convenient measure of the hole asymmetry, which
allows us to quantitatively compare the center and side hole shapes. 
This is shown in Fig.~\ref{fig:CompareCenterSideShape}(c) for a number
of different settings of $T_{\mathrm{10MHz}}$ and
$T_{\mathrm{17MHz}}$. We see a clear proportionality,
$\alphaI^{(\mathrm{side})}(\Delta_0) = 0.72\cdot
\alphaI^{(\mathrm{carrier})}(\Delta_0)$. 

We expect $\alphaI^{(\mathrm{carrier})}(\Delta_0)$ and
$\alphaI^{(\mathrm{side})}(\Delta_0)$ to be equal, since from
Eq.~\eqref{eq:MB_linear_phase} the phase shifts, of the carrier
$\phic$ and sideband $\phis$ are proportional to
$\alphaI^{(\mathrm{carrier})}(\Delta_0)$ and
$\alphaI^{(\mathrm{side})}(\Delta_0)$, respectively, and with a closed
laser stabilization feedback loop we must have zero error signal with
$\phic = \phis$. The reason for the slope not being unity is unknown.
We have thus shown that the laser may drift linearly, while the
feedback loop is still locked. We base this on the direct observations
of the drift, as exemplified in Fig.~\ref{fig:Drift_Zigzag_Stable}(a),
on the proportionality in Fig.~\ref{fig:CompareCenterSideShape}(c),
and on the fact that the electronic error signal is small.
\begin{figure}[t]
  \centering
  \includegraphics{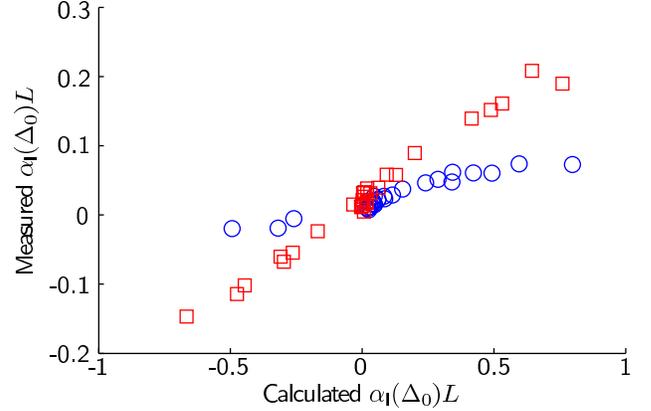}
  \caption{(Color online) Comparison of the measured and calculated
    imaginary center hole absorption depth, $\alphaI(\Delta_0)L$, for
    different settings of the RF pumping times $T_{\mathrm{17MHz}} =
    10 \cdot T_{\mathrm{10MHz}}$.  On the $y$-axis
    $\alphaI(\Delta_0)L$, as inferred from the measured
    $\alphaR(\omega)$ via the Kramers-Kr\"onig relations in
    Eq.~\eqref{eq:Kramers-Kronig} is plotted. On the $x$-axis
    $\alphaI(\Delta_0)$ calculated from Eq.~\eqref{eq:alphaI_drift}
    with the experimentally inferred values of $\alpha_0 L$,
    $\Gammaholec$, $\dholec$, $\Trg = \Tgr = T_{\mathrm{10MHz}}$, and
    $\beta$, while $T_1 = 150$ {\micro}s and $\ber = 0.5$ is plotted. 
    The blue circles and red squares correspond to the data shown in
    Fig.~\ref{fig:Threshold_plots}(a) and
    Fig.~\ref{fig:Threshold_plots}(c), respectively. These two cases
    differ by 1 GHz in the position on the inhomogeneous profile,
    giving background absorptions, $\alpha_0 L$, of 0.44 and 0.66 for
    the blue circles and red squares, respectively.} 
  \label{fig:CompareDriftShape}
\end{figure}

We may also wish to correlate the measured hole asymmetry,
$\alphaI(\Delta_0)$, with measured drift rate, $\beta$, since these
should be related by Eq.~\eqref{eq:alphaI_drift}, at least for small
drift rates when the first-order theory is valid. This equation
assumes a three-level model, and for our real \PrYSO ions we observe
that the $\pm\frac{1}{2},\pm\frac{3}{2} \rightarrow
\pm\frac{1}{2},\pm\frac{3}{2}$ transitions with typical branching
ratios around 0.5 and hole lifetimes, given mostly by
$T_{\mathrm{10MHz}}$, are reasonably representative. Hence, we use
$\Trg = \Tgr = T_{\mathrm{10MHz}}$ and $\ber = 0.5$. Also, from the
experimentally measured hole shape parameters,
$\Gammahole^{(\mathrm{meas})}$ and $\dhole^{(\mathrm{meas})}$, we can
estimate $\Gammahole$ and $\dhole$ from
Eq.~\eqref{eq:relate_gamma_d_meas}.  Inserting these parameters
together with the experimentally determined drift rate, $\beta$, into
Eq.~\eqref{eq:alphaI_drift} we can calculate, to first order, the
value of $\alphaI(\Delta_0)$ for the center hole. This is plotted on
the abscissa in Fig.~\ref{fig:CompareDriftShape}. On the ordinate
$\alphaI(\Delta_0)$ is plotted, inferred by the Kramers-Kr\"onig
relations in Eq.~\eqref{eq:Kramers-Kronig} from the measured
$\alphaR(\omega)$. The result shows a proportionality between these
two and hence we have demonstrated a clear correlation between
measured hole shape asymmetry and measured drift rate as suggested by
Eq.~\eqref{eq:alphaI_drift}. The fact that the constant of
proportionality is not unity is not alarming since we used a
simplified model with only three atomic levels. Also, when the laser
is drifting the first-order theory should not be sufficient.  The
figure shows results from two different experimental runs at two
different positions on the inhomogeneous profile. These have a
different slope, which is not understood. From the results in
Fig.~\ref{fig:CompareDriftShape} we have demonstrated that the
predictions of Eq.~\eqref{eq:alphaI_drift} show the correct order of
magnitude when compared to experiment. It should be borne in mind that
a full understanding of the data requires more than our first-order
calculations.
%
%
%
%
\begin{figure}[t]
  \centering
  \includegraphics{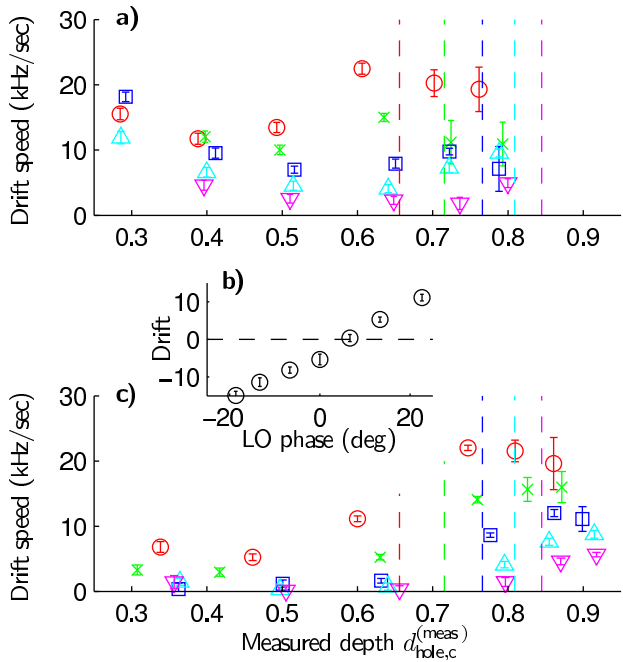}
  \caption{(Color online) Measured drift rates versus measured hole
    depth. The hole shapes are changed by varying $T_{\mathrm{10MHz}}$
    and $T_{\mathrm{17MHz}}$. Red circles, $m = 0.56$; green crosses,
    $m = 0.40$; blue squares, $m = 0.28$; cyan triangles up, $m =
    0.20$, purple triangles down, $m = 0.14$. The vertical dashed
    lines indicate the values of the corresponding thresholds shown in
    Fig.~\ref{fig:thresholds}. The two graphs differ in that the data
    shown in {\bf (a)} were collected before {\bf (c)} and between the
    measurements the laser frequency was moved 1 GHz on the
    inhomogeneous profile, changing the background absorption,
    $\alpha_0 L$, from 0.44 to 0.66. In addition, as shown in inset
    {\bf (b)}, the local oscillator phase (relative to the value used
    in {\bf (a)}) to demodulate the error signal, was adjusted in
    order to minimize the drift rate for shallow holes.}
  \label{fig:Threshold_plots}
\end{figure}

Let us now turn to the measured drift rate, $\beta$, for various
parameter settings. The results of these measurements are shown in
Fig.~\ref{fig:Threshold_plots}. The light intensity is kept constant
with the saturation parameter $s_0 \approx 0.09$. The modulation
index, $m$, has the values 0.14, 0.20, 0.28, 0.40, and 0.56, and each
color in the figure corresponds to one of these values. The hole
shapes are controlled by employing $T_{\mathrm{10MHz}} =
\frac{1}{10}\cdot T_{\mathrm{17MHz}}$ at the six values 2 ms, 4 ms, 8
ms, 20 ms, 40 ms, and 80 ms. In the figure the six corresponding data
points are plotted for each color from left to right since the hole
depth $\dholec^{(\mathrm{meas})}$ increases with $T_{\mathrm{10MHz}}$. 

The results in Fig.~\ref{fig:Threshold_plots} are divided into two
parts. In Fig.~\ref{fig:Threshold_plots}(a) we see that the drift
rate is lowest for hole depths around 0.5, corresponding to
$T_{\mathrm{10MHz}}= 8$ ms. For shorter timescales (towards the left)
the drift rate tends to increase and this could be caused, for
example, by offset errors in the electronics, or by the small signal
from the inhomogeneous background if the phase of the local oscillator
is set incorrectly. Note, that if the error signal has an offset of
$10^{-3}$ times the full-scale value, we would expect the laser to be
displaced by $10^{-3}\Gammahole$ in frequency. With a hole lifetime of
$\Trg$ we can then estimate the drift as $\beta \approx
10^{-3}\Gammahole/\Trg$. For our shortest timescales $\Trg \approx 2$
ms, we have $\Gammahole/2\pi \approx 40$ kHz leading to an estimate of
$\beta/2\pi \approx 20$ kHz/s. This is not far from the experimental
values, which we find to be typically a little less than 20 kHz/s and
we thus concluded that our relative offset errors are smaller than
$10^{-3}$ for the data shown in Fig.~\ref{fig:Threshold_plots}(a). For
longer timescales (towards the right) we can also see an increase in
drift rate, which can be explained by the fact that we are
approaching the maximum hole depth, discussed around
Eq.~\eqref{eq:drift_criterion} and Fig.~\ref{fig:thresholds}. For each
color in Fig.~\ref{fig:Threshold_plots} this threshold is shown as a
vertical dashed line. 

The above conclusions become more apparent when we consider
Fig.~\ref{fig:Threshold_plots}(c). Before collecting these data we
noted that the drift rate for a shallow hole depends on the position
on the inhomogeneous profile, which was then adjusted by 1 GHz.  In
addition, as shown in the figure inset (b), adjusting the local
oscillator phase slightly also has an impact on the drift rate, which
we fine tuned to a low value. Fig.~\ref{fig:Threshold_plots}(c) shows
low drift rates in the left part of the figure (the lowest measured
being below 0.5 kHz/s). However, the higher drift rates in the
right-hand part of the figure remain almost unchanged. A slight change
in error signal offset cannot change the fact that zero drift is an
unstable solution if the criterion in Eq.~\eqref{eq:drift_criterion}
is not met. The increase in drift rate on the right-hand side of
Fig.~\ref{fig:Threshold_plots}(c) is now consistent with the vertical
lines representing the threshold.  The highest drift rate is still
below 25 kHz/s which is fairly good. 

A very rough order of magnitude estimate of the drift rate can be
obtained by noting the fact that the calculations in
Sec.~\ref{sec:laser-drift} rely on the series expansion in the
dimensionless parameter, $\xi = \frac{\beta\Trg}{\Gammahole}$. To
calculate the threshold condition for $\beta = 0$ being a stable
solution it is sufficient to assume that $\xi \ll 1$. However, to
calculate the actual drift rate when $\beta = 0$ is unstable requires
higher order theory, where $\xi$ cannot be small.  Thus, if we then
take for our right-most data points the typical values $\Trg = 80$ ms,
$\Gammahole/2\pi \approx 100$ kHz, and $\beta/2\pi = 25$ kHz/s, we
obtain $\xi \approx 0.02$, which is actually quite low compared to
unity.  This is an indication that the fraction of ions resonant on
the $\pm\frac{5}{2}\rightarrow\pm\frac{5}{2}$ transition with ten-fold
longer hole lifetime, $\Trg = T_{\mathrm{17MHz}}$, and ten-fold higher
$\xi$ plays a role in limiting the measured drift rate to a low
value.  Without having completely searched the very large
multi-dimensional parameter space, our experience tells us that
$T_{\mathrm{10MHz}} = \frac{1}{10}\cdot T_{\mathrm{17MHz}}$ is a good
choice.  We have observed examples of much higher drift rates than 25
kHz/s, see e.g.~Fig.~\ref{fig:Drift_Zigzag_Stable}(a) where, in
addition to a high drift rate of 160 kHz/s, we also saw that the
direction changed occasionally.

\subsection{Short-term phase stability}
\label{sec:phase_stability}
\begin{figure}[t]
  \centering
  \includegraphics{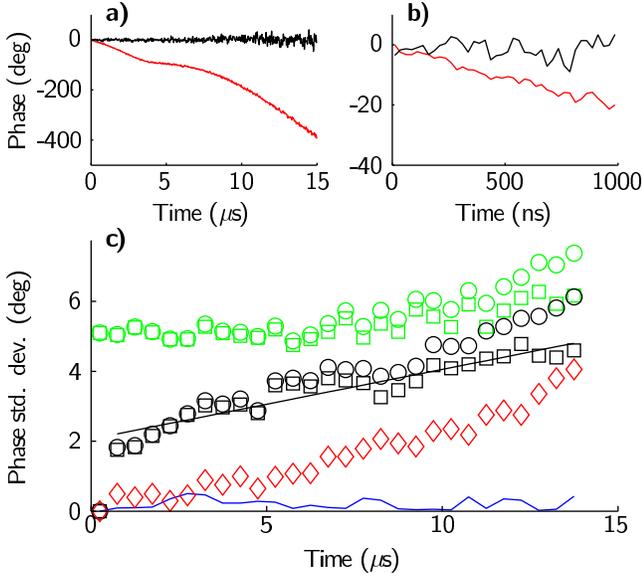}
  \caption{(Color online) {\bf (a)} A typical single-shot example of
    the measured FID phase with our stabilization system on (black)
    and off (red). {\bf (b)} An enlargement of the first $\micro$s of
    the plot shown in (a). {\bf (c)} Phase standard deviation versus
    time, collected in time bins of 500 ns.  Red diamonds represent the
    experimental uncertainty in the measurements, given by
    Eq.~\eqref{eq:phase_stab_meas_uncertainty}.  Green circles show
    the shot-to-shot variations, $\std(x_i)$. The black circles
    describe the variation of the phase with respect to the phase of
    the first time bin, $\std(x_i - x_1)$.  The green and black
    squares show the results when the measurement noise has been
    subtracted. The blue line shows the absolute value of the mean
    phase compared to the first time bin.  The black line is a linear
    fit to the black squares, giving $\delta\phi[\mathrm{deg}] = 2.0 +
    2.0\cdot 10^5 t[\mathrm{s}]$.}
  \label{fig:exp_phase_stab}
\end{figure}
Fig.~\ref{fig:exp_phase_stab}(a,b) shows the phase evolution of the
filtered FID signal for the case when the laser stabilization system
is on (black) and off (red). In the latter case, the dye laser is
stabilized to a reference cavity by its commercial feedback system.
The clear effect of our locking system can be seen, and below we
analyze this phase evolution quantitatively.

We note from Fig.~\ref{fig:illustrate_FID} that our method based on
detection of the FID signal works in practice for times up to around
15 {\micro}s.  The SNR when determining the phase
will decrease with increasing time. Also, referring to
Fig.~\ref{fig:illustrate_FID}(c), the variation in the phase of the
laser is present in both the FID signal ``3'' and in the heterodyning
pulse ``4''. In order to derive the actual phase stability of the
laser we repeated the measurements shown in black in
Fig.~\ref{fig:exp_phase_stab}(a) 100 times, divided the measured phase
into time bins of length 500 ns, and performed statistical operations
on these measured phases. Note, that when discussing the phase of the
FID signal we always mean the phase difference between the reference
signal and the FID signal (shown in green and red in
Fig.~\ref{fig:illustrate_FID}(a,b), respectively).

In the following, $x_{i,j}$ denotes the measured phase in time bin $i$
for repetition $j = 1,\ldots, 100$. For each time bin $i$ we may
calculate the mean value $\bar{x}_i = \frac{1}{n}\sum_j x_{i,j}$ and
variance $\var(x_i) = \frac{1}{n-1}\sum_j(x_{i,j} -\bar{x}_i)^2$,
where $n=100$. In Fig.~\ref{fig:exp_phase_stab}(c) $\abs(\bar{x}_i)$
is shown by the blue line and $\std(x_i) = \sqrt{\var(x_i)}$ by the
green circles. We see that the average phase deviates less than 1
degree over time, which simply means that there is no phase drift in
one particular direction. The $\std(x_i)$ values describe the
repeatability of the experiment, which is within roughly 5 degrees and
increases with time. To characterize the measurement noise we now
assume a model where the measured values can be written $x_{i,j} =
\phi_{i,j} + n_{i,j}$, where $\phi_{i,j}$ is the actual phase of the
FID signal and $n_{i,j}$ is a stochastic variable describing the noise
in the $i$'th time bin and $j$'th measurement.  Since technical laser
noise in general is very limited at the heterodyne beating frequency
45 MHz, we assume the noise level to be shot noise or electronic noise
in the detector. This is broadband (white) noise and we may assume all
the individual $n_{i,j}$ are independent. The mean value is assumed to
be equal to zero, and the standard deviation $\std(n_i)$ will depend
on $i$ due to the decaying FID signal.  In order to estimate the
standard deviation of $n_{i,j}$ we calculate the variance of adjacent
time bins for our experimental data:
\begin{equation}
\label{eq:phase_stab_meas_uncertainty}
  \begin{split}
  \var(x_i - x_{i-1}) &= \var(\phi_i - \phi_{i-1}) + \var(n_i-n_{i-1}) \\
      &\approx \mathrm{const} + 2\cdot\var(n_i). 
  \end{split}
\end{equation}
In the second step we use the fact that $\var(\phi_i - \phi_{i-1})$
must be time independent since the laser phase is in a steady-state
condition.  We also use the approximation $\var(n_i -n_{i-1}) =
\var(n_i) + \var(n_{i+1}) \approx 2\var(n_i)$, since we assume the
detection noise to vary slowly over time. Plotting the variance
$\var(x_i - x_{i-1})$ versus time (i.e.~plotting
Eq.~\eqref{eq:phase_stab_meas_uncertainty} as a function of $i$) will
give the measurement noise, $\var(n_i)$, and the constant,
$\var(\phi_i - \phi_{i-1})$. In Fig.~\ref{fig:exp_phase_stab}(c)
$\std(n_i) = \sqrt{\var(n_i)}$ is shown by red diamonds, and is seen
to increase from zero to less than 4 degrees at long times.

This noise level can now be used to correct the shot-to-shot variance,
shown by the green circles, since $\var(x_i) = \var(\phi_i) +
\var(n_i)$.  Subtracting the measurement noise, the green squares show
the standard deviation, $\std(\phi_i)$, of the FID signal itself.

When characterizing laser performance it is more interesting to
measure the phase evolution over time, i.e.~how much the phase of the
$i$'th time bin, $\phi_i$, deviates from the first one, $\phi_1$.  To
this end we first calculate $\var(x_i - x_1) = \var(\phi_i -\phi_1) +
\var(n_i - n_1)$ from the experimental data. The square root of this
is $\std(x_i-x_1)$, which is plotted in
Fig.~\ref{fig:exp_phase_stab}(c) as black circles. Since $\var(n_i -
n_1) = \var(n_i) + \var(n_1) = \var(n_i)$, we can subtract the
measurement noise again to obtain $\std(\phi_i - \phi_1)$, which is
plotted as black squares. We see that on short timescales the phase
error is around 2 degrees, and it increases linearly with time.  The
black line in the figure is a linear fit to the black squares, and it
can be used to convert the phase stability into a characteristic
frequency stability as a function of time. Since the relation between
phase and frequency is $\phi[\mathrm{deg}] = 360\cdot
f[\mathrm{Hz}]\cdot t + \phi_0$, the standard deviations of these will
be connected by:
\begin{equation}
\label{eq:exp_linewidth_short_time}
  \std(f) = \frac{\std(\phi)}{360\cdot t} 
          = \frac{2.0 + 2.0\cdot 10^5 t}{360\cdot t} 
\end{equation}
In the second step, we inserted the experimental linear fit shown as
the black line in Fig.~\ref{fig:exp_phase_stab}(c). For large $t$ the
characteristic frequency standard deviation, $\std(f)$, approaches 0.6
kHz, but we are not allowed to extend $t$ further than the measured 14
{\micro}s.  Note, for $t=10$ {\micro}s, $\std(f) = 1.1$ kHz.

The difference between the shot-to-shot standard deviation, plotted in
green in Fig.~\ref{fig:exp_phase_stab}(c), and the time separation
standard deviation, shown in black in
Fig.~\ref{fig:exp_phase_stab}(c), is approximately 3 degrees. The
extra noise in the shot-to-shot data arises from the fact that the
laser phase also fluctuates during the setup of the coherence in the
atoms (pulse ``2'' in Fig.~\ref{fig:illustrate_FID}). The practical
phase memory time is approximately 10 {\micro}s, corresponding to the
decay time of the FID signal in Fig.~\ref{fig:illustrate_FID}.
Averaged over 10 {\micro}s, the black data points of
Fig.~\ref{fig:exp_phase_stab} actually suggest approximately 3 degrees
fluctuation in the laser phase. Thus the data shown in green and black
in Fig.~\ref{fig:exp_phase_stab} are consistent with each other.

The experimental data shown in Fig.~\ref{fig:exp_phase_stab}(c) were
obtained with $T_{\mathrm{10MHz}} = \frac{1}{10}\cdot
T_{\mathrm{17MHz}} = 4$ ms, modulation index $m = 0.20$, and
saturation parameter $s_0 \approx 0.09$.  This setting is also
represented in Fig.~\ref{fig:Threshold_plots}(c) giving a drift of
$(0.34\pm 0.76)$ kHz/s. Hence, we have shown that a slow drift and low
phase variations can be obtained at the same time. Of the settings
investigated the best general phase stability performance is found for
the above RF-eraser time, which gives rise to a measured hole depth,
$\dholec^{(\mathrm{meas})}$, around 0.5 to 0.6 in
Fig.~\ref{fig:Threshold_plots}. This hole depth corresponds to $x =
\frac{\Gammahole}{\Gammah}\approx 2$, which is the best choice
according to the $f$-function~\eqref{eq:f_function} shown in
Fig.~\ref{fig:f_function}. At other settings we still find asymptotic
values of Eq.~\eqref{eq:exp_linewidth_short_time} less than 2.5 kHz,
which is only four times worse than the value of 0.6 kHz in the
example above.

We have also investigated a few different positions on the
inhomogeneous profile. These measurements indicate that a transmission
above 30\% is a good choice. Varying in the local oscillator phase
between the values shown in Fig.~\ref{fig:Threshold_plots}(b) does not
change the performance on short timescales.

The performance of the non-stabilized laser, i.e.~when the dye laser
is operated only with its built-in stabilization unit, is exemplified
in Fig.~\ref{fig:exp_phase_stab}(a,b). We see that in this case it
takes roughly 13 {\micro}s to change the phase by $360^{\circ}$, which
can be translated into a characteristic frequency of 75 kHz.
Repeating this measurement 100 times shows that these values are
typical. In our experience, cooling the laser dye affects this
frequency considerably (we typically use temperatures in the range
8\degreecelsius\: to 12\degreecelsius).

The above observation does not mean that the unstabilized laser
line-width is 75 kHz, it only means that on short timescales of the
order of 10 {\micro}s, the frequency deviations are around 75 kHz. In
\cite{Zhu1993} a typical dye laser line-width of 450 kHz is reported,
but it was also indicated that the frequency deviations had a
bandwidth $B \ll 190$ kHz, i.e.~the deviations occurred on a timescale
much longer than 5 {\micro}s. Our observations are consistent with
this behavior.

As discussed in Sec.~\ref{sec:system-design}, the group delay of our
stabilization feedback loop is approximately 100 ns. To understand how
this limits the phase stability of our laser we assumed that the dye
laser introduces a frequency error of 75 kHz. It takes 100 ns for the
stabilization loop to discover this and start counteracting it, and
after this time the phase has evolved roughly $360^{\circ}\cdot
75\:\mathrm{kHz}\cdot 100\:\mathrm{ns} = 2.7^{\circ}$. This
hand-waving argument is consistent with a standard deviation of the
order of 2 degrees, as shown by the data (black squares) in
Fig.~\ref{fig:exp_phase_stab}(c) as the limiting case for short
timescales. Another way of saying this is that in
Fig.~\ref{fig:exp_phase_stab}(b) the typical stabilized phase
deviation (shown in black) corresponds to the actual deviation of
the non-stabilized laser (shown in red) after a time of the order of
the delay time in the feedback loop.

\subsection{Overall laser performance}
\label{sec:general_laser_performance}
In summary, we can say that we achieved 1 kHz line-widths on the
timescale of 10 {\micro}s, and drift stabilities also around 1 kHz on
1 s timescales. We were not able to thoroughly examine the stability
on intermediate timescales. If we had had two identical lasers with
which to calculate the Allen deviation from beating experiments, our
reported line widths should be multiplied by $\sqrt{2}$ for short
timescales (with random shifts) and by 2 for long timescales (with
linear drift). Then we can compare our laser performance with others
reported in the literature.

Laser stabilization using a spectral hole has been reported for
semiconductor lasers in several publications \cite{Pryde2001,
  Pryde2002, Strickland2000, Sellin2001, Bottger2001, Sellin1999,
  Bottger2003}. In these cases the reported Allen deviation on a 10
{\micro}s timescale is typically 10 kHz, and around 3 kHz in
\cite{Strickland2000}.  For long timescales a linear drift rate of 10
kHz/s has been observed in pure transient systems
\cite{Strickland2000}. Stabilities better than our reported 1 kHz/s
can be obtained with permanent spectral holes (25 kHz/min)
\cite{Sellin1999}, ($\approx 1$ kHz/min) \cite{Bottger2003} or by
incorporating the permanent reference consisting of the inhomogeneous
profile (7 kHz/min) \cite{Strickland2000}, (1.4 kHz/min, no hole
burning) \cite{Bottger2007}.

The most important steps in obtaining our short-term phase stability
of down to 2 degrees, consist of (1) constructing fast analog
electronics to obtain short delays and low noise, (2) using a large
sample crystal diameter (19 mm) in order to be able to use a high
light power without too high an intensity, and (3) cooling the laser
dye in order to improve the passive stability. We should also mention
that the optical table is placed on legs with pneumatic vibration
dampers, that an improved dye jet nozzle (Radiant Dyes RDSN 02) and an
improved nozzle holder and pump mirror adjustment unit (Radiant Dyes
RDU 10) have been installed, and that the entire laser system is
located inside a clean room (from Terra Universal).  In order to
obtain the low frequency drift of 1 kHz/s the important steps are: (1)
being aware of the drift problem discussed in
Secs.~\ref{sec:laser-drift} and~\ref{sec:Drift_meas} and (2) carefully
constructing the electronics so as to ensure low offsets.

\subsubsection{Impacts on quantum gate experiments}
\label{sec:Impact_quant_exp}
As we discussed in the introduction in Sec.~\ref{sec:introduction},
our main reason for frequency stabilizing the laser is to enable
quantum computation experiments. One experimental approach utilizes
spectral hole burning to isolate strongly interacting qubits, as
described in \cite{Nilsson2004, Rippe2005}.  Such qubit structures
will typically have a spectral width of $\approx 200$ kHz, and can be
prepared by several hundred cycles of optical pumping. The entire
preparation process may take hundreds of milliseconds, and it is
important that the laser drift is much less than 200 kHz during this
time. With typical drift rates of 1 kHz/s we have fulfilled this
requirement.

Let us now consider the effects of laser phase fluctuations on the
fidelity of qubit operations. We use two ground state hyperfine levels
to represent a qubit, e.g.~$\ket{0} = \ket{\pm\frac{1}{2}}$ and
$\ket{1} = \ket{\pm\frac{3}{2}}$. Single qubit operations can be
performed by interaction with an excited state $\ket{e}$
\cite{Roos2004} where $\pi$-pulses are used to transfer the population
from $\ket{0}$ or $\ket{1}$ to $\ket{e}$ and back again.  Such basic
operations are also building blocks for more complicated gates,
e.g.~the CNOT gate \cite{Ohlsson2002, Roos2004}. Here we will estimate
the impact of a phase error in the laser on such a simple operation.

Consider the following scenario. The qubit is in an initial state
$\ket{\psi_0} = \alpha\ket{0}+\beta\ket{1}$, then a perfect laser
drives a $\pi$-pulse on the $\ket{1}\rightarrow\ket{e}$ transition
such that $\ket{\psi_0} \rightarrow \alpha\ket{0}+\beta\ket{e}$. The
laser phase is now assumed to change by an amount $\phi$, and we
finally drive the excited state back to the final state $\psif =
\alpha\ket{0}+\beta e^{i\phi}\ket{1}$. We have thus gathered all the
phase fluctuations in a single step. If there is no phase change
($\phi = 0$) we arrive back at the initial state $\ket{\psi_0}$,
and the deviation from this can be used to estimate the fidelity of
the operation given $\phi$.

However, with an unknown random phase change, $\phi$, the correct
description of the final state is given by the density matrix:
\begin{equation}
  \begin{split}
   \densf &= \int\ketbra{\psif}{\psif} f(\phi)  d\phi 
         = |\alpha|^2\ketbra{0}{0} + |\beta|^2\ketbra{1}{1} \\
    &\quad + \int\left(\alpha\beta^* e^{-i\phi}\ketbra{0}{1}
          +\alpha^*\beta e^{i\phi}\ketbra{1}{0}\right)f(\phi)d\phi,
  \end{split}
\end{equation}
where $f(\phi)$ is the probability distribution of the phase change,
$\phi$, where $\int f(\phi)d\phi = 1$. Given an initial state
$\ket{\phi_0}$ the fidelity of the operation is the overlap:
\begin{equation}
  \label{eq:fidelity_vs_phase}
  \begin{split}
  F &= \bra{\psi_0}\densf\ket{\phi_0} = |\alpha|^4 + |\beta|^4 
   + 2|\alpha|^2|\beta|^2\negthickspace
      \int\negthickspace f(\phi) \cos\phi d\phi \\
    &= 1 - 2|\alpha|^2|\beta|^2\epsilon
    \qquad\text{with }
    \epsilon = 1 - \int\negthickspace f(\phi) \cos\phi d\phi. 
  \end{split}
\end{equation}
The parameter $\epsilon$ is zero when there is no phase change
($f(\phi) = \delta(\phi)$) and varies to second order in the phase
deviation. To see this clearly we assume a simple top-hat distribution
function, $f(\phi) = \frac{1}{2\sqrt{3}\phi_0}$ if $-\sqrt{3}\phi_0 <
\phi < \sqrt{3}\phi_0$ and zero elsewhere. This distribution function
corresponds to a root mean square phase $\phi_0$. Inserting this in
Eq.~\eqref{eq:fidelity_vs_phase} gives $\epsilon =
\frac{1}{2}\phi_0^2$ and $F = 1-|\alpha|^2|\beta|^2\phi_0^2$ when
$\phi_0 \ll 1$.

The phase fluctuation plays no role if the initial state is either of
the basis states, $\ket{0}$ or $\ket{1}$. The impact is maximal if the
initial state is an even superposition with $|\alpha|^2 = |\beta|^2 =
\frac{1}{2}$. In the latter case, for a phase error with a standard
deviation as high as 10 degrees, the fidelity is 98.5\%.

More detailed analysis is naturally required to accurately calculate
the fidelity loss of entire quantum gate experiments given our laser
stability, which is outside the scope of this paper. However, with the
above investigations, we are confident that the laser stability is
sufficient for our intended quantum gate experiments.



\section{Outlook}
\label{sec:outlook}
As discussed in Sec.~\ref{sec:Impact_quant_exp}, we have a laser in a
sufficiently stable condition for quantum information experiments,
which was the main reason for the entire stabilization project.
Compared to previous laser stabilization studies on pure transient
hole-burning systems, we have achieved better stability, and the
understanding of laser drift was very important in this process.
However, some compromises had to be made since the absence of linear
drift does not necessarily coincide with the maximization of error
signals. The presence of both the center and side holes is responsible
for this effect, and a number of measures can be taken. (1) An
interferometric setup can be considered where only the beam \emph{not}
passing the atoms is modulated in the Pound-Drever-Hall scheme. This,
however, requires a very stable or an otherwise stabilized
interferometer, since the error signal from the spectral hole locking
can not conveniently distinguish whether phase changes occur relative
to the atoms or because of instabilities in the interferometer. (2)
Moving the modulation side bands outside the inhomogeneous profile.
This is inconvenient in our case with \PrYSO since the inhomogeneous
broadening is around 5 GHz. However, the system under consideration in
\cite{Bottger2007} is interesting in this respect. Note, that the
drift problem decreases when $\omegam/\Gammainh$ approaches unity, so
a combination of the inhomogeneous profile for permanent stability
together with hole burning effects for short-term stability may be
possible without suffering too strong constraints from the drift
criterion when $\omegam\approx\Gammainh$.

In any case, our work has also shown that offset levels and systematic
effects must be carefully controlled to achieve low drift rates in
pure transient systems.  For this reason, the incorporation of
permanent effects seems to be the most convenient and competitive
solution for obtaining a slow long-term drift. It is our hope that the
work presented here will help to further improve the already advanced
field of laser stabilization.


\section{Conclusions}
\label{sec:conclusion}
We have for the first time stabilized a dye laser to a spectral hole
in \PrYSO, obtaining 1 kHz frequency stability on a 10 {\micro}s
timescale, and a long-term drift rate below 1 kHz/s. The use of RF
fields to control the hyperfine level populations allowed us to
optimize the spectral hole parameters for best laser performance. The
stabilities obtained were sufficient for high-fidelity quantum
information experiments.

We have contributed to the theoretical understanding of laser
stabilization using spectral holes to an extent that we hope will
enable other scientists to further improve existing technology.
Although a few experimental observations reported in this paper are
not fully understood in a quantitative manner, we have provided strong
experimental support to the theory describing laser drift.


\section*{Acknowledgments}
We are grateful to Mike Jefferson and Pete Sellin for sharing their
detailed knowledge on laser stabilization.  We also wish to thank
Krishna Rupavatharam for introducing the coherent readout technique
shown in Fig.~\ref{fig:illustrate_deconvolution}. This work was
supported by the European Commission through the ESQUIRE project and
the integrated project QAP under the IST directorate, by the Knut and
Alice Wallenberg Foundation, and the Swedish Research Council.
B.~Julsgaard is supported by the Carlsberg Foundation.


\bibliographystyle{prsty}
\bibliography{bibfile} 

\end{document}